%% file: main.tex
\documentclass[12pt,a4paper]{article}
\usepackage[utf8]{inputenc}
\usepackage{scalerel,stackengine,amsmath}
\usepackage{amsfonts}
\usepackage{amsmath}
\usepackage{amssymb}
\usepackage{hyphenat}
\usepackage{enumitem}
\usepackage{dsfont}
\usepackage[left=2cm,right=2cm,top=2.5cm,bottom=2.5cm]{geometry}
\usepackage{cite}
\usepackage{bbold}
\usepackage{slashed}
\usepackage{hyperref}
\usepackage{graphicx}
\usepackage{caption}
\usepackage{float}
\usepackage{subcaption}
\usepackage{soul}
\usepackage[labelformat=simple]{subcaption}
\usepackage{xcolor}
\usepackage{framed}
\usepackage[english]{babel}
\newtheorem{theorem}{Theorem}[section]
\newtheorem{corollary}{Corollary}[theorem]

\colorlet{shadecolor}{cyan!10} 

\usepackage{authblk}
\hypersetup{
	unicode=true,          
	pdftoolbar=true,        
	pdfmenubar=true,        
	pdffitwindow=false,     
	pdfstartview={FitH},    
	pdftitle={DoFsLowScaleLepto},    
	pdfauthor={},     
	pdfsubject={},   
	pdfcreator={},   
	pdfproducer={Producer}, 
    pdfkeywords={Resonant Leptogenesis} {Baryon Asymmetry} {Transport Equations} {Degrees of Freedom},
    pdfnewwindow=true,      
	colorlinks=true,       
	linkcolor=blue,          
	citecolor=red,        
	filecolor=black,      
	urlcolor=blue           
}

\include{macros}

\renewcommand{\thesubsection}{\arabic{section}.\arabic{subsection}}
 \renewcommand{\theequation}{\arabic{section}.\arabic{equation}}

\usepackage{indentfirst}

\definecolor{ao}{rgb}{0.0, 0.0, 1.0}
\newenvironment{bl}[1]{{\color{ao}  #1}}

\newcommand*{\email}[1]{%
	\footnotesize\href{mailto:#1}{\bl{#1}}
}

\title{\textbf{Trajectories of Critical Unstable Qubits\\[1mm] {\em in} and {\em on} the Bloch Sphere}}

\author[]{Snehit Panghal$\,$\footnote{\email{snehit.panghal@postgrad.manchester.ac.uk}} } 
\author{Apostolos Pilaftsis$\,$\footnote{\email{apostolos.pilaftsis@manchester.ac.uk}} }

\affil[]{
\normalsize\textit{Department
 of Physics and Astronomy, University of Manchester,\newline Manchester, M13 9PL, United Kingdom}
}

\date{\empty}

\begin{document}
\numberwithin{equation}{section}
\setcounter{page}{1}

{\let\newpage\relax\maketitle}
\maketitle

\flushbottom
\vspace{-1cm}
\begin{abstract}
    
\noindent 
We extend previous studies on a novel class of unstable two-level systems which were called Critical Unstable Qubits (CUQs). In an appropriately defined co-decaying frame, the CUQs exhibit striking phenomena of indefinite anharmonic oscillations between two states and coherence-decoherence oscillations of mixed states. These features are distinct from the usual Rabi oscillations observed in the Hermitian counterpart of two-level systems, which are harmonic and preserve the coherence of the quantum state. We~employ the density matrix formalism to study these phenomena for mixed states and delve into the nature of the trajectory traversed by these states in the Bloch sphere by studying the time evolution of the Bloch vector that describes the quantum state of the unstable~qubit.
In particular, we provide for the first time explicit geometric constructions to obtain trajectories of both pure and mixed CUQs {\em in} and {\em on} the Bloch sphere. This enables us to identify the stationary points of CUQs, at which the states do not evolve in time in the co-decaying frame. The potential implications of our findings for particle cosmology and quantum simulations of non-Hermitian Hamiltonians are discussed. 

\end{abstract}
\begin{tabbing}
~~~~~~~{\small {\sc Keywords:}} \= {\small Critical Unstable Qubit, Non-Hermitian Hamiltonian, PT symmetry,} \\[-1mm]
\> {\small Exceptional Point}
\end{tabbing}

\newpage
\tableofcontents
\newpage

\section{Introduction}\label{sec:Intro}

The Standard Model successfully describes the dynamics of all fundamental particles observed in collider experiments. Barring electrons, electron neutrinos, and protons, all fundamental particles and their composites (mesons and baryons) are unstable particles with a short lifetime. Of these composites, neutral particles are an interesting class, as we observe oscillations between the neutral particle and its antiparticle as they pair up, forming a qubit. An example of these pairs is the neutral $K$, $B$, and $D$ mesons\cite{ParticleDataGroup:2024cfk}. The neutral particle-antiparticle oscillations for mesons are a mix of particle oscillation and particle decay. Such phenomena can be understood within the framework of unstable two-level oscillating systems.

For stable two-level particle systems, the oscillations between particles or states are commonly known as Rabi oscillations\cite{Rabi:1937dgo}. However, as mentioned above, the oscillations of an unstable two-level system differ from the usual Rabi oscillations, because the former decay with time. The Hamiltonian of an unstable two-level system or {\em unstable qubit} can be effectively described in the Weisskopf-Wigner (WW) approximation\cite{Weisskopf1930} by a non-Hermitian Hamiltonian
\begin{equation}
   \label{non_hermitian Hamiltonian}
    \mathrm{H}_{\text{eff}}\: =\: \mathrm{E}-\dfrac{i}{2}\Gamma\, .
\end{equation}
In fact, the WW approximation provides an effective approach to describe particle decay. Then, Lee, Oehme, and Yang\cite{Lee:1957qq} were the first to take this approach into account to study CP violation in the neutral $K$-meson system. It can be used to model the dynamics of decays with  particle mixing, which could take place in the SM or in theories of new physics, resulting in a resonantly enhanced phenomenon of CP violation, also known as Resonant CP violation~\cite{Pilaftsis:1997dr}.

In this paper, we extend previous studies of a special class of two-level quantum systems described by the effective (non-Hermitian) Hamiltonian $\mathrm{H}_{\text{eff}}$ in~\eqref{non_hermitian Hamiltonian}, which were called Critical Unstable Qubits (CUQs)\cite{Karamitros:2022oew,Karamitros:2025azy}. 
After first decomposing the two-by-two matrices~$\text{E}$ and~$\Gamma$ as
\begin{equation}
    \label{eq:defEGamma}
        \mathrm{E}\: =\: \mathrm{E}_\mu \sigma^\mu\,,\qquad 
        \Gamma\: =\: \Gamma_\mu \sigma^\mu\,,
\end{equation}
where $\sigma^\mu = (\mathbb{1},\boldsymbol{\sigma})$ is the Pauli-matrix four-vector, and $\mathrm{E}^{\mu} \equiv ({\rm E}_0, {\bf E})$ and $\mathrm{\Gamma}^{\mu} \equiv (\Gamma_0, {\bf \Gamma})$ are real four-vectors, the effective Hamiltonian $\mathrm{H}_{\text{eff}}$ can be rewritten as
\begin{equation}
    \mathrm{H}_{\text{eff}}\: =\: \left(\mathrm{E}_0-\dfrac{i}{2}\Gamma_0\right)\mathbb{1}_2\, -\, \left(\mathbf{E}-\dfrac{i}{2}\boldsymbol{\Gamma}\right)\!\cdot \boldsymbol{\sigma}\; .
    \label{nonHerm_Hamiltonian_Paulibasis}
\end{equation}
The CUQ is a specific case of this Hamiltonian, where 
\begin{equation}
    \text{(i)}\quad\! \mathbf{E}\perp\boldsymbol{\Gamma}  \qquad \text{and }\qquad \text{(ii)}\quad\! r\:\equiv\: \dfrac{|\boldsymbol{\Gamma}|}{2|\mathbf{E}|}\ <\ 1\,.\label{CUQcondition}
\end{equation}
If we work in what will be defined as a $\textit{co-decaying}$ frame~\cite{Karamitros:2022oew} in Section~\ref{sec:MixStatesolgen}, then the CUQ will oscillate between the two quantum states indefinitely, which will be the analog of Rabi oscillations for a {\em stable} qubit described by a Hermitian Hamiltonian.
The extremal scenario of a CUQ in the limit $r\to 1$ in \eqref{CUQcondition} was studied originally in~\cite{Pilaftsis:1997dr} and was termed at the time
an {\em anomalously degenerate} two-level quantum system. This extremal CUQ scenario corresponds to a Jordan-form Hamiltonian whose complex energy eigenvalues are exactly degenerate, forming an {\em exceptional point} in the parameter space, as termed later in the literature~\cite{Heiss:2012dx}.

CUQ oscillations between two quantum states are not simply the non-Hermitian equivalent of Rabi oscillations in Hermitian systems, but they have two important features that distinguish them.
\begin{enumerate}
    
    \item[(i)] The periodic oscillations for pure state CUQs are anharmonic in the aforementioned co-decaying frame, whereas Rabi oscillations are harmonic.
    
    \item[(ii)] The mixed state CUQs, along with anharmonic periodic oscillations, exhibit periodic coherence-decoherence oscillations. In contrast, the coherence of states is maintained for mixed state Rabi oscillations.

\end{enumerate}
Previous work on CUQs made the above two features explicit for pure states and maximally mixed states\cite{Karamitros:2022oew,Karamitros:2025azy}. 

Another motivation behind studying CUQs\cite{Brody:2012nxf,Sergi:2013kbw,Kawabata:2017gkk,KOWALSKI2019167955,Rembielinski:2021adg} lies within the context of non-Hermitian systems with $\mathcal{PT}$ (Parity-Time) symmetry\cite{Bender:1998gh,Bender:1998ke,Bender:2007nj}, where latter articles study the non-linear description of the Schrödinger equation. The works on non-Hermitian systems\cite{Brody:2012nxf,Kawabata:2017gkk} characterise coherence-decoherence oscillations as information gain/loss in the $\mathcal{PT}$-symmetric regime. Although implicit in their calculation, the anharmonicity feature of CUQs has not been adequately discussed. Thus, in some parts of the existing literature, CUQs have been equivalently represented as quantum two-level non-Hermitian systems with $\mathcal{PT}$-symmetry. In the context of high-energy physics, in addition to previous work on CUQs\cite{Pilaftsis:1997dr,Karamitros:2022oew,Karamitros:2025azy}, some recent studies have used the density matrix formalism to investigate the effective dynamics of mesons and the quantum decoherence of these particles in open systems~\cite{Alok:2024amd,Cheng:2025zcf}.

In this paper, we carry out a systematic analysis of CUQs that are initially in a pure or mixed quantum state. In particular, we analyse the Bloch-sphere trajectories of general mixed states, thereby extending previous work in this topic\cite{Brody:2012nxf,KOWALSKI2019167955,Sergi:2013kbw,Rembielinski:2021adg,Rembielinski:2023irg,Kowalski:2020vrx}.
We present for the first time analytic geometric constructions of 
CUQ trajectories {\em in} and {\em on} the Bloch sphere and explore their physical consequences on particle cosmology and quantum simulations. 

The paper has the following layout. In Section~\ref{sec:MixStatesolgen}, we analytically solve the differential equations describing the dynamics of the CUQ by the effective Hamiltonian~\eqref{non_hermitian Hamiltonian} under the restriction~\eqref{CUQcondition}. For both pure and mixed CUQs, we consider the Bloch-sphere formalism to describe the unstable qubit using a Bloch vector $\mathbf{b}$.  In Section~\ref{sec:Stationarypnts}, we discuss the stationary points of CUQs in the Bloch sphere, where the Bloch vector $\mathbf{b}$ does not evolve with time. In~Section~\ref{sec:BlochTrajectory}, we discuss the analytical solution of CUQ dynamics in detail and present plots of the Bloch-sphere trajectories that the Bloch vector $\mathbf{b}$ traverses. In~Section~\ref{sec:BlochGeometry}, we present a geometric construction by which  the Bloch sphere trajectory of  $\mathbf{b}$ can be determined solely using geometry, without the need to solve the master equation describing the dynamics of unstable qubits. Finally, in Section~\ref{sec:Concl}, we conclude our study by briefly discussing recent efforts in quantum computing to simulate CUQs and their potential implications in particle cosmology. In Appendix~\ref{App:CUQgensol}, we provide complete analytic results for the time evolution of the Bloch vector in a two-level non-Hermitian Hamiltonian. In Appendix~\ref{App:CUQhamilton}, we present a linear-algebraic theorem that enables us to express  non-Hermitian Hamiltonians describing a multi-level CUQ system as a product of two Hermitian matrices.

\section{Time Evolution of CUQs}\label{sec:MixStatesolgen}

The time evolution of a CUQ with a generic initial density matrix state $\rho(0)$ at time $t=0$ can be evaluated for a time-independent Hamiltonian\cite{Moiseyev_2011} as
\begin{equation}
  \label{rhoevol}
    \rho(t)\ =\ e^{-i\mathrm{H}t}\,\rho(0)\,e^{i\mathrm{H}^\dagger t}.
\end{equation}
For Hermitian systems, $\mathrm{H} = \mathrm{H}^{\dagger}$, the trace of the density matrix $\text{Tr}\,\rho(t)$ is preserved and equal to~one. Specifically, for two-level Hermitian systems, \eqref{rhoevol} describes a typical Rabi oscillation. However, for non-Hermitian systems, $\text{Tr}\,\rho(t)$ 
is not conserved and evolves as
\begin{equation}
    \dfrac{d}{dt}\mathrm{Tr}\,\rho\: =\: -\,\mathrm{Tr}\big(\Gamma\rho\big)\,.
\end{equation}
Thus, dividing $\rho$ by its trace in \eqref{rhoevol} gives a trace-preserving density matrix, which we define as the {\em co-decaying} frame: 
\begin{equation}
    \hat{\rho} (t)\: =\: \dfrac{e^{-i\mathrm{H}t}\rho(0)\,e^{i\mathrm{H}^\dagger t}}{\mathrm{Tr}\big(e^{-i\mathrm{H}t}\rho(0)\,e^{i\mathrm{H}^\dagger t}\big)}\, .\label{rhoevol_codec}
\end{equation}
In the context of particle decay, it can be understood as an ensemble of $N$ particles evolving with time, interacting with some external fields and decaying. In a lab frame, after some time, we are left with $n<N$ particles that are evolving. In this lab frame, the evolution of such an ensemble is given by \eqref{rhoevol}. In the co-decaying frame, however, we work in the frame where we divide our observable by the total number of particles in the ensemble at the instant when the measurement is made. A physical example of the latter is an observable of CP asymmetry\cite{Karamitros:2025azy}. 

The complete solution of \eqref{rhoevol} for two-level non-Hermitian systems is given in Appendix~\ref{App:CUQgensol}, where the CUQ dynamics is the class of systems satisfying $\mathbf{E}\cdot\boldsymbol{\Gamma} = 0$, with $|\mathbf{E}|\neq0$, $|\mathbf{\Gamma}|\neq0$ and $r<1$ [cf.~\eqref{CUQcondition}]. To emphasise the distinct features of CUQ oscillations, we revisit the case of pure-state qubit oscillations in these non-Hermitian systems and compare it with Rabi oscillations in Hermitian systems. 

The density matrix is a Hermitian matrix that can also be decomposed in the Pauli basis for two-level systems as
\begin{equation}
    \rho_{\text{H}}(t)\: =\: \dfrac{1}{2}\,\Big(\mathbb{1}\, +\, \mathbf{b}_{\mathrm{H}}(t)\cdot\boldsymbol{\sigma}\Big)\,,\label{density matrix_Bloch}
\end{equation}
where $\mathbf{b}(t)$ is known as the Bloch vector and $\rho_H(t)$ is the  evolving density matrix of a Hermitian system. For non-Hermitian systems, the density matrix in the lab frame is not trace preserving and is hence given by
\begin{equation}
    \rho (t) = \dfrac{1}{2}\Big(a^0(t)\mathbb{1}\, +\, \mathbf{a}(t)\cdot\boldsymbol{\sigma}\Big). 
\end{equation}
In the co-decaying frame, the density matrix $\hat{\rho}(t)$ can be written in the equivalent form of \eqref{density matrix_Bloch}, where $\mathbf{b}(t) \equiv \mathbf{a}/a^0$.

We compare the time-evolution of the density matrix in Hermitian and non-Hermitian systems by studying the differential equations that give rise to the solution~\eqref{rhoevol}. These are given by
\begin{equation}
    \dfrac{d\rho_{\text{H}}}{dt}\: =\: -i[\mathrm{E}\,,\rho_{\text{H}}]\label{Schro_Hermitian}\,,
\end{equation}
for the Hermitian case, and
\begin{equation}
    \dfrac{d\rho}{dt}\: =\: -i[\mathrm{E}\,,\rho] + \dfrac{1}{2}\{\Gamma\,,\rho\}\label{Schro_nonHerm}\,,
\end{equation}
for the non-Hermitian case, where $\rho_{H}$ and $\rho$ denote the density matrix under a Hermitian and non-Hermitian system, respectively. In the co-decaying frame, or the so-called trace-preserving map, the differential equation \eqref{Schro_nonHerm} becomes
\begin{equation}
    \dfrac{d\hat{\rho}}{dt}\: =\: -i[\mathrm{E}\,,\hat{\rho}]\: -\: \dfrac{1}{2}\{\Gamma\,,\hat{\rho}\}\: +\: \mathrm{Tr}(\Gamma\hat{\rho})\,\hat{\rho}\label{Schro_nonHerm_codec}\,.
\end{equation}
These differential equations can be written in terms of the Bloch vectors, which give what we call the master equation of state evolution\cite{Karamitros:2022oew}.
The master equation of state evolution for two-level Hermitian systems is
\begin{equation}
    \dfrac{d\mathbf{b}_{\text{H}}}{dt}\: =\: -2|\mathbf{E}|\,(\mathbf{e}\times \mathbf{b}_{\text{H}}\label{RabiMasteq})\,,
\end{equation}
which is a linear equation that indicates that the time evolution of the Bloch vector $\mathbf{b}_{\text{H}}$ (subscript H to indicate Hermitian dynamics) always lies in a plane perpendicular to $\mathbf{e}\equiv {\bf E}/|{\bf E}|$, i.e.~$\frac{d}{dt}(\mathbf{b}_{\text{H}}\cdot\mathbf{e})=0$. Thus, any dynamics of interest for the Bloch vector $\mathbf{b}_{\text{H}}(t)$ is in the two-dimensional plane perpendicular to $\mathbf{e}$. We let that plane be described by two basis vectors: one being $\mathbf{e}_{\perp}$, an arbitrary unit vector perpendicular to $\mathbf{e}$, and the other unit vector being $\mathbf{e}\times\mathbf{e}_{\perp}$. We can then solve the above master equation \eqref{RabiMasteq} to get $\mathbf{b}_{\text{H}}(t)$ for any initial state  $\mathbf{b}_{\text{H}}(0) = b_x\mathbf{e} + b_y\mathbf{e}_{\perp} +b_z\mathbf{e}\times\mathbf{e}_{\perp}$. Since $\mathbf{b}_{\text{H}}$ does not evolve along the $\mathbf{e}$ component, we ignore it and instead focus on the time evolution of the initial state, $\mathbf{b}_{\text{H}}(0)=b_y\mathbf{e}_{\perp} +b_z\mathbf{e}\times\mathbf{e}_{\perp}$. Note that the eigenvectors of the Hamiltonian are $\mathbf{b}_{\text{H}\pm} = \pm \mathbf{e}$. It is for this reason that the Bloch vector does not evolve under that component.

To derive the master equation of the two-level non-Hermitian system in terms of ${\bf b}$ in the $\textit{co-decaying}$ frame, we start 
from the differential equations in the lab frame:
\begin{equation}
    \begin{aligned}
        \dfrac{da^0}{dt}\ &=\ -\,\Gamma_\mu a^\mu\,,\\
        \dfrac{d\mathbf{a}}{dt}\ &=\ -\,2|\mathbf{E}|\,(\mathbf{e}\times \mathbf{a})\: +\: |\boldsymbol{\Gamma}|\,\boldsymbol{\gamma}\,a^0\: -\: \Gamma^0\mathbf{a}\,.
    \end{aligned}\label{nonherm_master_equation}
\end{equation}
Then, in terms of $\mathbf{b} \equiv \mathbf{a}/a^0$, $\tau\equiv|\boldsymbol{\Gamma}|t$ and $r\equiv |{\bf \Gamma}|/(2|{\bf E}|)$, the {\em master evolution equation} is given by~\cite{Karamitros:2022oew}:
\begin{equation}
    \dfrac{d\mathbf{b}}{d\tau}\: =\: -\dfrac{1}{r}(\mathbf{e}\times \mathbf{b})\, +\, \boldsymbol{\gamma}\, -\, (\boldsymbol{\gamma}\cdot\mathbf{b})\,\mathbf{b}\,,\label{CUQ_Mastereq}
\end{equation}
where ${\bf e} \equiv {\bf E}/|{\bf E}|$ and $\boldsymbol{\gamma} \equiv {\bf \Gamma}/|{\bf \Gamma}|$ are unit vectors along ${\bf E}$ and ${\bf \Gamma}$, respectively. Observe that in the co-decaying frame, the master evolution equation~\eqref{CUQ_Mastereq}  has a non-linear term, $-(\boldsymbol{\gamma}\cdot\mathbf{b})\mathbf{b}$, which is usually responsible for the anharmonic dynamics in CUQs. Moreover, along with the linear term $\boldsymbol{\gamma}$, the above non-linear term gives rise in the second distinct feature of coherence-decoherence oscillations. We study the analytical solutions to~\eqref{nonherm_master_equation} and~\eqref{CUQ_Mastereq} to capture these features in detail.

\subsection{Pure CUQs\label{purestatedyn}}

\begin{figure}[t]
    \centering
    \begin{subfigure}{0.49\textwidth}
        \includegraphics[width=\linewidth]{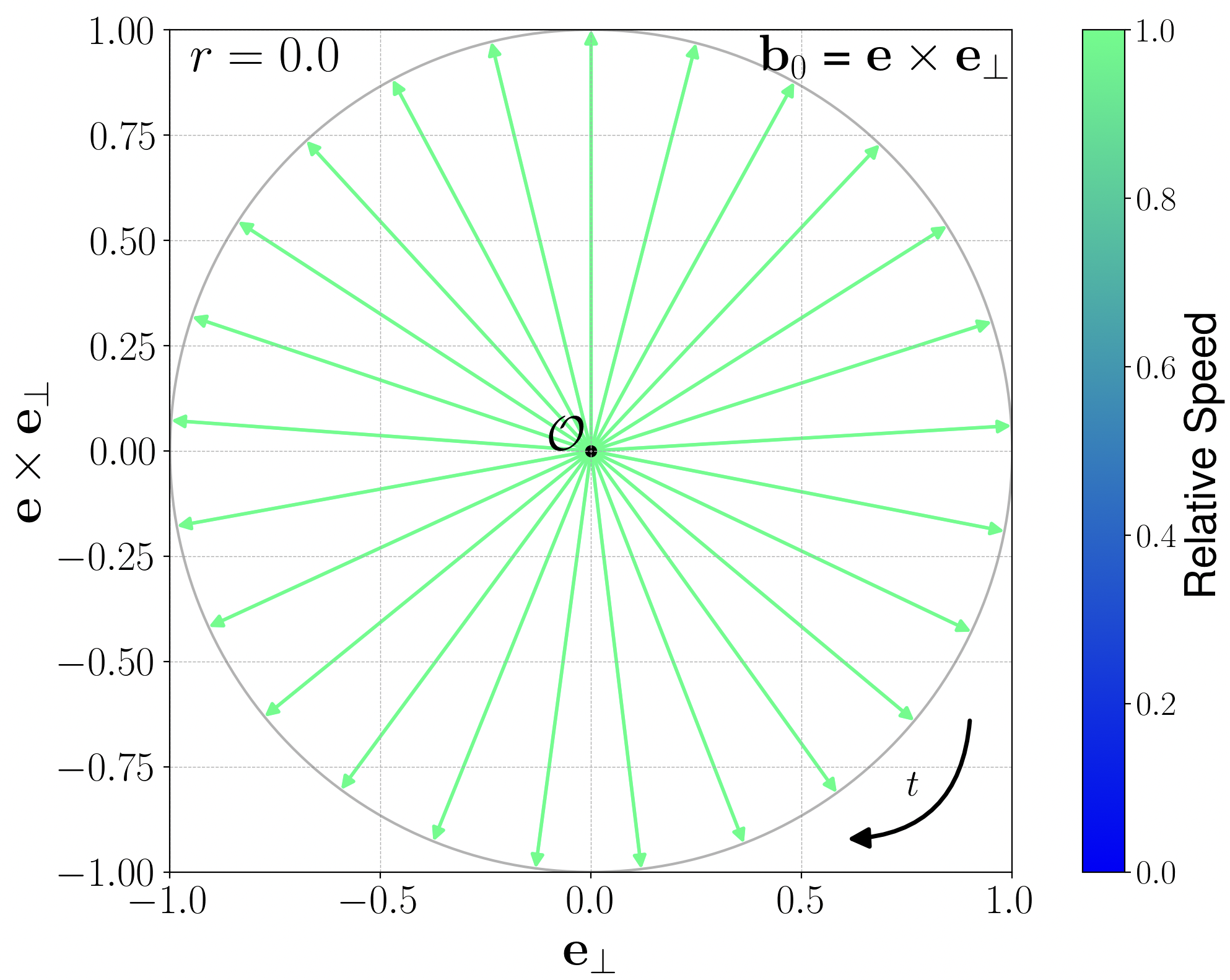}
        \caption{Rabi Oscillation}
    \end{subfigure}
    \begin{subfigure}{0.49\textwidth}
        \includegraphics[width=\linewidth]{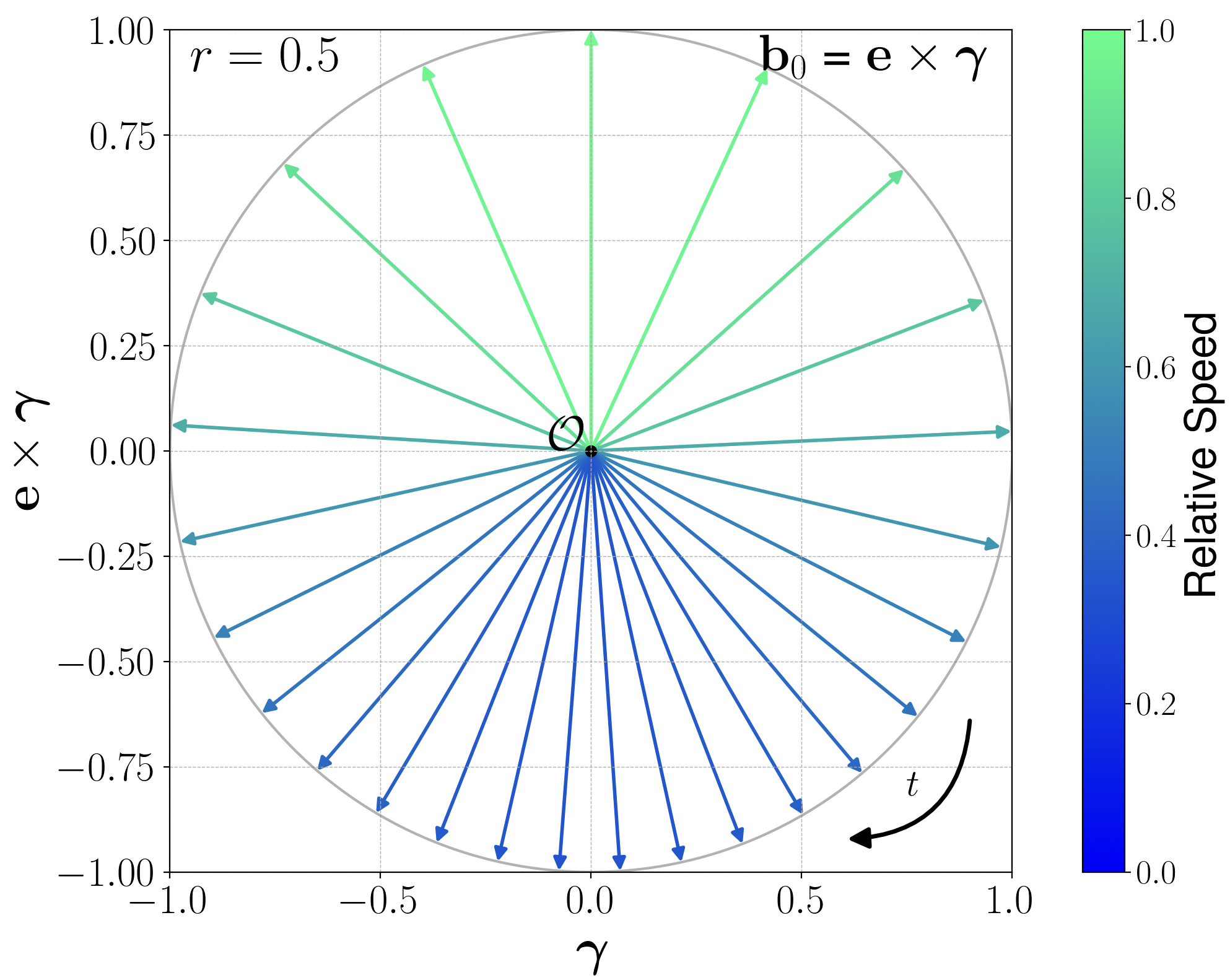}
        \caption{CUQ Oscillation}
    \end{subfigure}

    \caption{Comparative Bloch-sphere trajectories for pure state (a) Hermitian (Rabi) and (b) non-Hermitian (CUQ) oscillations. The vectors are the evolution of the initial state $\mathbf{b}_0$ captured in equal time intervals. The Bloch vector rotates in the clock-wise direction. The colour grading shows the speed of the moving Bloch vector (relative to the maximum speed in the trajectory) at that time interval.}\label{fig: Rabi vs CUQ pure}
\end{figure}

The time evolution of the Bloch vector ${\bf b}_{\rm H}$ of a qubit can easily be found by solving the master equation \eqref{RabiMasteq}. For a Bloch vector starting from the pure state, 
\begin{equation}
    \mathbf{b}_{\text{H}}(0)\:\equiv\: \mathbf{b}_0\: =\: \cos\varphi_0\,\mathbf{e}_{\perp} +\, \sin\varphi_0\, \mathbf{e}\times \mathbf{e}_{\perp}\,,\label{inihermitian}
\end{equation}
its time evolution is given by
\begin{equation}
    \mathbf{b}_{\text{H}}(t)\ =\ \cos\big(2|\mathbf{E}|t-\varphi_0 \big)\,\mathbf{e}_{\perp} -\, \sin\big(2|\mathbf{E}|t-\varphi_0\big)\, \mathbf{e}\times \mathbf{e}_{\perp}\,.\label{pureRabisol}
\end{equation}
This is a periodic rotating vector in uniform circular motion, where $\varphi(t) \equiv\varphi_0-2|\mathbf{E}|t$ is  the angle swept by the Bloch vector $\mathbf{b}_{\text{H}}$ at time $t$. The time evolution of $\mathbf{b}_{\text{H}}(t)$ is shown in the left panel of Figure~\ref{fig: Rabi vs CUQ pure}.

We may directly compare the above result with the pure state evolution of CUQs. To do so, we consider the dynamics of the Bloch vector perpendicular to the energy vector~$\mathbf{E}$, i.e.~with $\mathbf{b}\cdot\mathbf{e} = 0$. Solving the differential equation \eqref{CUQ_Mastereq} for the non-Hermitian Hamiltonian is not as straightforward as its Hermitian counterpart. Instead, we derive ${\bf b}(t)$ from the density matrix equation~\eqref{rhoevol_codec}. The method and general solution for a two-level non-Hermitian Hamiltonian are given in Appendix~\ref{App:CUQgensol}. The solution for a CUQ is given by~\eqref{mixCUQ_sol}. Since we are interested in the case of $\mathbf{b}(0)\perp\mathbf{E}$, we define the initial state in polar coordinates, as we did in~\eqref{inihermitian}, 
\begin{equation}
    \mathbf{b}(0)\: \equiv\: \mathbf{b}_0\ =\ \cos{\varphi_0}\,\boldsymbol{\gamma}\, +\, \sin{\varphi_0}\,\mathbf{e\times}\boldsymbol{\gamma}\label{inipure}\, .
\end{equation}
Plugging~\eqref{inipure} into~\eqref{mixCUQ_sol}, we get
\begin{equation}
    \mathbf{b}(t)\ =\ \dfrac{\sqrt{1-r^2}\sin{\big(2\theta(t)\:+\: \alpha_0\big)}\boldsymbol{\gamma}+\big[\cos{(2\theta(t)+\alpha_0)-r\big]\,\mathbf{e}\times \boldsymbol{\gamma}}}{1\,-\,r\cos{\big(2\theta(t)+\alpha_0\big)}}\; ,\label{pureCUQ_gensol}
\end{equation}
where $\theta(t) = |\mathbf{E}|\sqrt{1-r^2}\,t$, and the angle $\alpha_0$ is defined through
\begin{equation}
    \tan{\alpha_0}\: \equiv\: \dfrac{\sqrt{1-r^2}\,\cos{\varphi_0}}{r+\sin{\varphi_0}}\;.\label{alphdef}
\end{equation}
Thus, $\alpha_0$ contains the information of the initial state of the CUQ.

The length of the Bloch vector is preserved to be unity, i.e.~$|\mathbf{b}(t)| = 1$. Therefore, a pure state always remains pure, as can be shown using the master equation~\eqref{CUQ_Mastereq} \cite{Karamitros:2022oew}. The general form of the Bloch vector can thus be parameterised in polar form as
\begin{equation}
    \mathbf{b}(t)\: \equiv\: \cos{\varphi(t)}\,\boldsymbol{\gamma}\, + \, \sin{\varphi(t)}\,\mathbf{e\times}\boldsymbol{\gamma}\label{purestate}\,,
\end{equation}
where $\varphi(t)$ can be computed using
\begin{equation}
    \begin{aligned}
        \tan{\varphi(t)}\: & =\: \dfrac{\cos{\big(2\theta(t) +\alpha_0\big)} - r}{\sqrt{1-r^2}\,\sin{\big(2\theta(t)+\alpha_0\big)}}\; .
    \end{aligned}
    \label{phasesol_purestate}
\end{equation}
Note that the initial states of a CUQ that differ by angle $\alpha_0$ only produce a phase shift in their time evolution. This clearly means that the period for one full rotation of the qubit is the same for all initial states. Equation~\eqref{phasesol_purestate} reveals that the evolution of the phase of the Bloch vector is anharmonic, unlike the harmonic Rabi oscillations. The comparative features between Rabi and CUQ oscillations are presented in Figure~\ref{fig: Rabi vs CUQ pure}.

The application of the WW approximation to compute the dynamics of a pure state two-level unstable system is a standard calculation in particle physics for neutral mesons. The interested reader may consult the supplementary chapter on $B$-mesons in the PDG booklet\cite{ParticleDataGroup:2024cfk}. In this case, the computation of meson dynamics gives a combined decay and oscillatory dynamics, with the oscillations being harmonic (Rabi) in nature. The anharmonicity can only be observed in the co-decaying frame, where the state is always normalized. The CP-asymmetry measure\cite{Karamitros:2025azy,Alok:2024amd} is an example of an observable that is measured in the co-decaying frame.

\subsection{Oscillations of Mixed CUQs}

Before considering the CUQ scenario, let us first discuss the case of mixed stable qubit, with $b_{\text{H},0} \equiv|{\bf b}_{\rm H}(0)| \in [0,1)$. Starting with a mixed initial state written in polar form as
\begin{equation}
    \mathbf{b}_{\rm H,0}\: =\: b_{\text{H},0}\Big(\cos\varphi_0\,\mathbf{e}_{\perp} +\, \sin\varphi_0\, \mathbf{e}\times \mathbf{e}_{\perp}\Big)\,,
\end{equation}
it is not difficult to solve~\eqref{RabiMasteq} to find
\begin{equation}
    \mathbf{b}_{\rm H}(t)\: =\: b_{\text{H},0}\,\Big(\cos\big(2|\mathbf{E}|t-\varphi_0 \big)\,\mathbf{e}_{\perp} -\, \sin\big(2|\mathbf{E}|t-\varphi_0 \big)\, \mathbf{e}\times \mathbf{e}_{\perp}\Big)\, .\label{mixRabisol}
\end{equation}
Like in the pure-state case, the Bloch vector of the mixed qubit exhibits the same uniform circular motion, but now with a radius $b_{\text{H},0}$. We note that the coherence of the mixed state is preserved.

\begin{figure}[!t]
    \centering
    \begin{subfigure}{0.49\textwidth}
        \includegraphics[width=\linewidth]{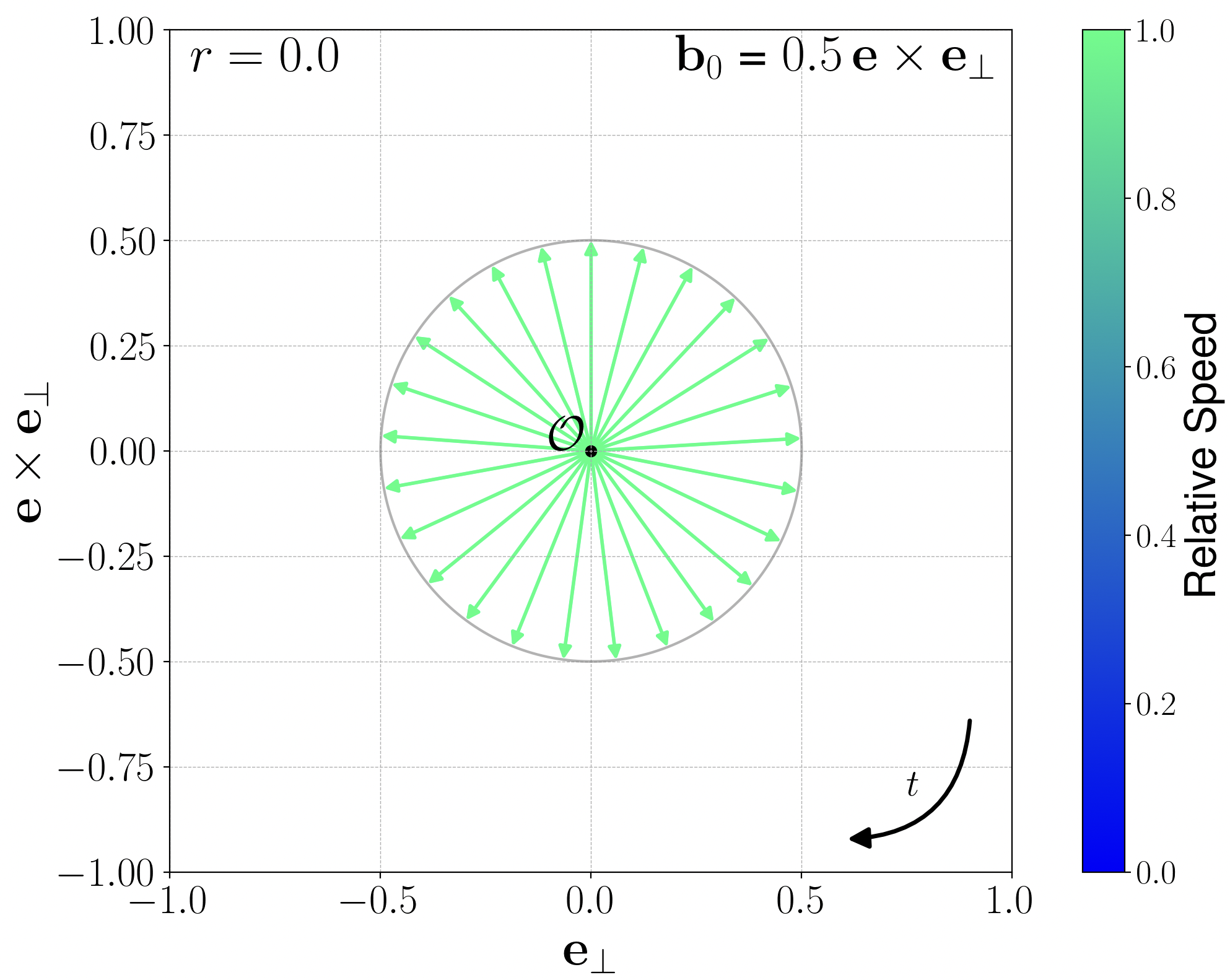}
        \caption{Rabi Oscillation}
    \end{subfigure}
    \begin{subfigure}{0.49\textwidth}
        \includegraphics[width=\linewidth]{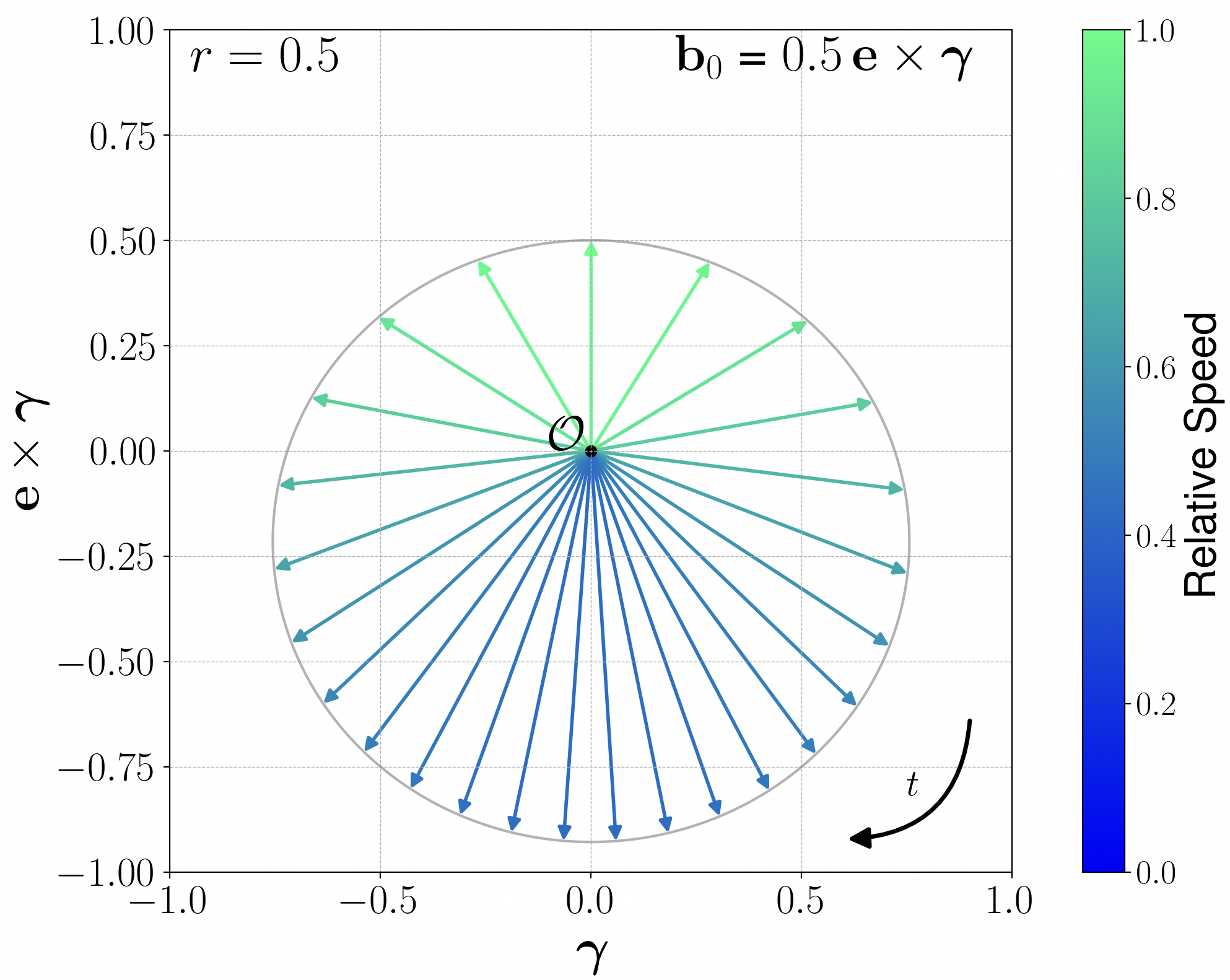}
        \caption{CUQ oscillation}
    \end{subfigure}

    \caption{Comparative Bloch-sphere trajectories for mixed state Rabi and CUQ oscillations. The vectors are the evolution of the initial state $\mathbf{b}_0$ captured in equal time intervals. The colour grading shows the speed of the moving Bloch vector at that instant of time.}\label{fig: Rabi vs CUQ}
\end{figure}

We now turn to analysing CUQ oscillations starting from an initial mixed state, where $0\leq b_0\equiv \mathbf{|\mathbf{b}}(0)|<1$.
To be precise, the initial Bloch vector has the form
\begin{equation}
    \mathbf{b}_0\: \equiv\: {\bf b}(0)\: =\: b_0\cos{\varphi_0}\,\boldsymbol{\gamma}\, +\,b_0\sin{\varphi_0}\,\mathbf{e}\times \boldsymbol{\gamma}\,.\label{initial_genCUQ}
\end{equation}
Proceeding as in Section~\ref{purestatedyn}, we may simplify our solution by redefining the initial conditions using some algebra. The Bloch vector of CUQ at a time $t$ can be expressed as
\begin{equation}
    \mathbf{b}(t)\ =\ \dfrac{\sqrt{1-r^2}\,\sin{\big(2\theta(t)+\beta_0\big)\cos{\delta_0}}\,\boldsymbol{\gamma}\: +\: \big(\cos{\big(2\theta(t)+\beta_0\big)\cos{\delta_0}\,-\,r\big)\,\mathbf{e}\times\boldsymbol{\gamma}}}{1\: -\: r\cos{\big(2\theta(t)+\beta_0\big)}\cos{\delta_0}}\;,\label{genCUQ_gensol}
\end{equation}
where $\beta_0$ and $\delta_0$ are defined as follows:
\begin{eqnarray}
   \label{paramtrans1}
        \tan{\beta_0} \!&\equiv&\! \dfrac{b_0\sqrt{1-r^2}\,\cos{\varphi_0}}{r+b_0\sin{\varphi_0}}\;,\\
    \label{paramtrans2}
 \tan{\delta_0} \!&\equiv&\! \left[\dfrac{(1-b_0^2)\,(1-r^2)}{(1+b_0 r\sin{\varphi_0})^2 -\, (1-b_0^2)\,(1-r^2)}\right]^{1/2}.
\end{eqnarray}
As in~\eqref{pureCUQ_gensol} in Section~\ref{purestatedyn}, we may write the solution~\eqref{genCUQ_gensol} in polar form
\begin{equation}
    \mathbf{b}(t)\: =\: |\mathbf{b}(t)|\,\Big(\cos{\varphi(t)}\,\boldsymbol{\gamma}\, +\,\sin{\varphi(t)}\,\mathbf{e}\times \boldsymbol{\gamma}\Big)\label{final_genCUQ}\,.
\end{equation}
Comparing the two forms enables us to compute the length of the Bloch vector $|\mathbf{b}(t)|$ and its phase angle $\varphi(t)$
as
\begin{eqnarray}
   \label{genlengthCUQ}
    |\mathbf{b}(t)|^2 \!&=&\! 1\, -\, \dfrac{(1-r^2)\sin^2{\delta_0}}{[1-r\cos{(2\theta(t)+\beta_0)}\cos{\delta_0}]^2}\;, \\
   \label{genangleCUQ}
    \tan{\varphi(t)} \!&=&\! \dfrac{\cos{(2\theta(t)+\beta_0)\cos{\delta_0}}-r}{\sqrt{1-r^2}\sin{(2\theta(t)+\beta_0)\cos{\delta_0}}}\;.
\end{eqnarray}
In addition to the coherence-decoherence oscillations for general mixed states in \eqref{genlengthCUQ}, \eqref{genangleCUQ} again describes the anharmonicity in the evolution of the quantum state. These qualitative differences in the stable qubit and CUQ cases are illustrated in Figure~\ref{fig: Rabi vs CUQ}.

\begin{figure}[ht]
    \centering
    \includegraphics[width=0.5\linewidth]{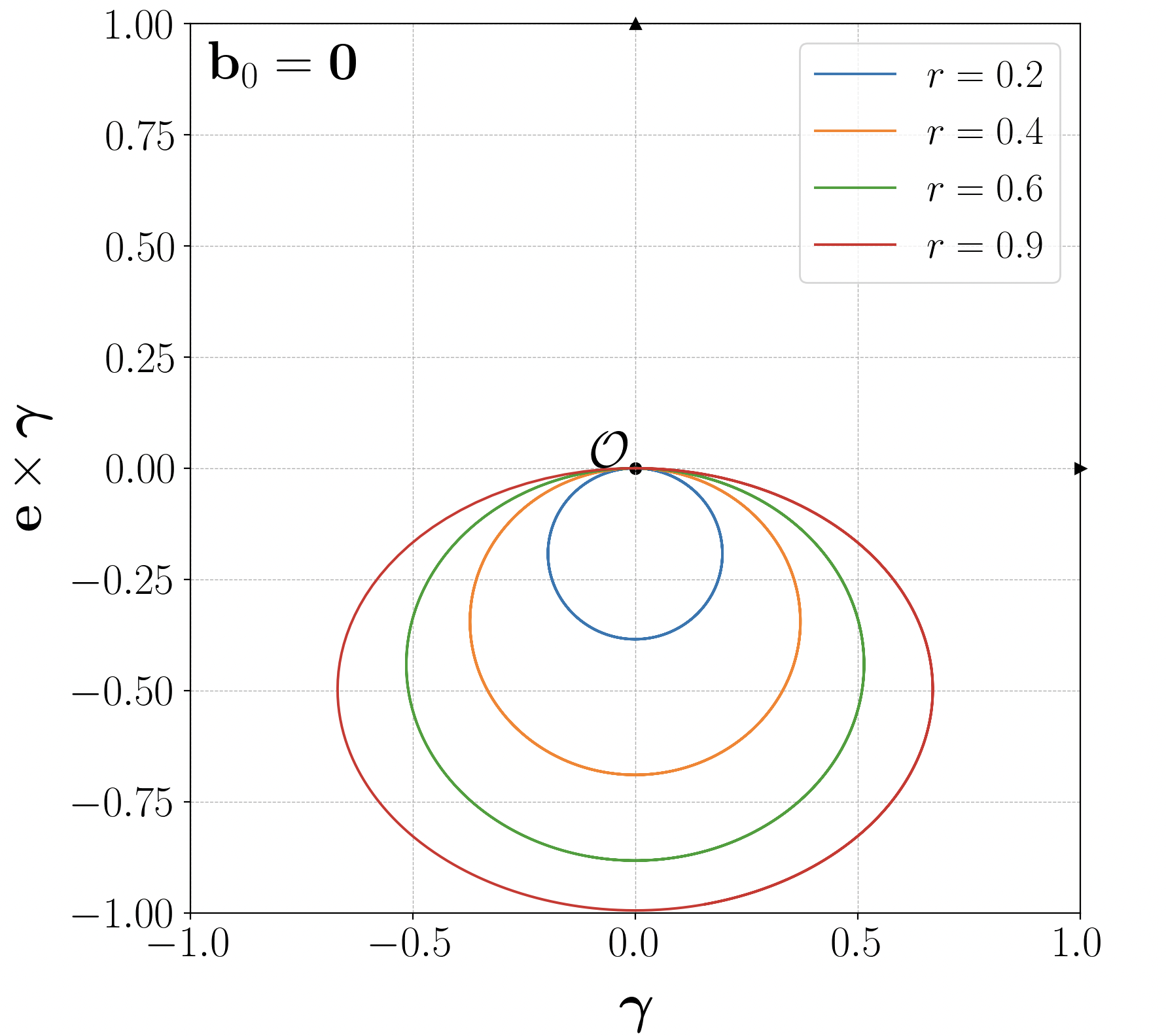}
    \caption{Bloch-sphere trajectories $\mathbf{b}(t)$ of the maximally-mixed CUQ for different values~$r$.}
    \label{fig: Bellstate}
\end{figure}

For the maximally-mixed CUQ state, we have $b_0 = 0$, which implies $\beta_0 = 0$ and $\cos{\delta_0} = r$ from~\eqref{paramtrans1} and~\eqref{paramtrans2}, respectively. Substituting these values into~\eqref{genlengthCUQ} yields
\begin{equation}
    |\mathbf{b}(t)|^2\ =\ 1\,-\, \dfrac{(1-r^2)^2}{\big[1-r^2\cos{2\theta(t)}\big]^2}\;.
\end{equation}
The above result is obtained for an initial state $\mathbf{b}(0) =\mathbf{0}$, in agreement with\cite{Karamitros:2022oew}. Figure~\ref{fig: Bellstate} shows the Bloch-sphere trajectory of the maximally-mixed state for different values of $r$. For a pure state, $b_0 = 1$, we have $\delta_0 = 0$, which in turn gives $|\mathbf{b}(t)| = 1$ for all times $t$ on account of~\eqref{genlengthCUQ}. Moreover, we have $\beta_0 = \alpha_0$ 
[cf.~\eqref{alphdef}], and since $\cos{\delta_0} = 1$, the general solution~\eqref{genangleCUQ} reduces to~\eqref{phasesol_purestate}, as should be.

We note that $\beta_0$ and $\delta_0$ can be related to each other as they are transformations of the initial parameters $b_0$ and $\varphi_0$ through~\eqref{paramtrans1} and~\eqref{paramtrans2}. Unlike in the previous section, the way~$\beta_0$ and~$\delta_0$ are defined does not correspond to a one-to-one mapping to $b_0$ and $\varphi_0$, as there is one point in the parameter space of the domain that remains undetermined, which is $\mathbf{b}_0 = -r\,\mathbf{e}\times \boldsymbol\gamma$. In the polar form, this point corresponds to $b_0 = r$ and $\varphi_0=-\pi/2$. It is the only point that cannot be mapped in terms of $\beta_0$ and $\delta_0$. The physical significance of this point
will be discussed in detail in the next section.

\section{Stationary Points of Unstable Qubits}\label{sec:Stationarypnts}

In this section, we analyse the general conditions under which unstable qubits do not oscillate, including CUQ scenarios. This is an aspect that has not been adequately investigated in previous studies\cite{Brody:2012nxf,Rembielinski:2021adg,Karamitros:2022oew,Karamitros:2025azy}. 
We should emphasise that our interest is in examining unstable qubits that are not overdamped, i.e.~unstable qubits for which $r<1$.

\begin{figure}[t!]
    \centering
    \begin{subfigure}{0.32\textwidth}
        \includegraphics[width=\linewidth]{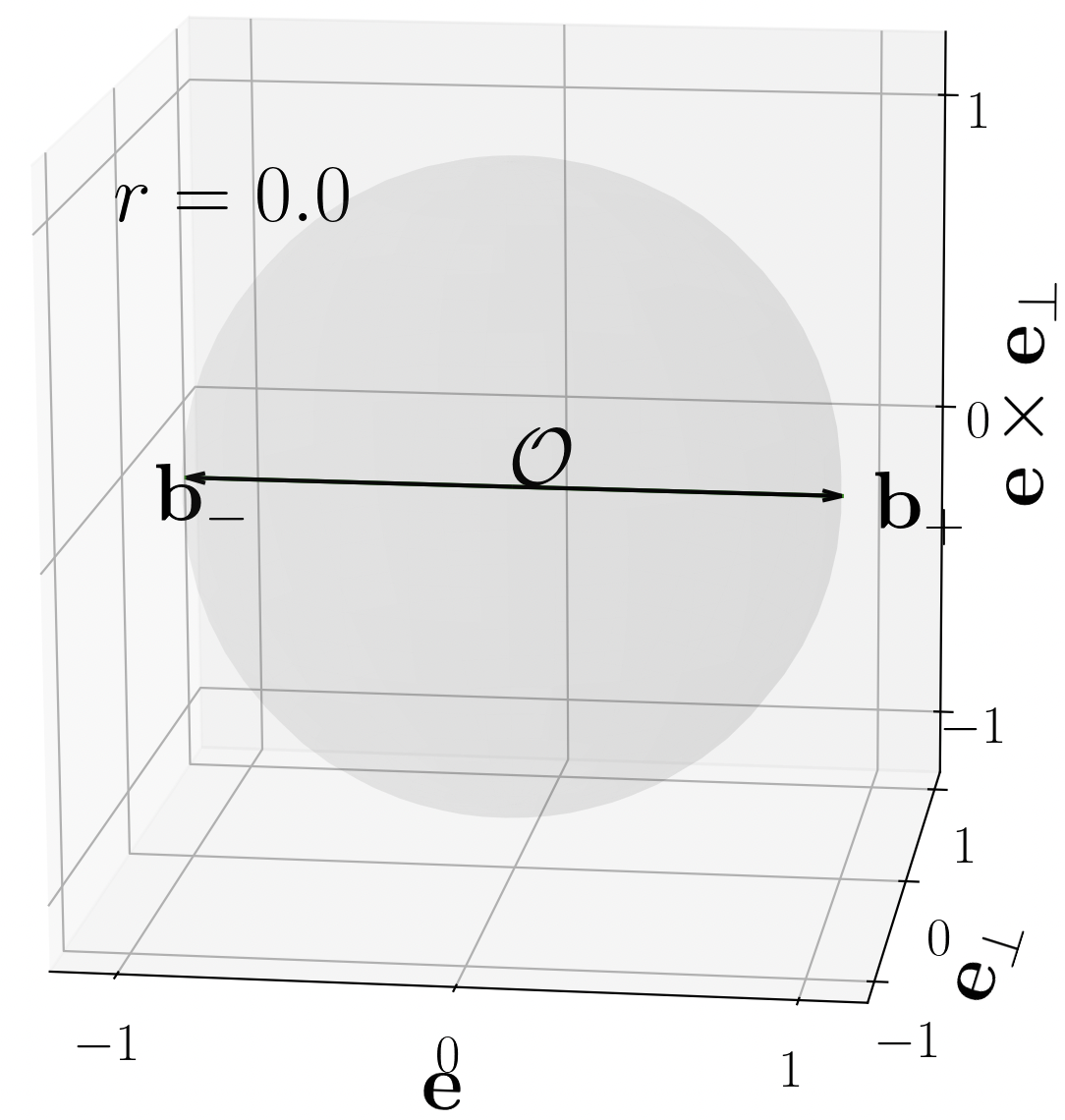}
        \caption{}\label{stationarya}
    \end{subfigure}
    \begin{subfigure}{0.32\textwidth}
        \includegraphics[width=\linewidth]{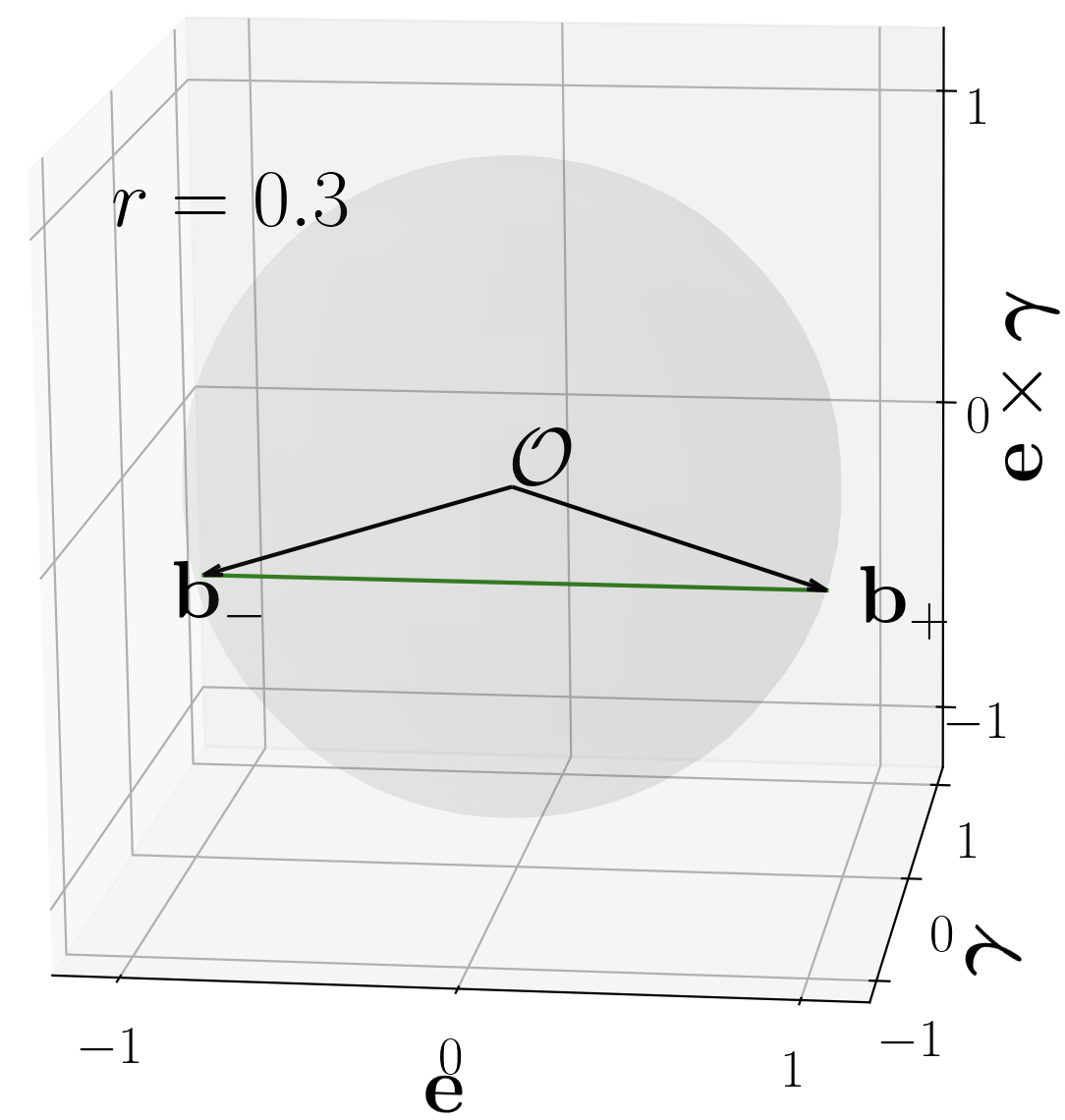}
        \caption{}\label{stationaryb}
    \end{subfigure}
    \begin{subfigure}{0.32\textwidth}
        \includegraphics[width=\linewidth]{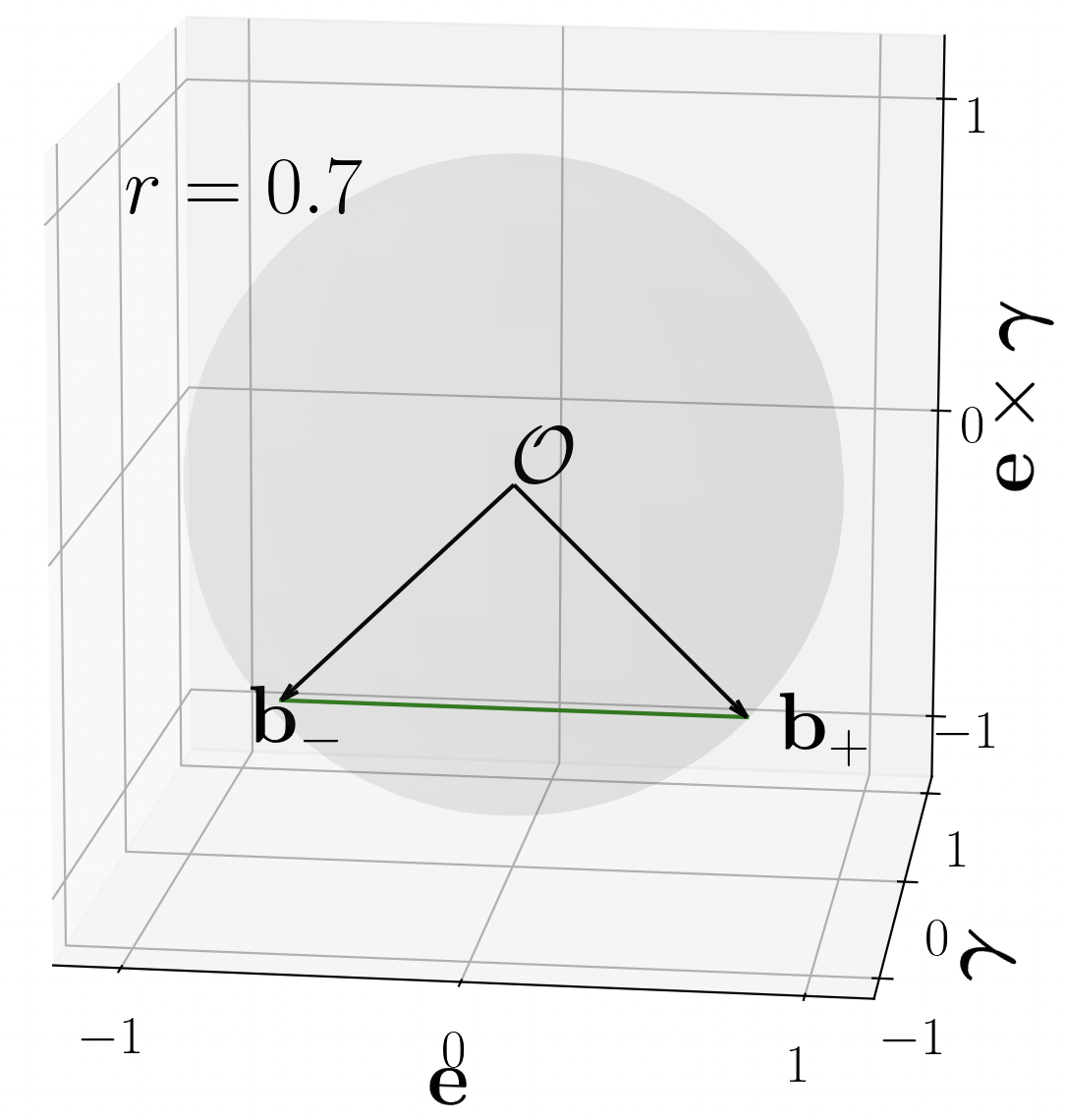}
        \caption{}\label{stationaryc}
    \end{subfigure}

    \caption{Eigenvectors of the CUQ Hamiltonian on the Bloch sphere}\label{fig: eigenvecCUQ}
\end{figure}

Before discussing the case of unstable qubits, it is instructive to first consider stable qubits whose dynamics are governed by the Hermitian Hamiltonian  
\begin{equation}
   \label{eq:Hstable}
  \mathrm{H}\: \equiv\: \text{E}\ =\ \mathrm{E}_0\mathbb{1}\, -\, \mathbf{E}\cdot \boldsymbol{\sigma}\,,  
\end{equation}
where $\mathbf{E} = |\mathbf{E}|\,\mathbf{e}$ is the Bloch energy vector (see our discussion in Section~\ref{sec:Intro}). The energy eigenvalues of $\mathrm{H}$ are given by $\lambda_{\pm} = \mathrm{E}_0\pm |\mathbf{E}|$. In the density matrix formalism, the eigenvectors of $\mathrm{H}$, $\rho_{\mathrm{H},\pm} \equiv \dfrac{1}{2}(\mathbb{1}\pm\mathbf{b}_{\mathrm{H},\pm}\cdot\boldsymbol{\sigma})$, obey the eigenvalue equations 
\begin{equation}
    \Big(\mathrm{H}\, -\, \lambda_{\pm}\mathbb{1}\Big)\,\rho_{\mathrm{H},\pm}\: =\: \rho_{\mathrm{H},\pm}\Big(\mathrm{H}\, -\, \lambda_{\pm}\mathbb{1}\Big)\: =\: 0\,.\label{eigenveceq}
\end{equation}
From the above, we can determine the Bloch eigenvectors to be $\mathbf{b}_{\mathrm{H},\pm} = \pm \mathbf{e}$, i.e. 
\begin{equation}
    \rho_{\mathrm{H},\pm}\: \equiv\: \ket{\pm}\bra{\pm}\ =\ \frac{1}{2}\,\Big(\mathbb{1}\, \pm\, \mathbf{e}\cdot\boldsymbol{\sigma}\Big)\,,
\end{equation}
with $\mathbf{e} \equiv \mathbf{E}/|\mathbf{E}|$. 
From~\eqref{eigenveceq}, it is easy to see that $d\rho_{\mathrm{H},\pm}/dt = [\mathrm{H},\rho_{\mathrm{H},\pm}] = 0$. This means that the eigenvectors are the stationary points in the Bloch-vector dynamics. As a result, all quantum states of the form 
\begin{equation}
  \label{eq:rhoHstar}
\rho_{\text{H},*}(s)\: =\: s\,\rho_{\text{H},+}\, +\, (1-s)\,\rho_{\text{H},-}\,,\quad \mbox{or}\quad \mathbf{b}_{\text{H},*}(s)\: =\: (2s-1)\,\mathbf{e}
\end{equation}
will be stationary points, with $0\leq s\le 1$. In the Bloch sphere, all these stationary points can be viewed as a set in a straight line joined by the two eigenstates, which in this case is a line on the $\mathbf{e}$-axis, as demonstrated in Figure~\ref{stationarya}. This is another way to understand why the motion of the Bloch vector ${\bf b}_\text{H}(t)$  is perpendicular to $\mathbf{e}$ in the case of Rabi oscillations.

For a generic unstable qubit described by the effective Hamiltonian~$\text{H}_{\rm eff}$~\eqref{non_hermitian Hamiltonian}, there are only two stationary points given by the two eigenvectors of $\text{H}_{\rm eff}$. Any linear superposition of the density matrices corresponding to these two eigenvectors 
does not describe a stationary point. For a CUQ, the situation 
differs. To see this, we  consider the Hamiltonian \eqref{nonHerm_Hamiltonian_Paulibasis} in the Pauli basis
\begin{equation}
\begin{aligned}
    \mathrm{H}_{\text{eff}}\ =\ \left(\mathrm{E}_0 -\frac{i}{2}\Gamma_0\right)\,\mathbb{1}\: -\: |\mathbf{E}|\,(\mathbf{e} - ir\boldsymbol{\gamma})\cdot\boldsymbol{\sigma}\,,
\end{aligned}
\label{nonHermPaulidec}
\end{equation}
for which the eigenvalues $\lambda_\pm$ for $0<r<1$ are
\begin{equation}
    \lambda_{\pm}\: \equiv\: \mathrm{E}_\pm -\dfrac{i}{2}\Gamma_{\pm}\ =\ \mathrm{E}_0\pm |\mathbf{E}|\sqrt{1-r^2}\:-\:\frac{i}{2}\Gamma_0\,,\label{CUQeigenvalues}
\end{equation}
and $\Gamma_\pm = \Gamma_0$. In other words, the decay widths of the two CUQ eigenvectors are equal, while $r\ne 0$.

Similarly to what we did in the stable qubit case, we can use the eigenvalue equations,
\begin{equation}
   \label{eq:Heffstat}
    \Big(\mathrm{H}_\text{eff}\, -\, \lambda_{\pm}\mathbb{1}\Big)\,\rho_{\pm}\: =\: \rho_{\pm}\Big(\mathrm{H}^\dagger_{\rm eff}\, -\, \lambda_{\pm}\mathbb{1}\Big)\: =\: 0\,,
\end{equation}
to calculate the eigenvectors of the CUQ. In the Bloch vector form, we denote the eigenvectors by~$\mathbf{b}_\pm$, which are found to be
\begin{equation}
   \label{eq:bplusminus}
    \mathbf{b}_\pm\ =\ \pm\sqrt{1-r^2}\,\mathbf{e}\: -\: r\,\mathbf{e}\times\boldsymbol{\gamma}\,.
\end{equation}
From the master equation \eqref{CUQ_Mastereq}, it can be checked that $d\mathbf{b}_\pm/d\tau =0$, confirming our finding in~\eqref{eq:bplusminus}. Like in the stable qubit case,
 the straight line joining the end-points of the two Bloch-vector eigenstates $\mathbf{b}_\pm$ represents an infinite set of stationary points in the Bloch sphere. This is shown in
Figures~\ref{stationaryb} and \ref{stationaryc}. The {\em infinite} set of stationary points is
\begin{equation}
   \label{eq:stationary}
    \rho_{*}(s)\ =\ s\,\rho_+\: +\: (1-s)\,\rho_-\,,\quad \text{or}\quad \mathbf{b}_{*}(s)\ =\ (2s-1)\,\sqrt{1-r^2}\, \mathbf{e}\: -\: r\,\mathbf{e}\times \boldsymbol{\gamma}\,,
\end{equation}
with $0\leq s\leq 1$. Evidently, the stationary point $\mathbf{b}_* (\frac{1}2) = -r\,\mathbf{e}\times \boldsymbol{\gamma}$ found in the previous section corresponds to the mixed state $\rho_{*}(\frac{1}2) = \frac{1}{2}\left(\rho_++\rho_-\right)$. Thus, for CUQs, the line of stationary points shifts from the $\mathbf{e}$-axis by a distance of $r$ in the negative $(\mathbf{e}\times\boldsymbol{\gamma})$-direction, remaining parallel to the $\mathbf{e}$-axis.

As mentioned above, the two eigenstates of a CUQ may have different energies $\text{E}_\pm$,
but they have equal decay widths~$\Gamma_\pm$\cite{Karamitros:2022oew}:
\begin{equation}
    \Delta \mathrm{E}\: \equiv\: \mathrm{E}_+-\mathrm{E}_-\ =\ 2|\mathbf{E}|\sqrt{1-r^2}\quad \text{and}\quad \Delta\Gamma\: \equiv\: \Gamma_+ - \Gamma_- \: =\: 0\,,
\end{equation}
with $|\boldsymbol{\Gamma}|\neq 0$, which implies that the decay matrix $\Gamma$ has non-zero off-diagonal elements. In the limit~${r\to 1}$, the period of CUQ oscillation approaches infinity, which we call an $\textit{extremal}$ CUQ. In this case, we have $\Delta \mathrm{E} = 0$, i.e.~degenerate states, and the eigenvectors of the Hamiltonian coalesce into a single degenerate eigenvector as well.
The first observation of such extremal CUQs was made some time ago in~\cite{Pilaftsis:1997dr}. In this case, CP violation is maximized for an unstable two-level mixing system, giving rise to resonantly enhanced CP violation, also known as resonant CP violation in the literature. In this respect, we should point out that in the condensed matter physics literature, an extremal CUQ is referred to as the exceptional point\cite{Heiss:2012dx} of the non-Hermitian Hamiltonian.

\section{Trajectories of CUQs on the Bloch Sphere}\label{sec:BlochTrajectory}

In this section, we give analytical expressions for the CUQ trajectories under general initial conditions: (i)~$\mathbf{b}(0)\perp \mathbf{e}$ and (ii)~$\mathbf{b}(0)\not\perp \mathbf{e}$. In particular, we present the basic geometric features of these trajectories for both pure and mixed CUQs. 

\begin{figure}[t!]
    \centering
    \begin{subfigure}{0.49\textwidth}
        \includegraphics[width=\linewidth]{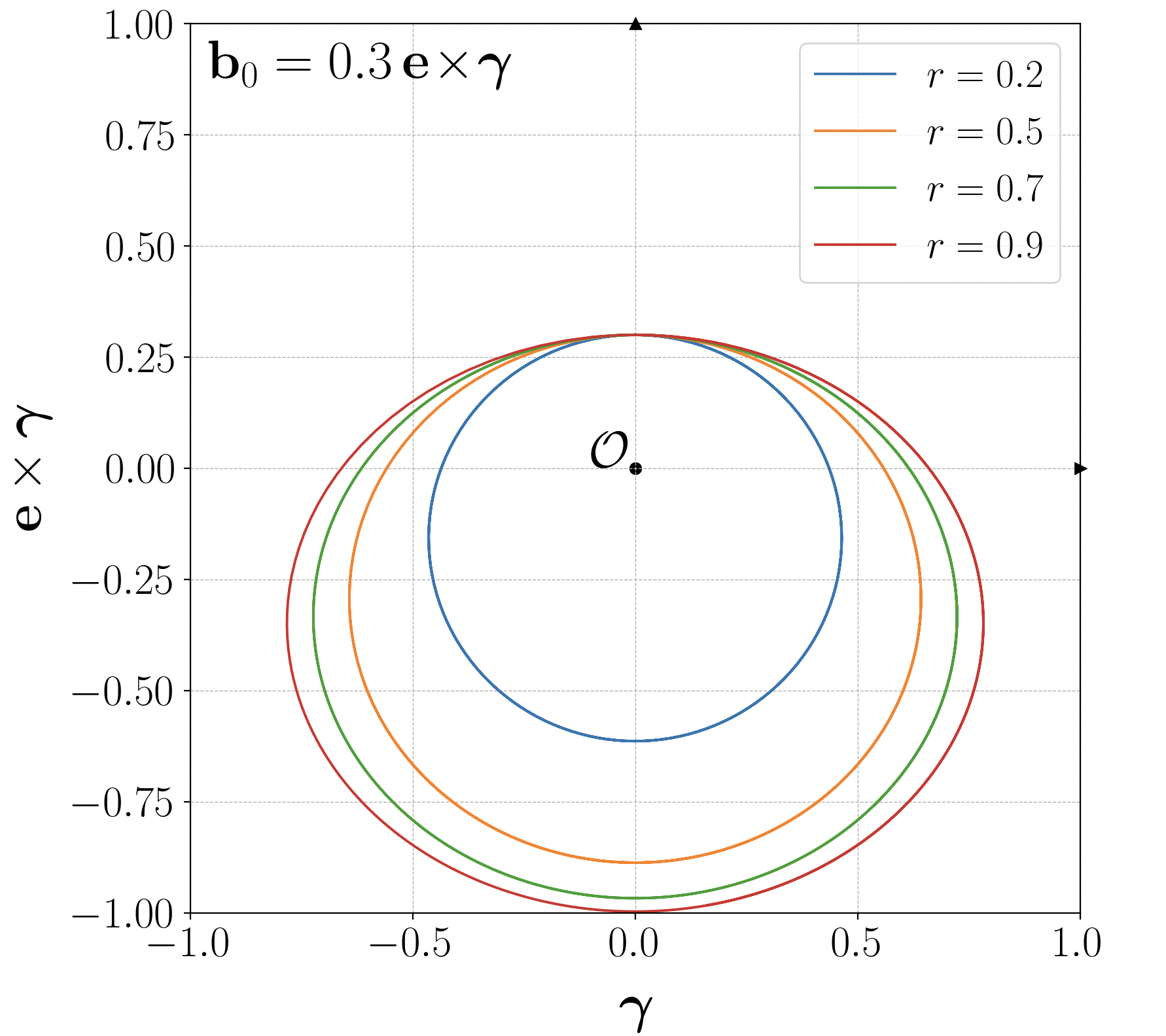}
        \caption{$\mathbf{b}_0\cdot\boldsymbol{\gamma}=0$}
    \end{subfigure}
    \begin{subfigure}{0.49\textwidth}
        \includegraphics[width=\linewidth]{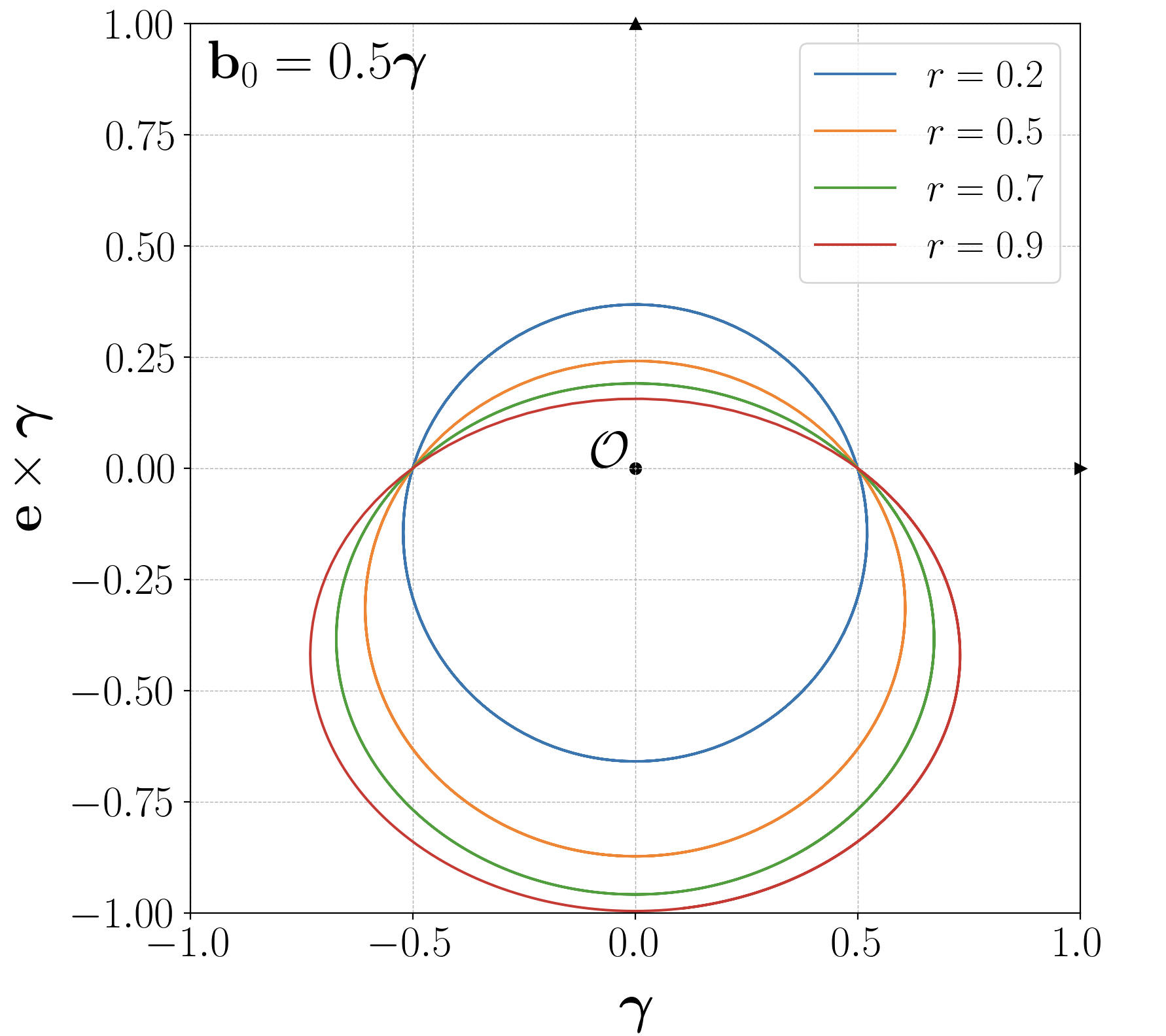}
        \caption{$\mathbf{b}_0\cdot(\mathbf{e}\times\boldsymbol{\gamma})=0$}
    \end{subfigure}
\\
    \begin{subfigure}{0.49\textwidth}
        \includegraphics[width=\linewidth]{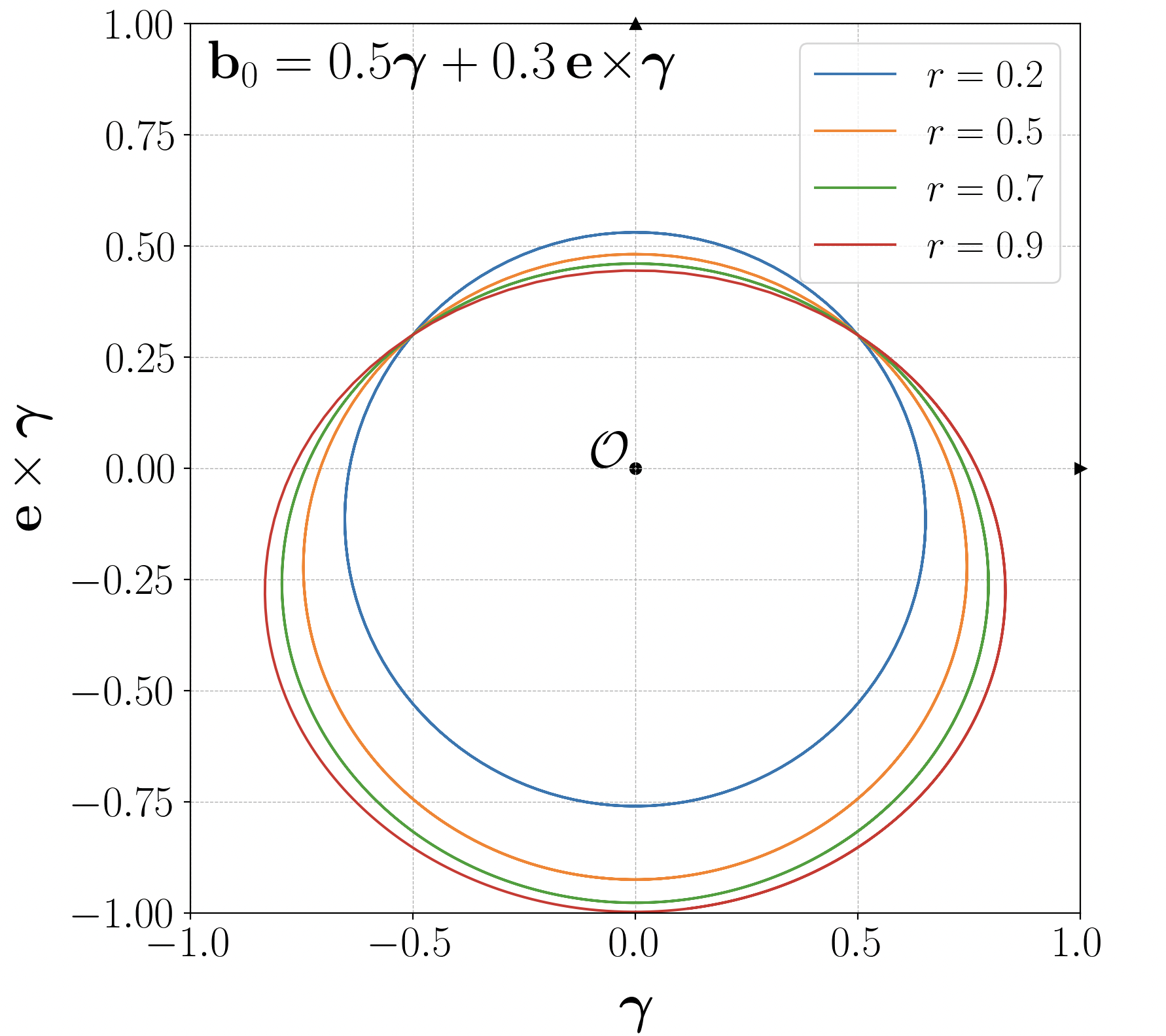}
        \caption{$\mathbf{b}_0\cdot(\mathbf{e}\times\boldsymbol{\gamma})> 0$}
    \end{subfigure}
    \begin{subfigure}{0.49\textwidth}
        \includegraphics[width=\linewidth]{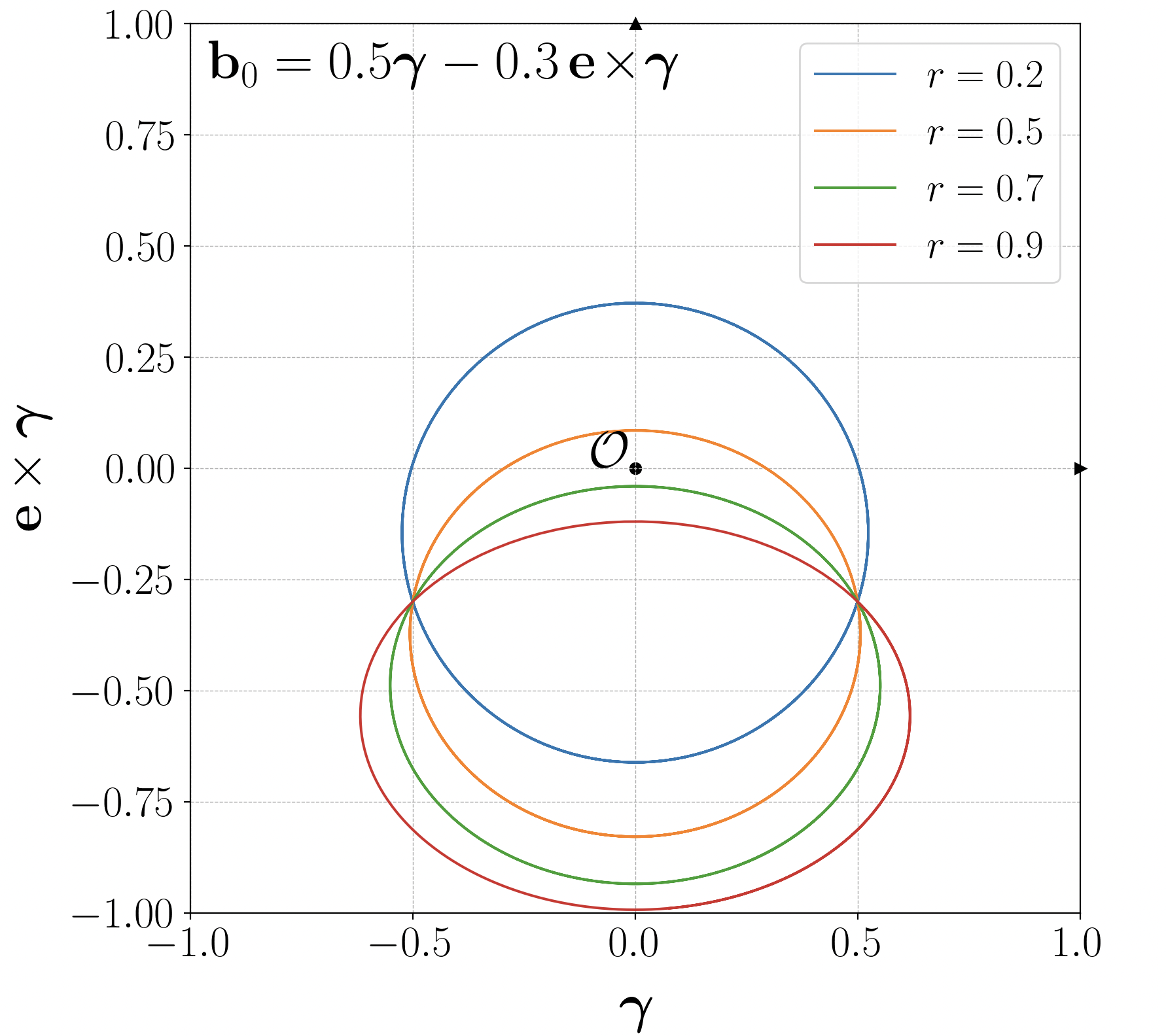}
        \caption{$\mathbf{b}_0\cdot(\mathbf{e}\times\boldsymbol{\gamma})< 0$}
    \end{subfigure}

    \caption{Plots of $\mathbf{b}(t)$ with an initial mixed state $\mathbf{b}_0$, for different values of $r$.}\label{fig: mixCUQplots_dr_1}
\end{figure}

\subsection{Bloch Sphere Orbits for $\mathbf{b}(0)\perp \mathbf{e}$}

Using the analytical result of the CUQ Bloch vector~${\bf b}(t)$ in~\eqref{genCUQ_gensol}, we can compute the equation of the curve traced by~$\mathbf{b}(t)$ in the $\{\boldsymbol{\gamma},\mathbf{e}\times\boldsymbol{\gamma}\}$ plane. To this end, we define
\begin{equation}
    X \:\equiv\: \dfrac{\sqrt{1-r^2}\sin{(2\theta+\beta_0)}\cos{\delta_0}}{1 - r\cos{(2\theta+\beta_0)}\cos{\delta_0}}\; ,\qquad Y\: \equiv\: \dfrac{\cos{(2\theta+\beta_0)\cos{\delta_0}-r}}{1 - r\cos{(2\theta+\beta_0)}\cos{\delta_0}}\;.
\end{equation}
With some algebra, we obtain the following relation between $X$ and $Y$:
\begin{equation}
  \label{genellipse_CUQ}
    \dfrac{X^2}{a^2}\: +\: \dfrac{\left(Y+e^2\right)^2}{a^2(1-e^2)}\ =\ 1\,,
\end{equation}
which is the equation for an {\em ellipse}. 
In~\eqref{genellipse_CUQ}, $a$ is the semi-major axis given by
\begin{equation}
    a\ =\ \left(\dfrac{1-r^2}{1-r^2\cos^2\delta_0}\right)^{1/2}\cos{\delta_0}\,,
\end{equation}
and $a\sqrt{1-e^2}$ is the semi-minor axis, conventionally denoted as $b$. In addition, $e$ is the eccentricity of the ellipse:
\begin{equation}
\begin{aligned}
    e\ &=\ \dfrac{r\sin{\delta_0}}{\sqrt{1-r^2\cos^2{\delta_0}}}\;.
\end{aligned} 
\end{equation}
For a CUQ mixed maximally with $\mathbf{b}_0 = \mathbf{0}$, the redefined parameters of the initial state~$\beta_0$ and~$\delta_0$ are $\beta_0 = 0$ and $\tan{\delta_0} = \sqrt{1-r^2}/r$. The eccentricity of the elliptical orbits of the maximally mixed state in the Bloch sphere is
\begin{equation}
    e\: =\: \dfrac{r}{\sqrt{1+r^2}}\;.
\end{equation}
Note that the ellipse is always symmetric about the $X$-axis, i.e.~along $\boldsymbol{\gamma}$, but shifted along the $Y$-axis
(or $\mathbf{e}\times\boldsymbol{\gamma}$-axis) by an amount $-e^2$.

\begin{figure}[t!]
    \centering
    \begin{subfigure}{0.49\textwidth}
        \includegraphics[width=\linewidth]{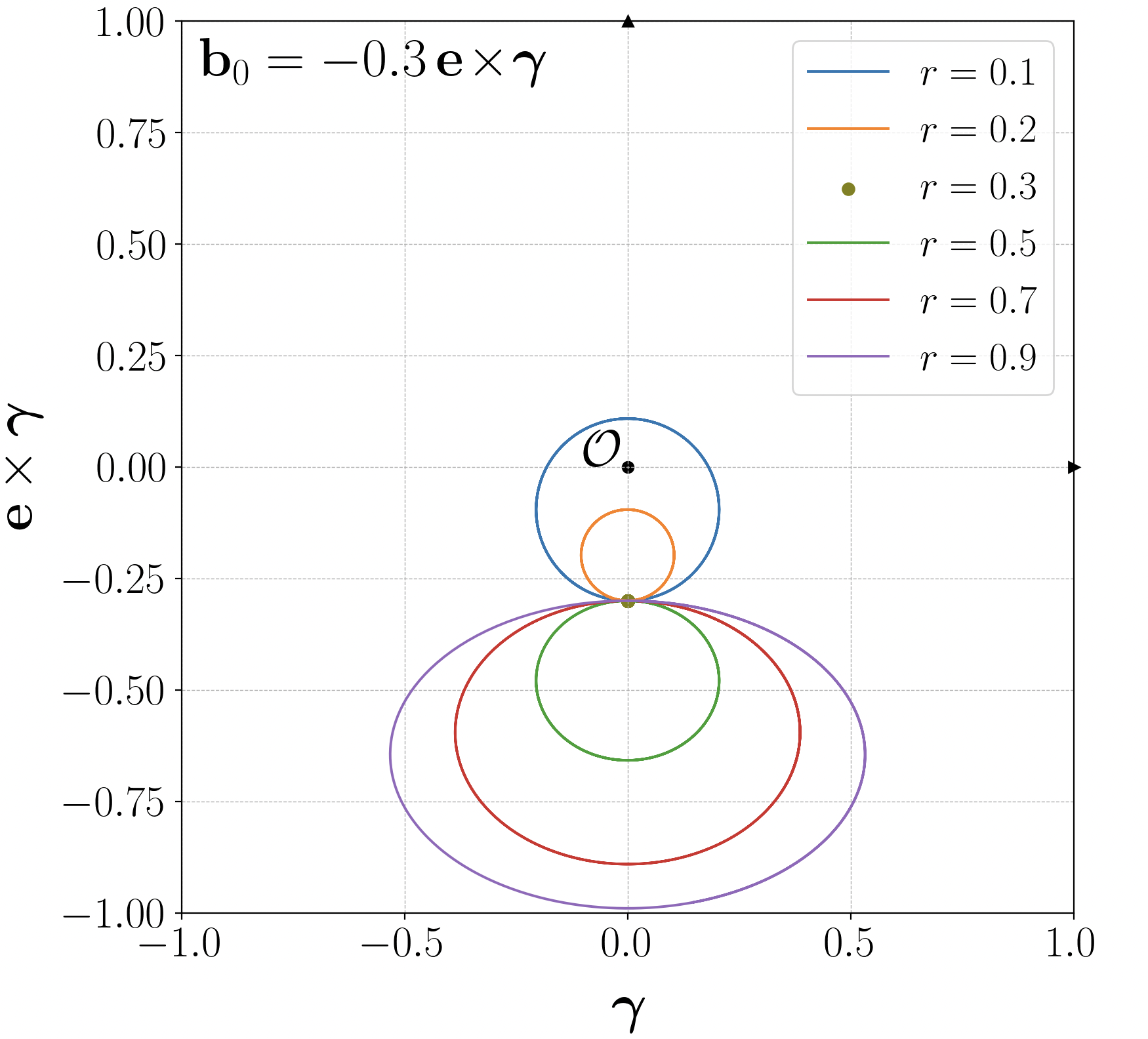}
    \end{subfigure}
    \begin{subfigure}{0.49\textwidth}
        \includegraphics[width=\linewidth]{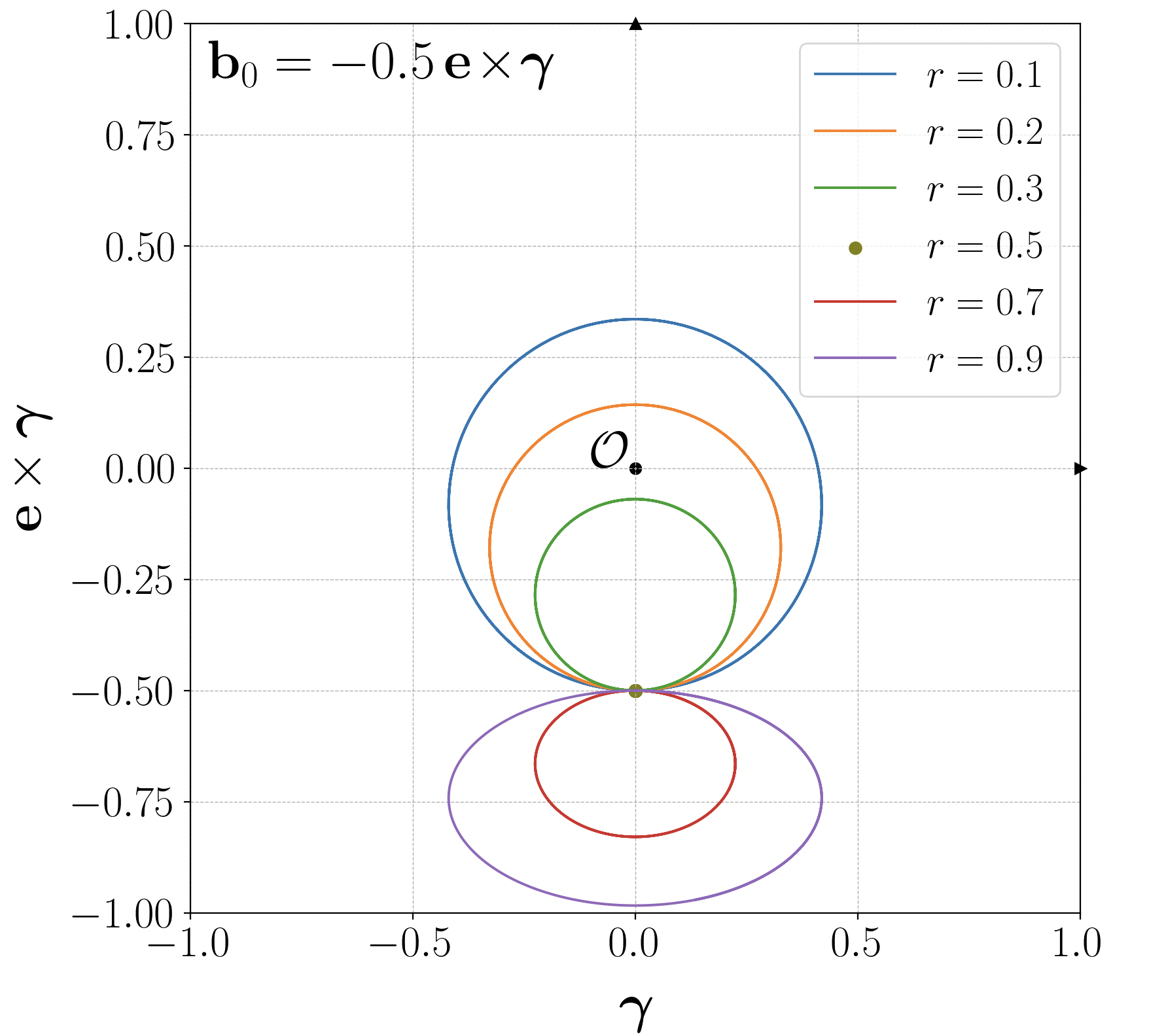}
    \end{subfigure}

    \caption{Plot of $\mathbf{b}(t)$ with initial mixed state of the form $\mathbf{b}_0\cdot\boldsymbol{\gamma}=0$ and $\mathbf{b}_0\cdot(\mathbf{e}\times\boldsymbol{\gamma})< 0$, for different values of $r$}\label{fig: mixCUQplots_dr_2}
\end{figure}

Figures \ref{fig: mixCUQplots_dr_1} and \ref{fig: mixCUQplots_dr_2} show the Bloch-sphere CUQ trajectories ${\bf b}(t)$ of initially mixed states ${\mathbf{b}_0 = {\bf b}(0)}$ for different values of $r$.  As mentioned above, we see that the curves are symmetric along the $\boldsymbol{\gamma}$-axis but asymmetric along the $\mathbf{e}\times\boldsymbol{\gamma}$-axis. The eccentricity of the elliptical orbit is higher for a larger value of $r$. For initially mixed CUQs of the form $-|c|\, \mathbf{e}\times \boldsymbol{\gamma}$, the trajectory of the Bloch vector is a circle for $r=0$ and stretches to become an ellipse as $r$ increases. However, as explained in Section~\ref{sec:Stationarypnts}, the size of the orbit reduces and reaches a stationary point when $r=|c|$, as can be seen in Figure~\ref{fig: mixCUQplots_dr_2}. The CUQ coherence-decoherence oscillation is resumed once again by increasing further $r$. For $r<|c|$, the trajectory in the Bloch sphere remains above the $-|c|\,\mathbf{e}\times\boldsymbol{\gamma}$ point and shifts below it when $r>|c|$. This transition is well exemplified in Figure~\ref{fig: mixCUQplots_dr_2}.

\begin{figure}[t!]
    \centering
    \begin{subfigure}{0.49\textwidth}
        \includegraphics[width=\linewidth]{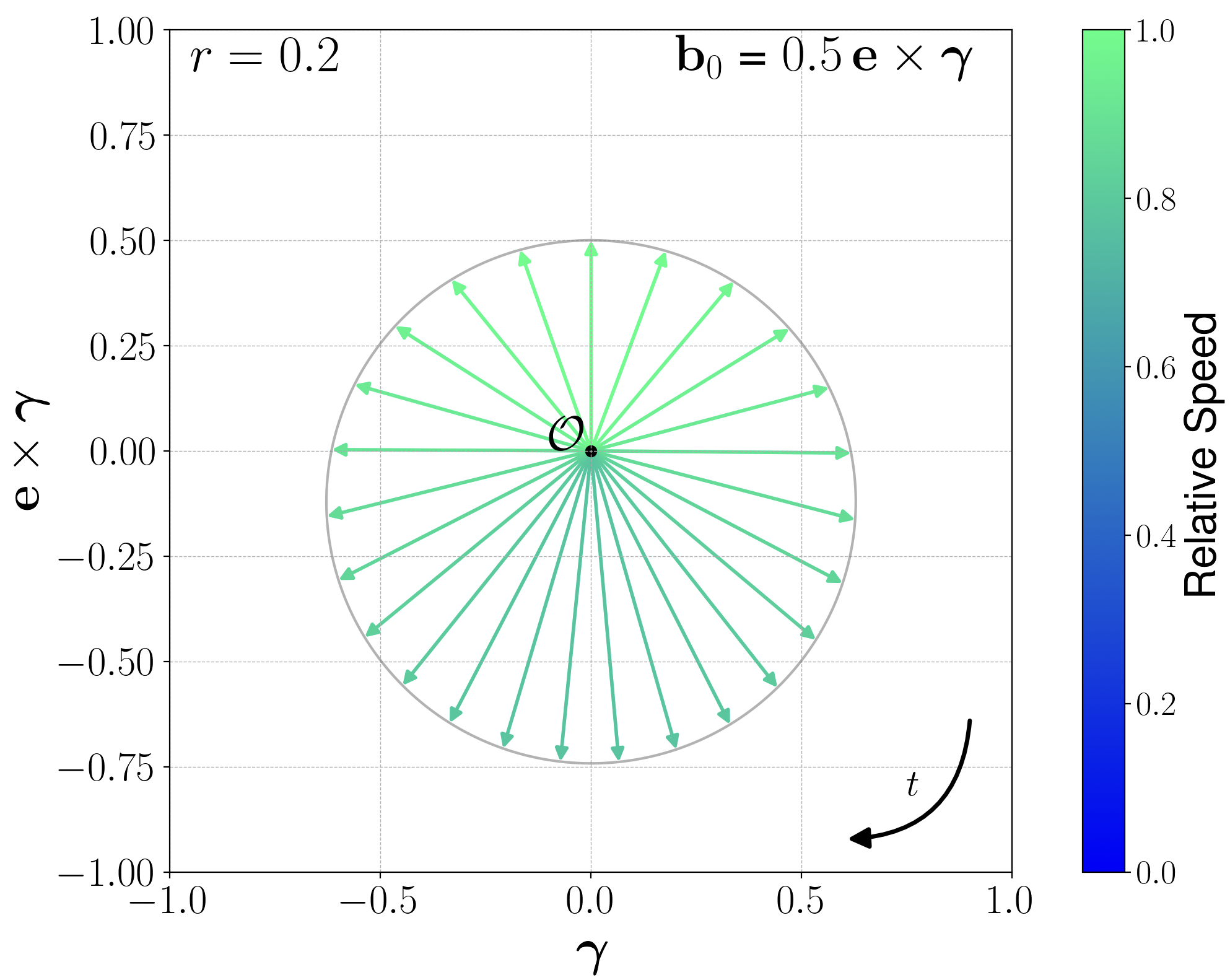}
    \end{subfigure}
    \begin{subfigure}{0.49\textwidth}
        \includegraphics[width=\linewidth]{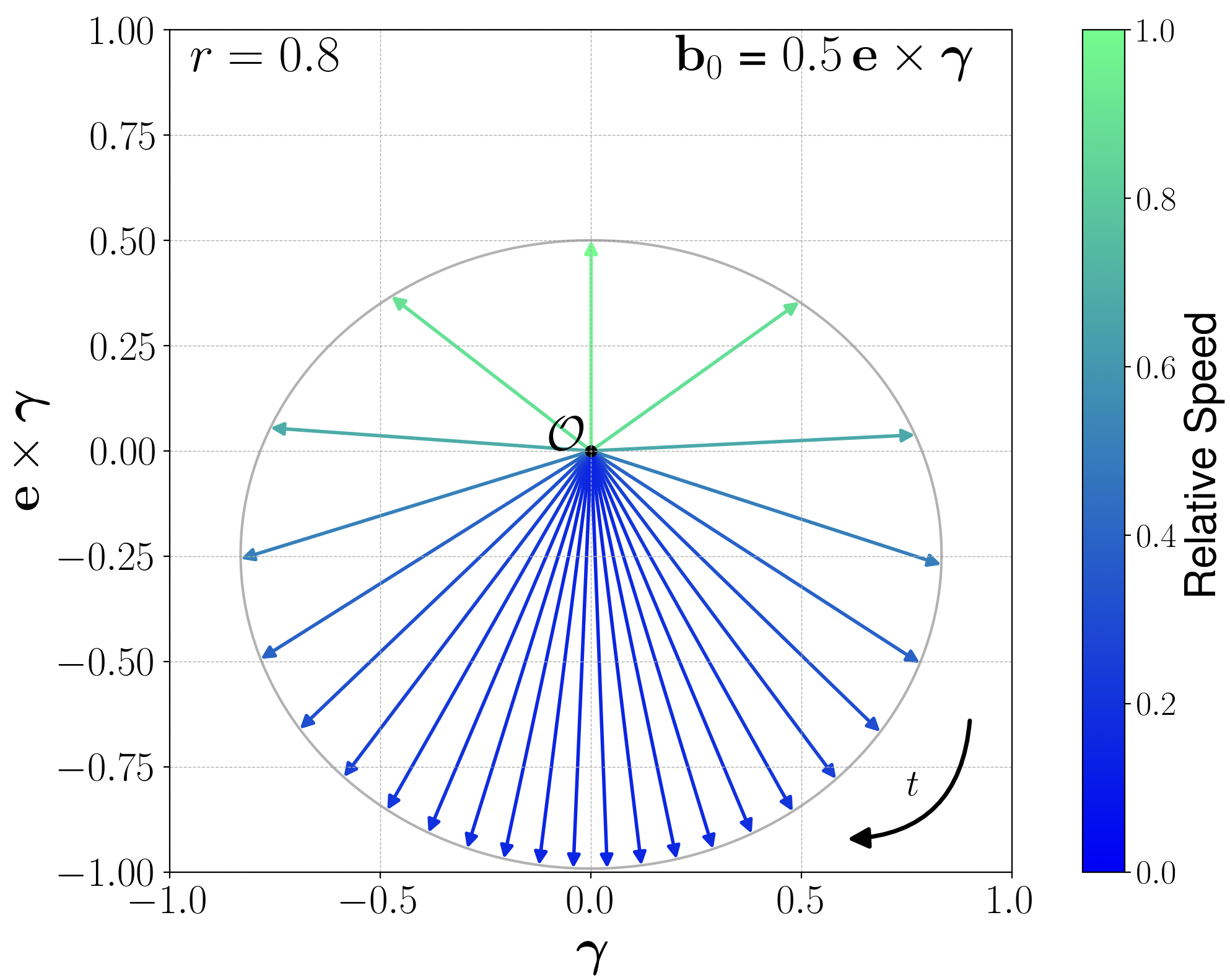}
    \end{subfigure}

    \caption{CUQ anharmonicity and coherence-decoherence oscillations for different values of $r$.}\label{fig: Anharm func r}
\end{figure}

As the value of $r$ increases, the degree of anharmonicity of the CUQ oscillation also increases. This can be established through a Fourier-series analysis of CUQ oscillations as done in\cite{Karamitros:2025azy}. In~Figure~\ref{fig: Anharm func r}, we show the anharmonic oscillation of the same initial mixed CUQ state for two different values of $r$: $r = 0.2$ and $0.8$. The evolution of the Bloch vector is captured in equal time intervals, and the color grading displays the relative speed of motion of the Bloch vector~${\bf b}(t)$. Evidently, the anharmonicity of the oscillations is greater for a larger value of $r$.

\subsection{Bloch Sphere Orbits for $\mathbf{b}(0) \not\perp \mathbf{e}$}
\begin{figure}[t!]
\centering
    \begin{subfigure}{0.49\textwidth}
        \includegraphics[width=\linewidth]{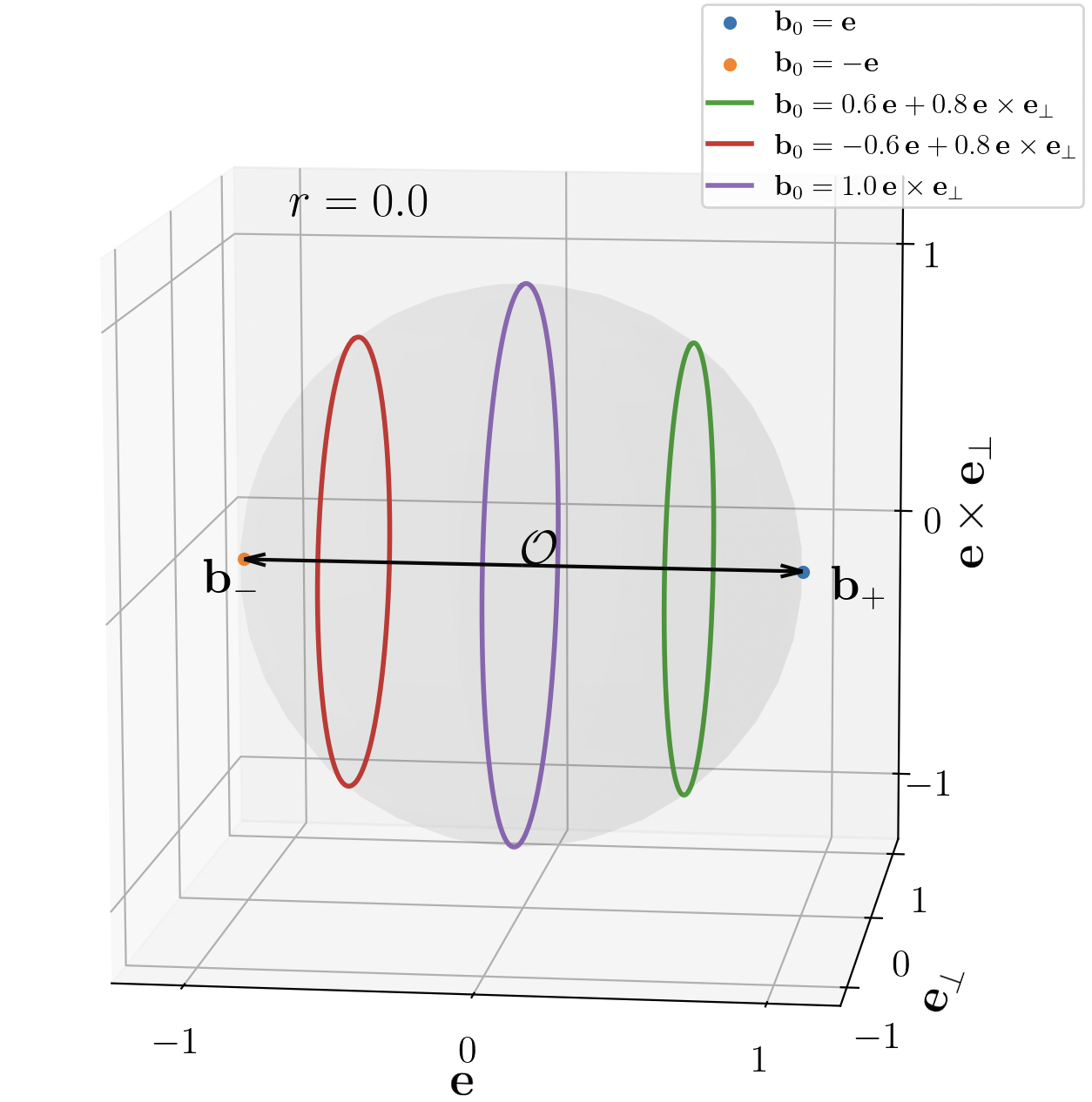}
        \caption{Rabi}
    \end{subfigure}
    \begin{subfigure}{0.49\textwidth}
        \includegraphics[width=\linewidth]{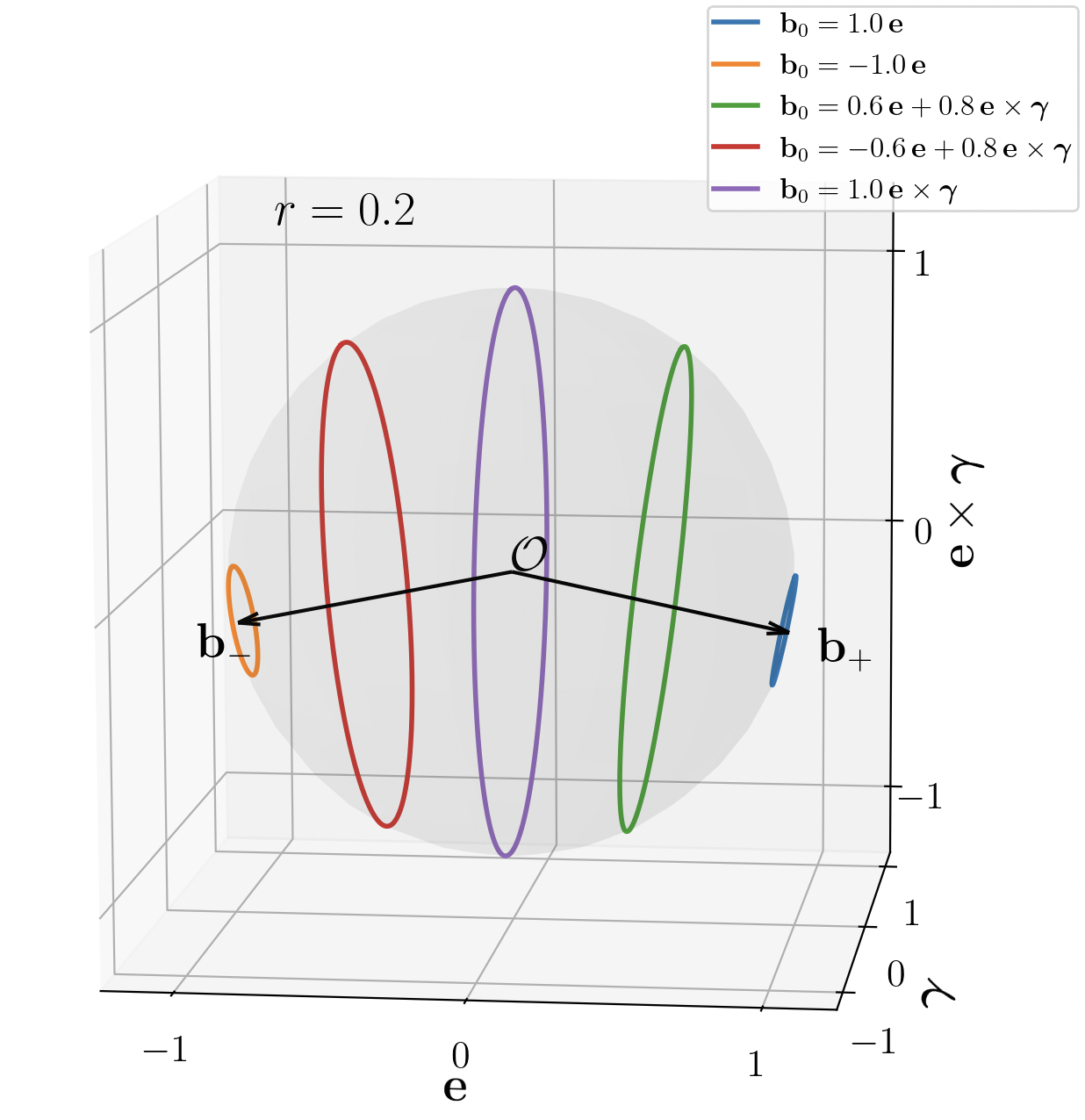}
        \caption{CUQ}
    \end{subfigure}

    \caption{Comparative Bloch-sphere trajectory for different pure state oscillations in (a) Hermitian and (b) Non-Hermitian systems. The two black vectors are the two Eigenstates of the Hamiltonian, $\mathbf{b}_{\pm}$.}\label{fig: Rabi_v_CUQ 3D}
\end{figure}

We now consider the more general case of a CUQ with an initial mixed state $\mathbf{b}_0\not\perp\mathbf{e}$, and study its Bloch-sphere trajectories. We take the initial mixed CUQ state to be
\begin{equation}
    \mathbf{b}_0\: \equiv\: {\bf b}(0)\ =\ b_e\,\mathbf{e}\: +\: b_0\cos{\varphi_0}\,\boldsymbol{\gamma}\: +\: b_0\sin{\varphi_0}\,\mathbf{e}\times \boldsymbol{\gamma}\, ,\quad \label{initialCUQmixpolar}
\end{equation}
with $0\leq (b_e^2 + b_0^2)^{1/2}\leq 1.$
As we did in Section~\ref{sec:MixStatesolgen}, we substitute the initial state ${\bf b}_0$ in \eqref{mixCUQ_sol} to compute the time evolution of CUQ. In this way, we find
\begin{equation}
    \mathbf{b}(t) = \dfrac{b_e(1-r\cos{\delta_0}\cos{\beta_0})\mathbf{e}+\sqrt{1-r^2}\sin{(2\theta+\beta_0)}\cos{\delta_0}\boldsymbol{\gamma}+[\cos{(2\theta+\beta_0)\cos{\delta_0}-r}]\mathbf{e}\times\boldsymbol{\gamma}}{1-r\cos{(2\theta+\beta_0)}\cos{\delta_0}}.
    \label{General mix state sol CUQ}
\end{equation}

\begin{figure}[t!]
    \centering
    \begin{subfigure}{0.49\textwidth}
        \includegraphics[width=\linewidth]{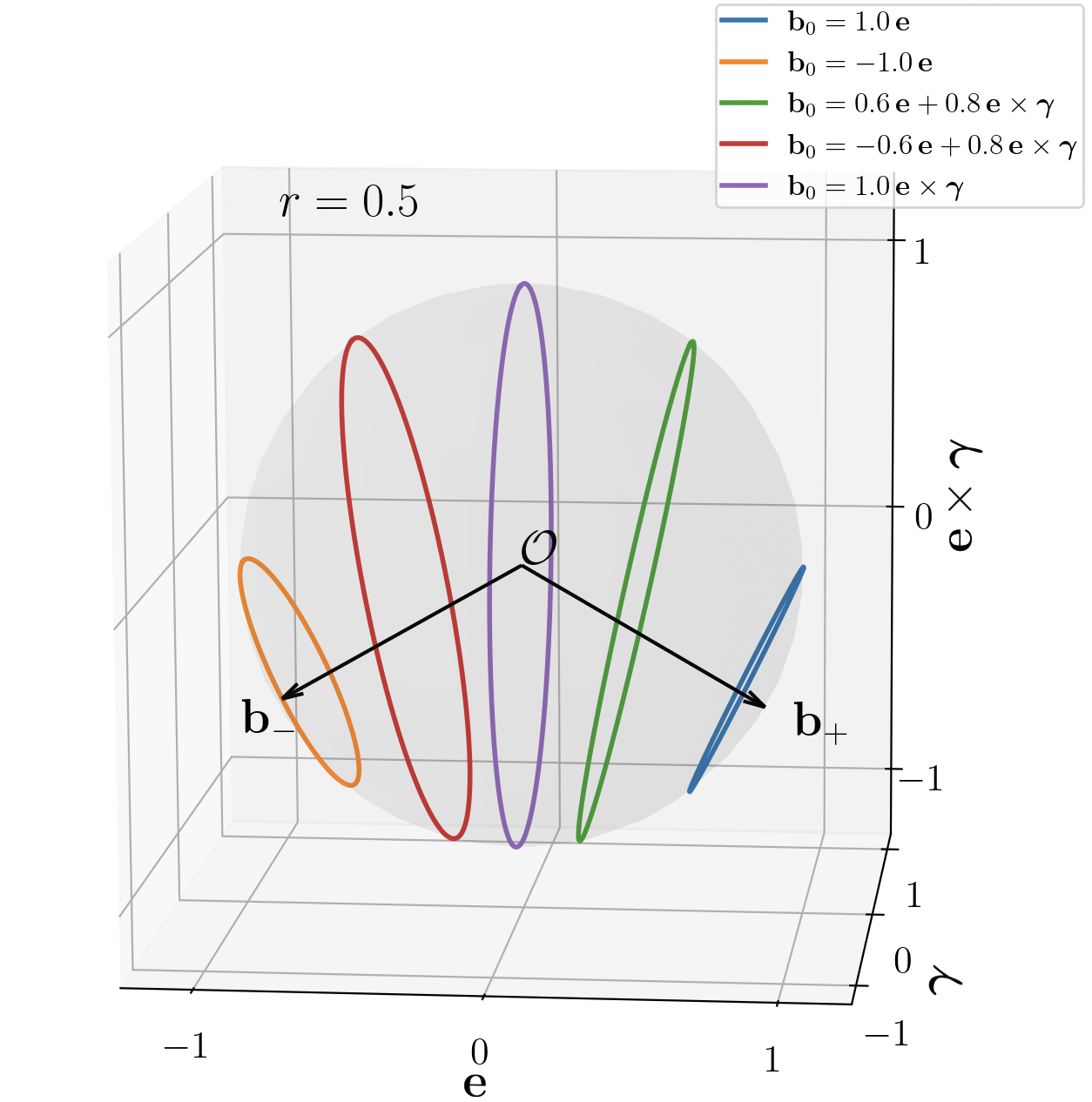}
    \end{subfigure}
    \begin{subfigure}{0.49\textwidth}
        \includegraphics[width=\linewidth]{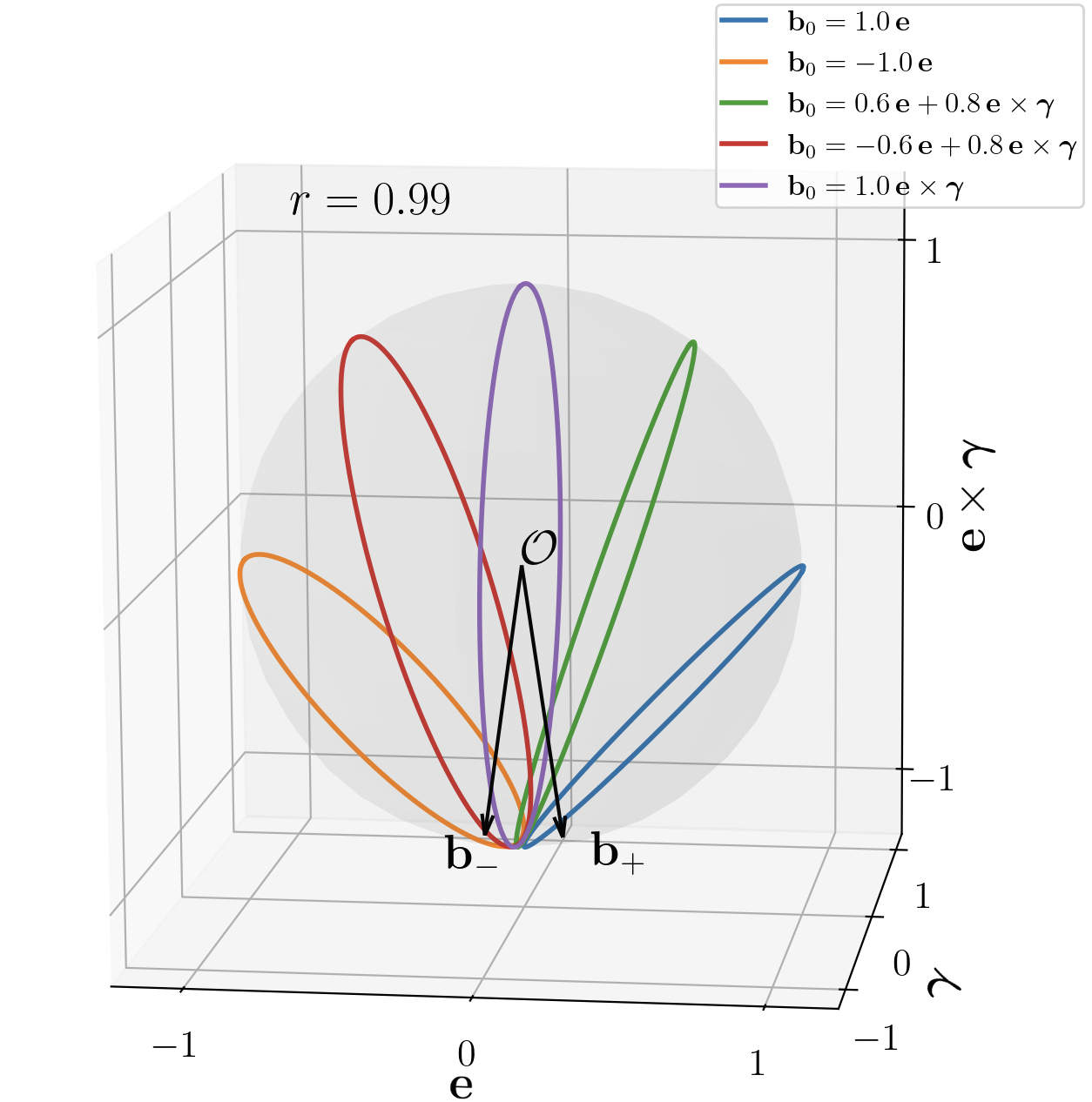}
    \end{subfigure}

    \caption{Bloch-sphere trajectories of pure states with $\mathbf{b}_0\cdot\mathbf{e}\neq 0$ for different values of $r$.}\label{fig: CUQ tilt r}
\end{figure}

If the CUQ is initially in a pure state, we use the polar form \eqref{initialCUQmixpolar} and ensure $|\mathbf{b}(t)|=1$ by setting $b_e = \sqrt{1-b_0^2}$.  Then, \eqref{initialCUQmixpolar} becomes
\begin{equation}
  \mathbf{b}_{0,\text{pure}}\ =\ \sqrt{1-b_0^2}\:\mathbf{e}\: +\: b_0\cos{\varphi_0}\,\boldsymbol{\gamma}\: +\: b_0\sin{\varphi_0}\,\mathbf{e}\times \boldsymbol{\gamma}\, .  
\end{equation}
As a consequence, \eqref{General mix state sol CUQ} further simplifies into
\begin{equation}
    \mathbf{b}_{\text{pure}}(t)\: =\: \dfrac{\sqrt{1-r^2}[\sin{\delta_0}\mathbf{e}+\sin{(2\theta+\beta_0)}\cos{\delta_0}\boldsymbol{\gamma}]+[\cos{(2\theta+\beta_0)\cos{\delta_0}-r}]\mathbf{e}\times\boldsymbol{\gamma}}{1-r\cos{(2\theta+\beta_0)}\cos{\delta_0}}\; .
    \label{General pure state sol CUQ}
\end{equation}
Figure~\ref{fig: Rabi_v_CUQ 3D} provides a comparison of the general solution of different pure state CUQ oscillations with Rabi oscillations. We observe a shift in the direction of the eigenstates of CUQ at a tilted angle compared to that of the stable qubit case. By comparing the same Bloch-vector trajectories in the two cases, we see the CUQ oscillations on a plane that lies at a $\textit{tilted}$ angle with respect to the $\mathbf{e}$-axis, as opposed to the Hermitian stable qubit case, where the plane on which the Bloch-vector orbits is perpendicular to $\mathbf{e}$. As shown in Figure~\ref{fig: CUQ tilt r}, this $\textit{tilt}$ of the planes of CUQ oscillations is more pronounced as $r$ increases and approaches $1$.

\section{Geometric Construction of CUQ Motion}\label{sec:BlochGeometry}

The time evolution of a stable qubit forming a two-level Hermitian system has a simple geometric solution on the Bloch sphere. Starting from an initial state $\mathbf{b}_0$, the trajectory of its evolution on the Bloch sphere is simply a circle on a plane orthogonal to the $\mathbf{e}$-axis. In the case of CUQs, however, the Bloch-Sphere trajectories can become rather complex for two reasons. First, the plane on which the CUQ orbit lies depends on the value of $r \equiv |{\bf \Gamma}|/2|{\bf E}|$  and on the initial co-decaying state $\mathbf{b}_0 \equiv {\bf b}(0)$. Second, the orbits are circular only for pure states; and are elliptic for mixed states, the eccentricity of which again depends on~$r$ and~$\mathbf{b}_0$. However, by studying the analytic solution closely, we find for the first time a non-trivial geometric construction to describe the trajectory of a CUQ {\em in} and {\em on} the Bloch sphere.

\subsection{Motion of Pure CUQ on the Bloch Sphere\label{PureGeometrysol}}

\begin{figure}[ht!]
    \centering
    \begin{subfigure}{0.49\textwidth}
     \includegraphics[width=\linewidth]{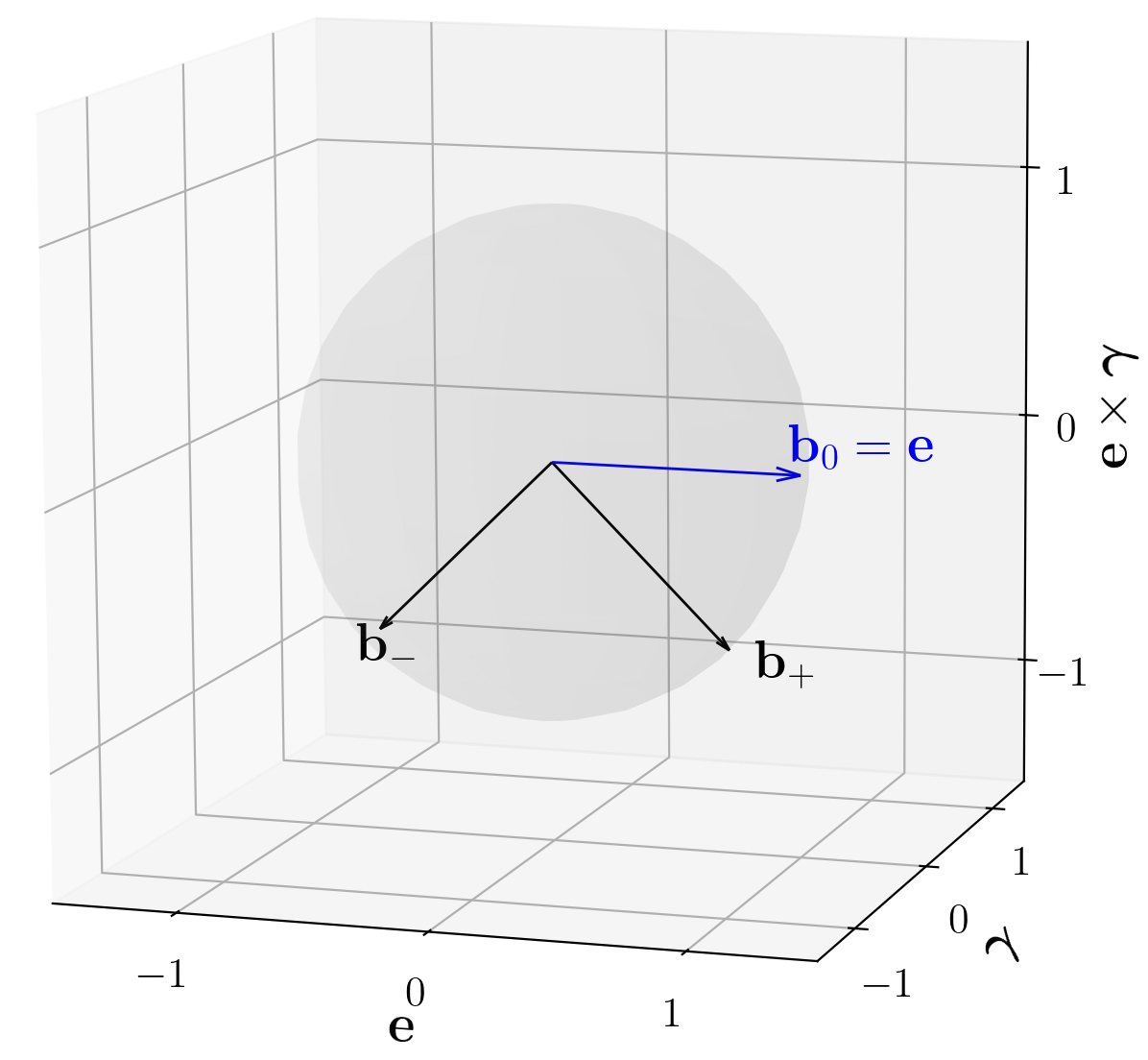}
        \caption{Initial pure state}\label{purestep0}
    \end{subfigure}   
    \begin{subfigure}{0.49\textwidth}
    \includegraphics[width=\linewidth]{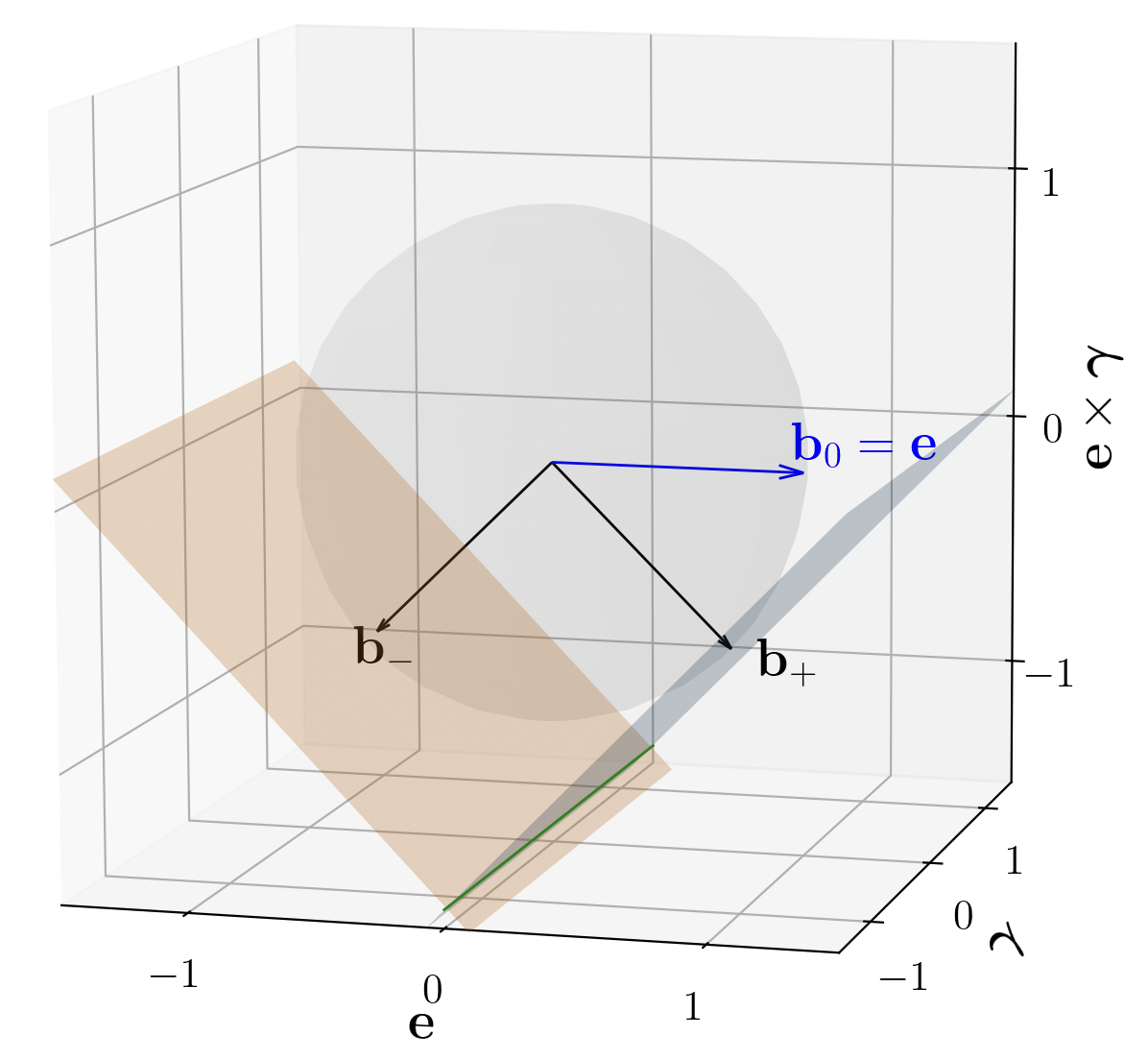}
        \caption{Step 1}\label{purestep1}
    \end{subfigure}
\\
    \begin{subfigure}{0.49\textwidth}
     \includegraphics[width=\linewidth]{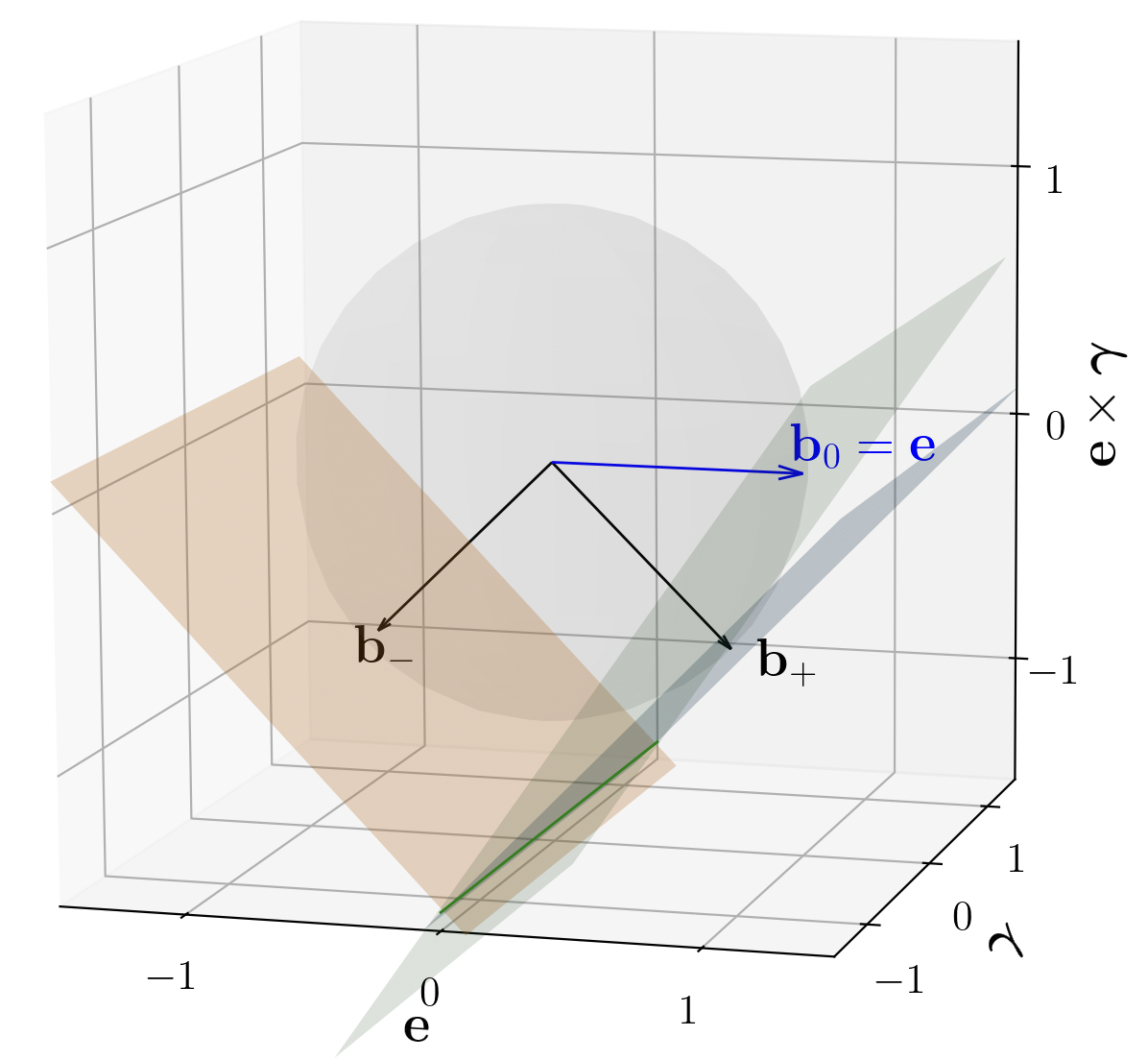}
        \caption{Step 2}\label{purestep2}
    \end{subfigure}   
    \begin{subfigure}{0.49\textwidth}
    \includegraphics[width=\linewidth]{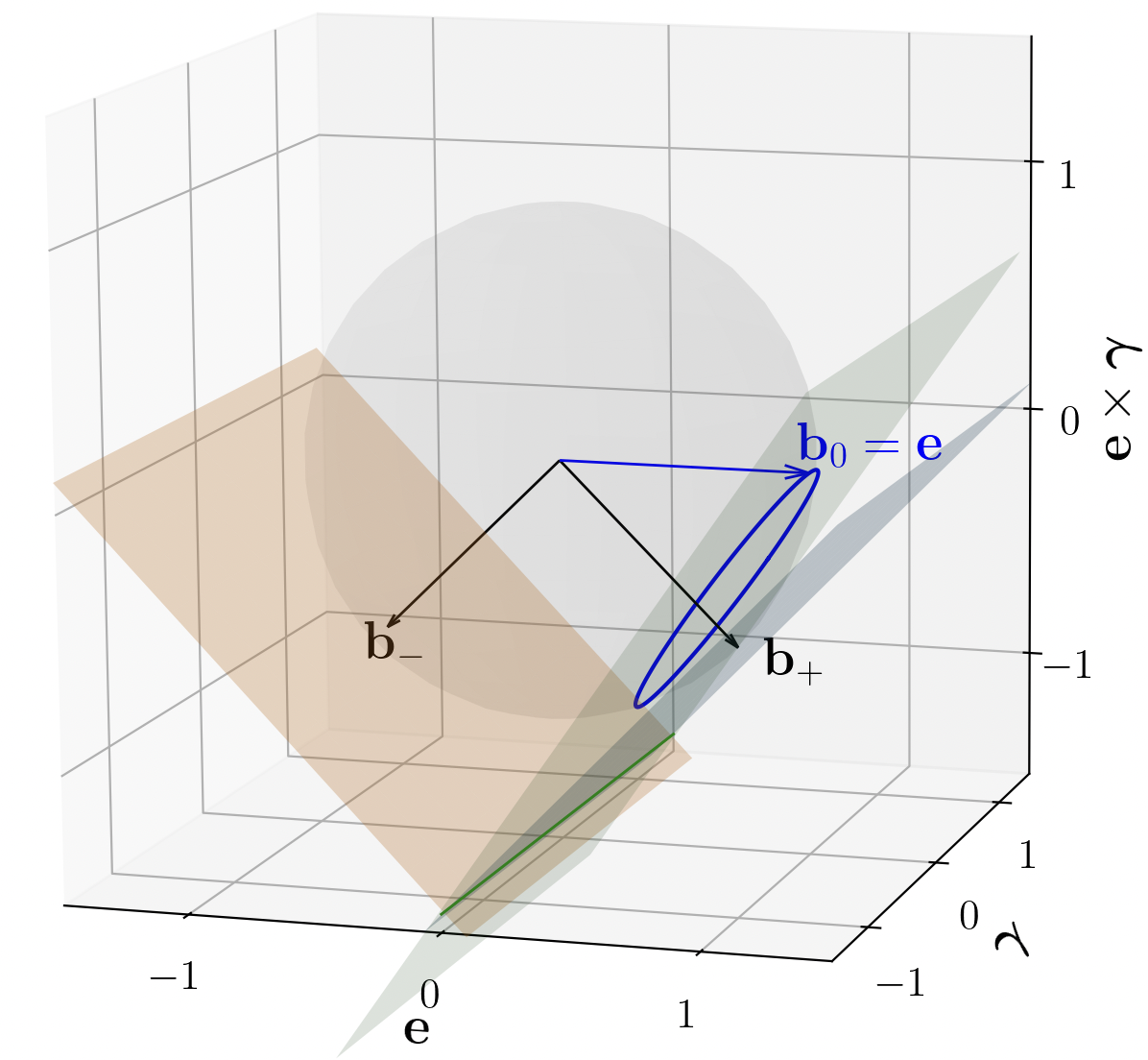}
        \caption{Step 3}\label{purestep3}
    \end{subfigure}

    \caption{Geometric construction of CUQ orbit ${\bf b}(t)$ for an initial state $\mathbf{b}_0 = \mathbf{e}$. The plane cuts the Bloch sphere along a circular patch. Since $\mathbf{b}_0$ is a pure state, the boundary of the circular patch is also the Bloch-sphere trajectory of this pure CUQ state.}\label{fig: Plane construction CUQ}
\end{figure}

We begin with the geometric construction of the orbit of a pure CUQ state on the Bloch sphere. As discussed in the previous section, we know that a pure CUQ Bloch vector ${\bf b}(t)$ will orbit along the surface of the Bloch sphere in a circular path, just like in the case of a stable qubit. The difference — other than the crucial anharmonicity — is that the plane on which the pure CUQ oscillates is not easily determined, but each time needs to be evaluated to obtain its trajectory. In Figure~\ref{fig: CUQ tilt r}, we see the pure states orbiting on different $\textit{tilted}$ planes. We should clarify that the planes are $\textit{tilted}$ at an angle with respect to the $\mathbf{e}$-axis only. The $\textit{tilt}$ in the plane of the Bloch-sphere trajectories can be computed analytically by performing a Frenet-Serret construction, as done in Appendix A of\cite{Karamitros:2025azy}. Nonetheless, there is a geometric construction that helps visualize the $\textit{tilted}$ plane in which the initial CUQ  Bloch vector traverses its path. This is presented schematically step-by-step in the geometric construction in Figure~\ref{fig: Plane construction CUQ}, which we outline below.

We begin with our initial state $\mathbf{b}_0$ on the Bloch sphere, along with the two eigenvectors $\mathbf{b}_\pm$ of the CUQ Hamiltonian [cf.~\eqref{eq:bplusminus}]. In Figure~\ref{purestep0}, the initial pure state considered is $\mathbf{b}_0 = \mathbf{e}$. As a first step, shown in Figure~\ref{purestep1}, we construct two tangent planes on the Bloch sphere that are tangent to the two end-points of the eigenvectors $\mathbf{b}_{\pm}$. The two tangent planes intersect each other along a line parallel to the $\boldsymbol{\gamma}$-axis, marked as the green line in Figure~\ref{purestep1}, at a distance $1/r$ in the negative $\mathbf{e}\times\boldsymbol{\gamma}$ direction. 

As shown in Figure~\ref{purestep2}, our second step consists of constructing a plane that passes through the green intersection line of the two tangent planes and the point of initial state $\mathbf{b}_0$ on the Bloch-sphere. This is the plane on which our initial state will evolve in the Bloch sphere, and as such we call the $\textit{solution plane}$ or the $\textit{orbit plane}$. The equation of the {\em intersection line}, which we denote by $\mathbf{I}$, on which the two tangent planes intersect, can be represented in vector form as
\begin{equation}
   \mathbf{I}\ =\  k\,\boldsymbol{\gamma}\: -\: \dfrac{1}{r}\,\mathbf{e}\times \boldsymbol{\gamma}\,, 
\end{equation}
with $k \in \mathbb{R}$.
For the initial state $\mathbf{b}_0 = \mathbf{e}$ considered in Figure~\ref{purestep0}, the solution plane shown in Figure~\ref{purestep2} can be imagined as slicing the Bloch sphere, creating a circular patch. The boundary of this circular patch is the Bloch-sphere trajectory that our pure state traverses, as shown in Figure~\ref{purestep3}. This is our third and last step in our geometric solution of the orbit of a pure CUQ. Notice that our solution encompasses the stable qubit limit $r\to 0$, in which case ${\bf I} \parallel \boldsymbol{\gamma}$ and the solution plane is also parallel to $\boldsymbol{\gamma}$, which implies that the orbit plane is perpendicular to $\mathbf{e}$ only. Consequently, we recover the Rabi oscillation orbits in Figure~\ref{fig: Rabi_v_CUQ 3D}.

\subsection{Motion of Mixed CUQ in the Bloch Sphere\label{MixGeometrysol}}

\begin{figure}[t!]
    \centering
    \begin{subfigure}{0.43\textwidth}
     \includegraphics[width=\linewidth]{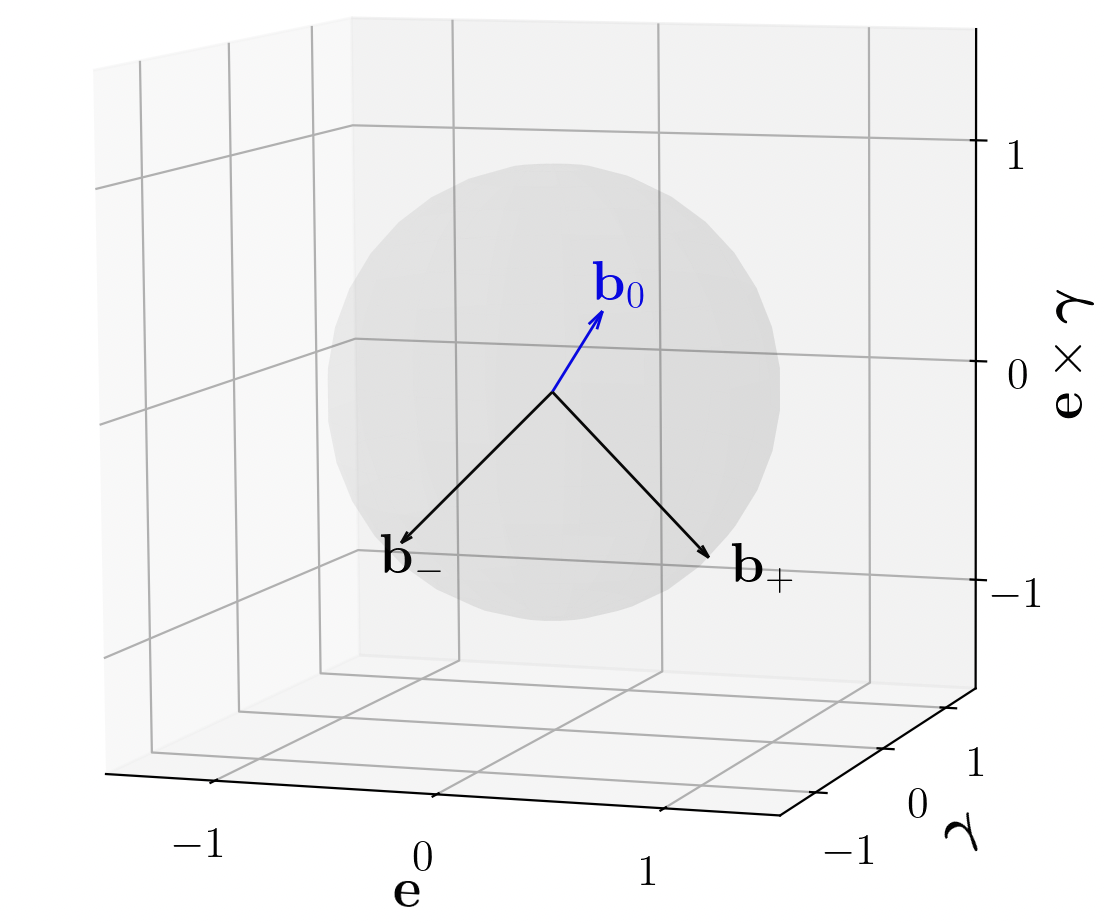}
        \caption{Initial mixed state}\label{mixstep0}
    \end{subfigure}   
    \begin{subfigure}{0.43\textwidth}
    \includegraphics[width=\linewidth]{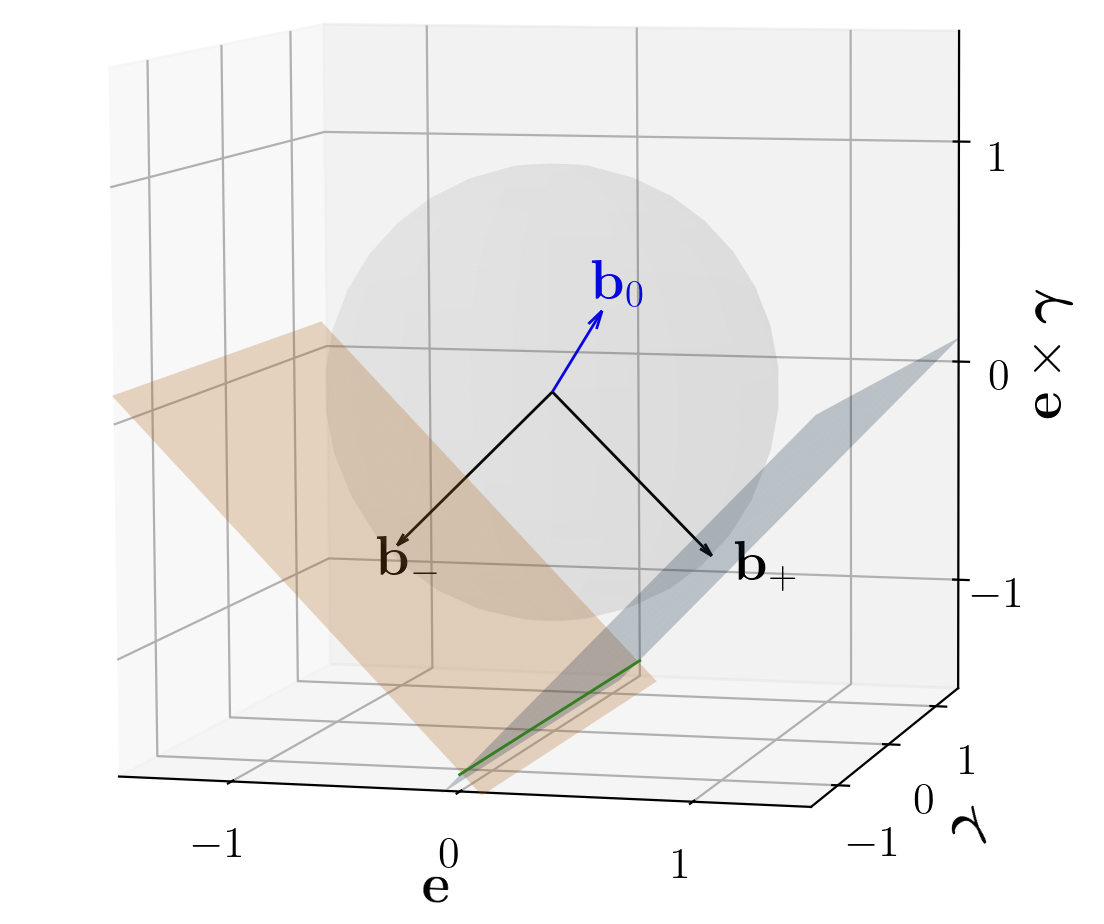}
        \caption{Step 1}\label{mixstep1}
    \end{subfigure}
\\
    \begin{subfigure}{0.43\textwidth}
     \includegraphics[width=\linewidth]{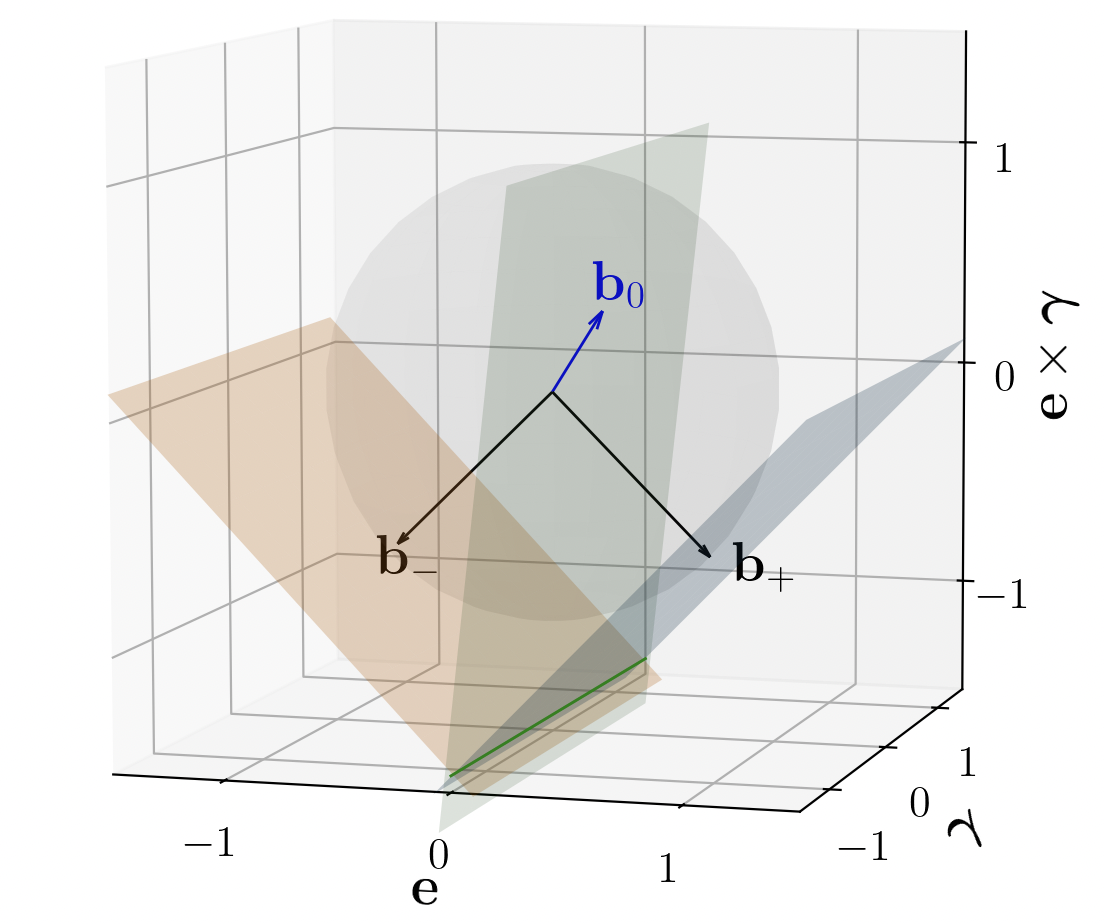}
        \caption{Step 2}\label{mixstep2}
    \end{subfigure}   
    \begin{subfigure}{0.43\textwidth}
    \includegraphics[width=\linewidth]{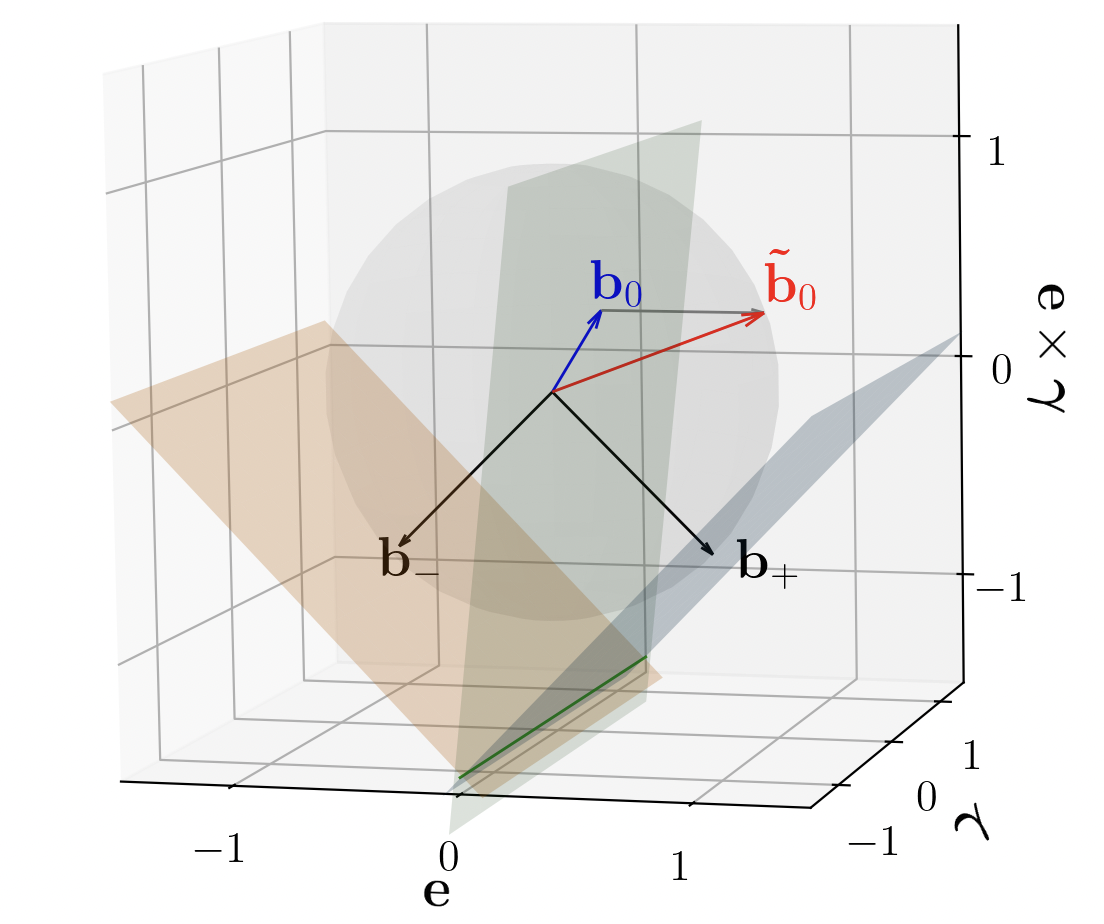}
        \caption{Step 3}\label{mixstep3}
    \end{subfigure}
\\
    \begin{subfigure}{0.43\textwidth}
     \includegraphics[width=\linewidth]{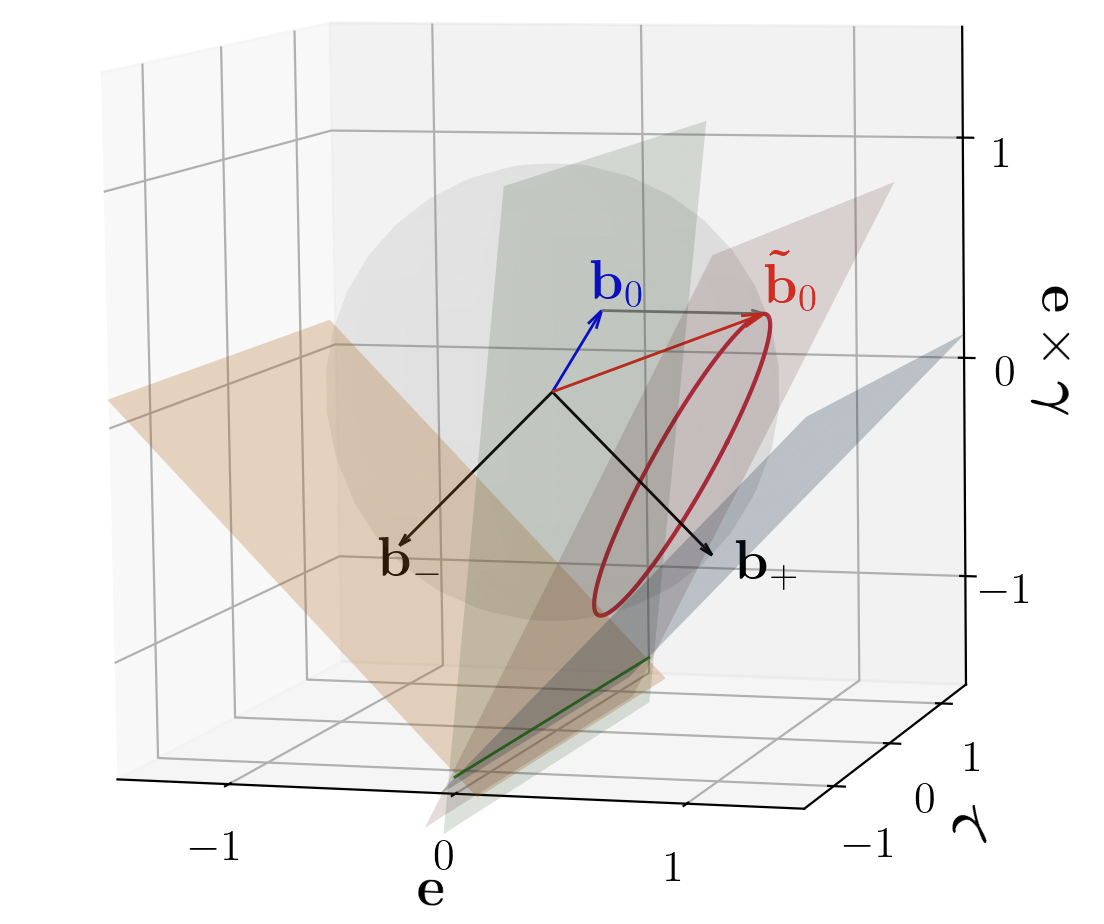}
        \caption{Step 4}\label{mixstep4}
    \end{subfigure}   
    \begin{subfigure}{0.43\textwidth}
    \includegraphics[width=\linewidth]{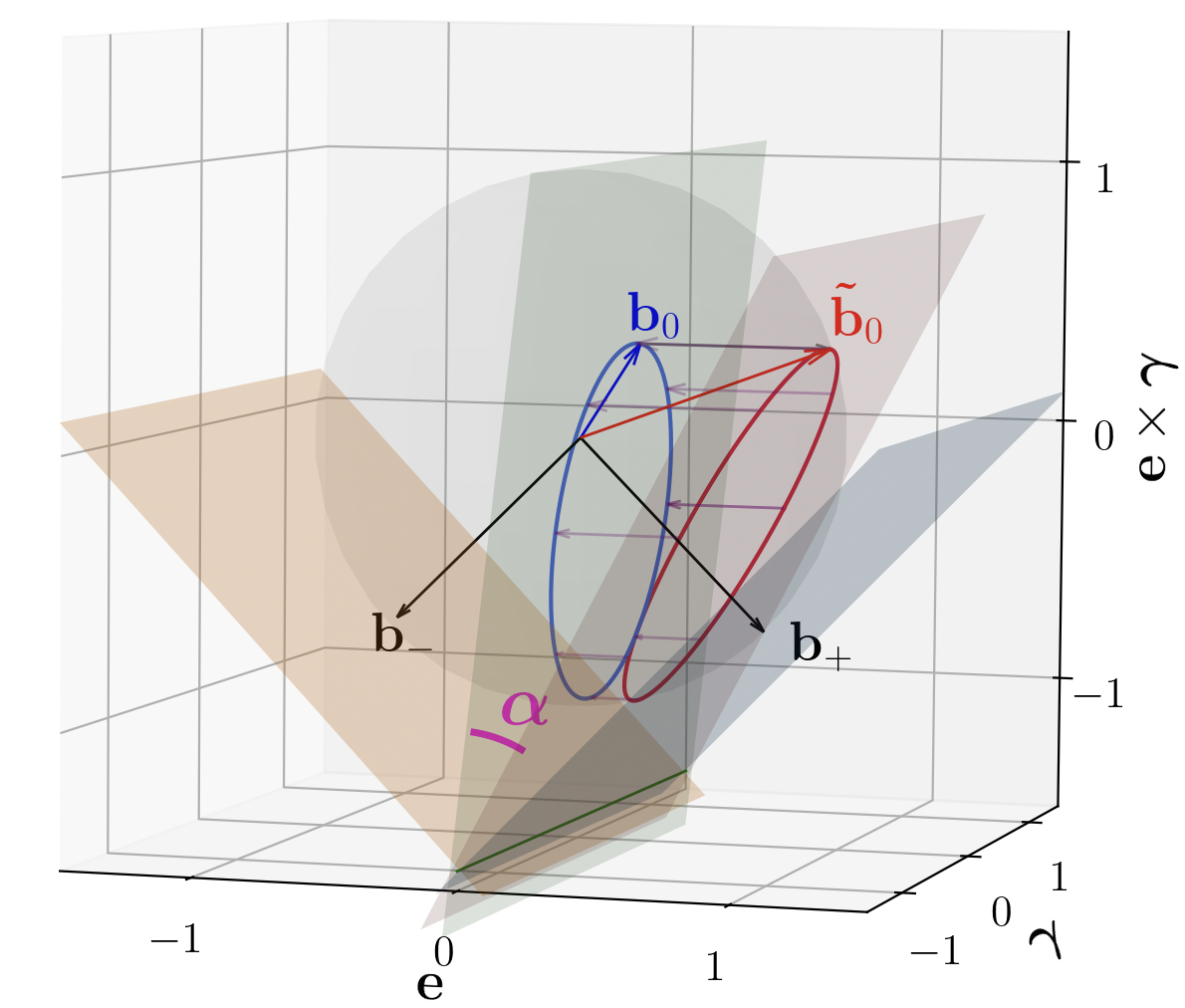}
        \caption{Step 5}\label{mixstep5}
    \end{subfigure}    
    \caption{Bloch-sphere trajectory of an initial mixed CUQ state $\mathbf{b}_0$, constructed geometrically.}
    \label{fig: Ellipse construction}
\end{figure}

The construction of the pure CUQ state orbit on the Bloch sphere in Section~\ref{PureGeometrysol} can also be extended to geometrically compute the elliptic paths traversed by the mixed CUQ states {\em within} the Bloch sphere. We follow the steps to construct the solution or orbit plane in Figure~\ref{fig: Plane construction CUQ} and extend it with additional steps, as illustrated in Figure~\ref{fig: Ellipse construction}. 

As in the previous subsection, we start with an initial mixed state $\mathbf{b}_0$ and eigenvectors~$\mathbf{b}_\pm$ [cf.~\eqref{eq:bplusminus}] of the CUQ Hamiltonian, as shown in Figure~\ref{mixstep0}. Similarly as in the case of pure CUQ states, we follow steps one and two in Figures~\ref{purestep1} and~\ref{purestep2} here for the mixed state~$\mathbf{b}_0$ by constructing the planes tangent to the endpoints of the eigenvectors $\mathbf{b}_\pm$ on the Bloch sphere, as illustrated in Figure~\ref{mixstep1}. Then, the plane passing through the intersecting line $\mathbf{I}$ of the tangent planes and the initial state $\mathbf{b}_0$ is our solution plane on which the orbit of the mixed state will lie, as depicted in Figure~\ref{mixstep2}. We now have to perform additional steps in our geometric construction to obtain the proper trajectory of our mixed state $\mathbf{b}_0$.

In the third step, shown in Figure~\ref{mixstep3}, we extend the end-point of our initial state Bloch vector $\mathbf{b}_0$ along the $\mathbf{e}$ direction until it touches the surface of the Bloch sphere. This new point on the sphere, which can be defined via a new auxiliary Bloch vector $\mathbf{\widetilde{b}}_0$, is a pure state. We can follow the steps of the pure state solution described in Section~\ref{PureGeometrysol} to obtain the orbit traversed by $\mathbf{\widetilde{b}}_0$. Thus, in our fourth step, we construct the solution plane for $\mathbf{\widetilde{b}}_0$. The solution plane of $\mathbf{\widetilde{b}}_0$ slices the Bloch sphere, forming a circular patch, the boundary of which is the Bloch trajectory of $\mathbf{\widetilde{b}}_0$, as shown in Figure~\ref{mixstep4}. 

In the final step, shown in Figure~\ref{mixstep5}, we project this circular orbit traversed by the pure auxiliary state $\mathbf{\widetilde{b}}_0$ back into the solution plane of $\mathbf{b}_0$ along the $\mathbf{e}$-direction. Since the plane is $\textit{tilted}$ at an angle with respect to the solution plane of $\mathbf{b}_0$, the projection of the circular path on it is an {\em ellipse}. This projected elliptical path is the proper CUQ trajectory of our initial mixed state $\mathbf{b}_0$ in the Bloch sphere. Thus, by computing the eigenvectors $\mathbf{b}_\pm$ for a value of $r$, we can geometrically construct the Bloch-sphere trajectory of any CUQ state $\mathbf{b}_0$. 

Our geometric construction also provides a simple way to compute the eccentricity $e$ of the elliptical orbit of a mixed state CUQ. Essentially, we note that the solution planes for the mixed state $\mathbf{b}$ and the extended point $\mathbf{\widetilde{b}}$ are at an angle, say $\alpha$. Since we project the circular orbit of~$\mathbf{\widetilde{b}}$ onto the solution plane of $\mathbf{b}$, the eccentricity $e$ of the elliptical orbit $\mathbf{b}(\tau)$ is
\begin{equation}
    e\: =\: \sin{\alpha}\, .
\end{equation}
We can algebraically compute the direction of the planes using the Frenet-Serret construction (as was done in \cite{Karamitros:2025azy}), and solve for the angle $\alpha$ in terms of $r$ and the initial state $\mathbf{b}_0$ (and its extended auxiliary pure CUQ state $\mathbf{\widetilde{b}}_0$) in polar form as defined in \eqref{initialCUQmixpolar}. This gives the eccentricity $e$ of the orbit of $\mathbf{b}(\tau)$ as
\begin{equation}
   \label{eq:eccentric}
    e\ =\ \dfrac{\big(1-r^2\big)r\sin{\delta_0}\:-\:b_er\big(1-r^2\big)^{1/2}\big(1-r\cos{\delta_0}\cos{\beta_0}\big)}{\sqrt{\big(1-r^2\cos^2{\delta_0}\big)\,\big[\big(1-r^2)^2+b_e^2 r^2\big(1-r\cos{\delta_0\cos{\beta_0}}\big)^2\big]}}\;.
\end{equation}
There is a good reason why this projection in Figure~\ref{mixstep5} works well. To understand it, we consider the master equation \eqref{nonherm_master_equation}. We study the evolution of $\mathbf{a}$ along the $\mathbf{e}$ component,
\begin{equation}    \dfrac{d}{dt} (\mathbf{a}\cdot\mathbf{e})\ =\: -\,\Gamma^0 (\mathbf{a}\cdot\mathbf{e})\,.\label{Mastereq_along_e}
\end{equation}
We compare it with the evolution of $\mathbf{a}$ along $\boldsymbol{\gamma}$ and $\mathbf{e}\times\boldsymbol{\gamma}$, where we have the differential equations
\begin{equation}
    \begin{aligned} \dfrac{d}{dt} (\mathbf{a}\cdot\boldsymbol{\gamma})\ &=\ 2|\mathbf{E}|\,\big[\mathbf{a} \cdot(\mathbf{e}\times\boldsymbol{\gamma})\big]\: -\: \Gamma^0 (\mathbf{a}\cdot\boldsymbol{\gamma})\: +\:|\mathbf{\Gamma}|\,a^0\,,\\
    \dfrac{d}{dt}\big[\mathbf{a}\cdot(\mathbf{e}\times\boldsymbol{\gamma})\big]\ &=\ -\,2|\mathbf{E}|\,(\mathbf{a}\cdot\boldsymbol{\gamma})\: -\: \Gamma^0\, \big[\mathbf{a}\cdot(\mathbf{e}\times\boldsymbol{\gamma})\big]\,.    
    \end{aligned}
\label{Mastereq_along_g_and_exg}
\end{equation}
We observe that \eqref{Mastereq_along_g_and_exg} is a coupled differential equation between the two components, $\mathbf{a}\cdot\boldsymbol{\gamma}$ and $\mathbf{a}\cdot(\mathbf{e}\times\boldsymbol{\gamma})$, whereas \eqref{Mastereq_along_e} is a simple linear differential equation, decoupled from the other two components. Thus, in the co-decaying frame, the component $\mathbf{b}\cdot\mathbf{e}$ evolves in an anharmonic, periodic fashion because of the $a^0$ term (which is the trace of the density matrix), or equivalently, the non-linear term in \eqref{CUQ_Mastereq}. This is what we have computed in~\eqref{General mix state sol CUQ}. Thus, projections of a CUQ trajectory along the $\mathbf{e}$-direction are valid solutions to the master equation \eqref{CUQ_Mastereq}.

\section{Conclusions}\label{sec:Concl}

Critical Unstable Qubits are unstable two-level quantum systems that exhibit remarkable phenomena in an appropriately defined co-decaying or trace-preserving frame [cf.~\eqref{rhoevol_codec}], such as indefinite anharmonic oscillations between two states and coherence-decoherence oscillations of mixed states~\cite{Karamitros:2022oew,Karamitros:2025azy}.
CUQs are described by a non-Hermitian Hamiltonian $\text{H}_{\rm eff}$ characterised by two conditions in the Pauli basis~\eqref{nonHerm_Hamiltonian_Paulibasis}: (i)~${\bf E}\perp {\bf \Gamma}$ and (ii)~$r\equiv |{\bf \Gamma}|/(2|{\bf E}|) < 1$. 
The time evolution of CUQs is best described by a Bloch vector ${\bf b}(t)$ defined in the co-decaying frame.

In this work, we have revisited and extended previous studies 
by analytically calculating the trajectories of the CUQ Bloch vector ${\bf b}(t)$ {\em in} and {\em on} the Bloch sphere. 
In particular, we find that the mixed CUQ states, much like the pure state CUQs, oscillate in the Bloch sphere an\-harmonically. In addition, the mixed CUQs evolve periodically in an elliptical path, which physically corresponds to coherence-decoherence oscillations of the mixed state. The degree of anharmonicity depends on the value of $r$, whereas the degree of coherence-decoherence oscillations depends on $r$ and the initial coherence of the CUQ ${\bf b}(0) = {\bf b}_0$. We contrast these properties of CUQs with the typical Rabi oscillations of ordinary stable qubits governed by a
Hermitian Hamiltonian in order to explicitly demonstrate these two major differences. 

Of equal importance is our geometric construction of CUQ trajectories in and on the Bloch sphere for the first time, for {\em any} values of $r<1$ and {\em any} initial CUQ coherence $0\le {\bf b}_0\le 1$. Our geometric construction requires only knowledge of the eigenvectors ${\bf b}_\pm$ of the non-Hermitian Hamiltonian $\text{H}_{\rm eff}$ and the value of ${\bf b}_0$. With the aid of these three vectors, we are able to appropriately define intersection planes that cut the Bloch sphere and project the orbit plane of the CUQs for both pure- and mixed-state CUQs. These geometric constructions are presented in Figures~\ref{fig: Plane construction CUQ} and~\ref{fig: Ellipse construction}, respectively. For a mixed CUQ, the geometric solution allows us to analytically compute the eccentricity $e$ of its elliptical path for {\em any} value of ${\bf b}_0$ in~\eqref{eq:eccentric}, which has not been determined in the existing literature. 

In Section~\ref{sec:Stationarypnts}, we have identified an {\em infinite} set of stationary points of CUQs in~\eqref{eq:stationary}, at which the states do not evolve in time in the co-decaying frame. In the original lab frame, these mixed CUQ states experience a universal exponential drop-off as functions of the zeroth component $a^0(t)$ of the Bloch four-vector $a^\mu(t)$. As a consequence, these mixed CUQs do {\em not} oscillate while decaying in the lab frame. In a cosmological setting, this feature could have profound implications for Particle Cosmology, especially for models of resonant baryogenesis (e.g.~see\cite{Karamitros:2023tqr}), where theoretical predictions of such scenarios with stationary initial conditions for the baryon asymmetry in the universe have not been adequately investigated thus far.

Another direction for further research would be to realise the set of solutions of a CUQ in a quantum computer simulation. This requires simulating a non-Hermitian Hamiltonian on a quantum computer, an area on which there have been quite a few recent advances\cite{Zhang:2020iqa,Dogra:2021omk,Koukoutsis:2024teg,Jebraeilli:2025zrm}. The usual approach for simulating a non-Hermitian Hamiltonian is to operate quantum gates on the main quantum state along with an {\em ancillary qubit}. The operations are such that both qubits evolve and entangle with each other. Then, partially tracing out the ancillary qubit simulates an open quantum system, the effective Hamiltonian of which is non-Hermitian. There have been mainly four broad approaches for simulating a non-Hermitian Hamiltonian in a quantum computer:  (i) Linear Combination of Unitaries~(LCU), (ii) Unitary Decomposition, (iii) Dilation, and (iv) Biorthogonal Representation\cite{Koukoutsis:2024teg}. The practical challenges of achieving this are beyond the scope of this paper. The references mentioned above are indicative, but they explain the methodologies of these simulations in detail. In some of these works, the CUQs have been independently simulated\cite{Zhang:2020iqa,Dogra:2021omk,Jebraeilli:2025zrm}, although the distinct properties of CUQs that we discussed here have not been explored in detail. A comparison of different methods to simulate a non-Hermitian Hamiltonian is discussed in \cite{Koukoutsis:2024teg}. The latter paper mainly discusses a novel way of simulating non-Hermitian systems using a biorthogonal representation of quantum mechanics\cite{Brody:2013axr}. The biorthogonal representation itself can be a useful approach for studying non-Hermitian systems in future works.

Our work provides a comprehensive solution to non-Hermitian two-level systems in the {\em critical regime} of unstable qubits. The novel features of anharmonic oscillations and coherence-decoherence oscillations of the CUQ Bloch vector result from working in a trace-preserving map or a co-decaying frame. In contrast, in the typical lab frame, as $t\rightarrow\infty$, then $\rho\rightarrow 0$ in a harmonic oscillatory fashion, like the usual Rabi oscillations, for an unstable qubit with~${r<1}$. However, the quantum coherence of a state via a Bloch-vector representation in the context of Quantum Information theory is understood only in the co-decaying frame where the trace of the density matrix is preserved to be unity. As demonstrated in~\cite{Karamitros:2025azy}, such a frame does have an actual physical meaning in particle physics and, as such, can become the basis for further exploration in other settings. 

Finally, it would be interesting to explore these critical regimes in $d$-level quantum systems, known as {\em qudits}. An analysis of {\em qutrits}, or three-level non-Hermitian systems, has been performed in\cite{Kowalski:2020vrx}. The~Bloch-sphere analysis has limitations for $d$-level quantum systems because the number of dimensions  of the sphere increases as $\mathcal{O}(d^2)$. Nevertheless, it would be interesting to see if our geometrical constructions can provide valuable insight into the higher dimensional solutions for $d$-level unstable systems.

\section*{Acknowledgements}
The authors thank Thomas McKelvey for his collaboration in the early stages of this project.
The work of SP is funded by the Faculty of Science and Engineering Bicentenary Scholarship from Manchester U., and AP's work is supported in part by the STFC research grant: ST/X00077X/1. 

\noindent

\newpage

\appendix
\renewcommand{\thesubsection}{\Alph{section}.\arabic{subsection}}
\renewcommand{\theequation}{\Alph{section}.\arabic{equation}}

\section{Analytic Results}\label{App:CUQgensol}

\subsection{Solution of the Master Equation}

The master equation \eqref{nonherm_master_equation} can be solved in the density matrix formalism for an initial state $\rho(0) \equiv \dfrac{1}{2}a_0^\mu\bar{\sigma}_\mu=\dfrac{1}{2}(\mathbb{1} + \mathbf{a}_0\cdot\boldsymbol{\sigma})$ by evolving it using \eqref{rhoevol}, with $a^0(0) = 1$ and ${\bf a}(0) = {\bf a}_0$. Then, the Bloch four-vector~$a^\mu$ can be computed using the identity $a^\mu(t) = \mathrm{Tr}[\rho(t) \sigma^\mu]$, which gives its components
\begin{eqnarray}
   \label{eq:a0t}
    a^0(t) \!&=&\! \frac{e^{-\Gamma_0 t}}{|\mathbf{H}_{\text{eff}}|\,|\mathbf{H}_{\text{eff}}|^*} 
    \bigg\{
    |\mathbf{H}_{\text{eff}}|^*|\mathbf{H}_{\text{eff}}| \cos (|\mathbf{H}_{\text{eff}}|^* t) \cos (|\mathbf{H}_{\text{eff}}| t) 
    \nonumber\\
    &&\
    +2 \, \text{Re} \Big[ i |\mathbf{H}_{\text{eff}}|^* \, (\mathbf{a}_0 \cdot \mathbf{H}_{\text{eff}}) \cos (|\mathbf{H}_{\text{eff}}|^* t) \sin (|\mathbf{H}_{\text{eff}}| t) \Big] \\
    &&\  +\Big[|\mathbf{E}|^2 (1 + r^2)+ \mathbf{a}_0 \cdot (\mathbf{E} \times \mathbf{\Gamma})\Big] \sin (|\mathbf{H}_{\text{eff}}|^* t)\sin (|\mathbf{H}_{\text{eff}}| t)
    \bigg\},
    \nonumber\\
   \label{eq:avect} 
    \mathbf{a}(t) \!&=&\! \frac{e^{-\Gamma^0 t}}{|\mathbf{H}_{\text{eff}}|\,|\mathbf{H}_{\text{eff}}|^*} 
    \bigg\{
    \mathbf{a}_0 |\mathbf{H}_{\text{eff}}|^*|\mathbf{H}_{\text{eff}}| \cos (|\mathbf{H}_{\text{eff}}|^* t) \cos (|\mathbf{H}_{\text{eff}}| t)
    \nonumber\\
    &&\ + 2 \, \text{Re} \Big[ |\mathbf{H}_{\text{eff}}|^* \, (i\mathbf{H}_{\text{eff}} + \mathbf{a}_0 \times \mathbf{H}_{\text{eff}}) \cos (|\mathbf{H}_{\text{eff}}|^* t) \sin (|\mathbf{H}_{\text{eff}}| t) \Big] \\
    &&\ +\frac{1}{2} \, \Big[(\mathbf{a}_0 \cdot \mathbf{\Gamma}) \, \mathbf{\Gamma} + 4 \, (\mathbf{a}_0 \cdot \mathbf{E}) \, \mathbf{E} - 2 |\mathbf{E}|^2(1+r^2) \, \mathbf{a}_0 - 2 \mathbf{E} \times \mathbf{\Gamma} \Big]  \sin (|\mathbf{H}_{\text{eff}}|^* t) \sin (|\mathbf{H}_{\text{eff}}| t)
    \bigg\},\nonumber
\end{eqnarray}
where the following definitions are used: 
\begin{equation}
|\mathbf{H}_{\text{eff}}| \equiv \sqrt{\mathbf{H}_{\text{eff}}\cdot \mathbf{H}_{\text{eff}}} \quad\text{ and } \quad|\mathbf{H}_{\text{eff}}|^* \equiv \sqrt{\mathbf{H}^*_{\text{eff}}\cdot \mathbf{H}^*_{\text{eff}}}\,,
\end{equation}
as well as the result,
\begin{equation}
\mathbf{H}_{\text{eff}}\cdot \mathbf{H}_{\text{eff}}^* = |\mathbf{E}|^2(1+r^2)\,. 
\end{equation}
We note that the expressions~\eqref{eq:a0t} and~\eqref{eq:avect} agree well with those presented before in~\cite{Karamitros:2022oew}, after some typographical errors have been corrected.  

\subsection{Unstable Qubits}

For completeness, we present here the general solution of \eqref{non_hermitian Hamiltonian} for the cases of an unstable qubit deviating from a CUQ. Analogous expressions can be found in~\cite{KOWALSKI2019167955}, but in a different coordinate frame. Unlike~\cite{KOWALSKI2019167955}, we also discuss the physics of these solutions in the context of unstable particles.

The eigenvalues of \eqref{non_hermitian Hamiltonian} are
\begin{equation}
    \lambda_{\pm}\:  =\ E_0-\frac{i}{2}\Gamma_0\: \pm\: |\mathbf{E}|\sqrt{1-r^2-2ir\mathbf{e}\cdot\boldsymbol{\gamma}}\label{gen_nonHerm_eigenvalues}\,,
\end{equation}
which, in the case of CUQ, reduces to \eqref{CUQeigenvalues}. For CUQs, the imaginary parts of the eigenvalues are degenerate, implying that the decay widths of the two eigenstates are equal.

\subsubsection{General Solution of a CUQ}

For a CUQ, we have
\begin{equation}
    \begin{aligned}
        |\mathbf{H}_{\text{eff}}|\ &=\ |\mathbf{H}_{\text{eff}}|^*\ =\ |\mathbf{E}|\sqrt{1-r^2}\, .        \end{aligned}
\end{equation}
For later convenience, we also define 
\begin{equation}
\theta(t)\: \equiv\: |\mathbf{E}|\sqrt{1-r^2}\,t\: \equiv\: \dfrac{\sqrt{1-r^2}}{2r}\,\tau\,.    
\end{equation}

For an arbitrary initially mixed CUQ state where $|\mathbf{b}(0)| \neq 0$, consider an initial condition
\begin{equation}
    \mathbf{b}(0)\: =\: \mathbf{b}_0\: =\: b_{x}\mathbf{e}\,+\, b_{y} \boldsymbol{\gamma}\, +\, b_{z} \mathbf{e}\times \boldsymbol{\gamma}\,,\label{initialcondition}
\end{equation}
with $0<|\mathbf{b}_0|<1$ and $\mathbf{b}_0\perp \mathbf{e}$.
By substituting~\eqref{initialcondition} into~\eqref{eq:a0t} and~\eqref{eq:avect}, the following result is obtained:
\begin{eqnarray}
        a^0(t) \!&=&\! \frac{e^{-\Gamma_0 t}}{1-r^2}\left[1-r^2\cos{2\theta + b_{y} r\sqrt{1-r^2}\sin{2\theta}}+b_{z}r(1-\cos{2\theta})\right],\\[3mm]
        \mathbf{a}(t) \!&=&\! \frac{e^{-\Gamma_0 t}}{1-r^2}\left[b_{x}(1-r^2)\mathbf{e}+\left(r\sqrt{1-r^2}\sin{2\theta}+ b_{y} (1-r^2)\cos{2\theta}+b_{z}\sqrt{1-r^2}\sin{2\theta}\right)\boldsymbol{\gamma} \right.\nonumber\\
        &&\quad\quad\quad\quad\quad -\left. \left(r(1-\cos{2\theta})+b_{y}\sqrt{1-r^2}\sin{2\theta}+ b_{z} (r^2-\cos{2\theta)}\right)\mathbf{e}\times\boldsymbol{\gamma}\right].
\end{eqnarray}
Employing the last two equations, we may conveniently write the co-decaying Bloch vector $\mathbf{b}(t) \equiv \mathbf{a}(t)/a^0(t)$ as
\begin{equation}
  \label{mixCUQ_sol} 
    \mathbf{b}(t)\: =\: \dfrac{A_e(t)\mathbf{e}+A_{\gamma}(t)\boldsymbol{\gamma}- A_{e\times\gamma}(t)\mathbf{e}\times\boldsymbol{\gamma}}{1-r^2\cos{2\theta + b_{y} r\sqrt{1-r^2}\sin{2\theta}}+b_{z}r(1-\cos{2\theta})}\;,
\end{equation}
where
\begin{equation}
\begin{aligned}    
    A_e(t)\ &\equiv\ b_x(1-r^2)\,,\\
    A_{\gamma}(t)\ &\equiv\ r\sqrt{1-r^2}\sin{2\theta}\,+\,b_{y}(1-r^2)\cos{2\theta}\,+\,b_{z}\sqrt{1-r^2}\sin{2\theta}\,,\\
    A_{e\times\gamma}(t)\ &\equiv\ r(1-\cos{2\theta})\,+\,b_{y}\sqrt{1-r^2}\sin{2\theta}\,+\,b_{z}(r^2-\cos{2\theta)}\,.
\end{aligned}
\end{equation}
Note that $A_e(t)$ is  constant in time and $\mathbf{b}\cdot\mathbf{e}$ evolves or oscillates with time, only because of the denominator, i.e.~$a^0(t)$. Thus, if we have $\mathbf{b}(0)\cdot\mathbf{e} = 0$, then the Bloch vector remains perpendicular to $\mathbf{e}$, i.e.~$\mathbf{b}(t)\cdot\mathbf{e}=0$ at all times. This is not true for the other components of the CUQ, which always evolve to be non-zero, as can be seen from the evolution of the maximally mixed state, i.e.~$\mathbf{b}(0)=\mathbf{0}$, which evolves in the $\{\mathbf{e},\mathbf{e}\times\boldsymbol{\gamma}\}$ plane. In this case, the Bloch vector will evolve with time as
\begin{equation}
    \begin{aligned}
        \mathbf{b}(t)\: =\: \frac{r}{1-r^2\cos{2\theta}}\left[\sqrt{1-r^2}\sin{2\theta}\boldsymbol{\gamma}\, -\, (1-\cos{2\theta})\mathbf{e}\times \boldsymbol{\gamma}\right]\,.
    \end{aligned}\label{max_mix_CUQ_sol}
\end{equation}

\subsubsection{$\mathbf{E}\perp\mathbf{\Gamma}$ and $r>1$: Damped CUQ}

For the CUQ case, but with $r>1$, the oscillating feature of CUQ is lost, as has been discussed in\cite{Karamitros:2022oew}. The eigenvalues of the Hamiltonian in this case are
\begin{equation}
    \lambda_{\pm}\: =\ E_0\: -\: \frac{i}{2}\left(\Gamma_0 \mp 2|\mathbf{E}|\sqrt{r^2-1}\right).
\end{equation}
The situation here flips compared to that in CUQ, where the two eigenstates are degenerate with equal energy eigenvalues but have different decay widths. The eigenstates of the effective Hamiltonian corresponding to these eigenvalues are
\begin{equation}
    \mathbf{b}_{\pm}\: =\ \pm\dfrac{\sqrt{r^2-1}}{r}\,\boldsymbol{\gamma}\:-\:\dfrac{1}{r}\,\mathbf{e}\times\boldsymbol{\gamma}\,.
\end{equation}
With different decay widths, the two-level system becomes overdamped and simply evolves to the state with a higher half-life, or smaller decay width, which here is the $\mathbf{b}_+$ state. This is also the asymptotic solution calculated in~(2.16) of\cite{Karamitros:2022oew}. In the lab frame, the evolution of the density matrix $\rho(t)$ displays no oscillations, as expected.

For the case of $r>1$, we can start with an initial state $\mathbf{b_0} = b_x\mathbf{e}+b_y\boldsymbol{\gamma}+b_z\mathbf{e}\times\boldsymbol{\gamma}$ and plug it into~\eqref{eq:a0t} and~\eqref{eq:avect}. In addition, we have
\begin{equation}
    |\mathbf{H}_{\text{eff}}|\: =\: |\mathbf{H}_{\text{eff}}|^*\: =\: i|\mathbf{E}|\sqrt{r^2-1}\,.
\end{equation}
Hence, the solution for the case $r>1$ is
\begin{eqnarray}
        a^0 \!&=&\! \dfrac{e^{-\Gamma_0 t}}{r^2-1}\left[r^2\cosh{2\theta_d}-1+b_yr\sqrt{r^2-1}\sinh{2\theta_d}+b_z r(\cosh{2\theta_d}-1)\right]
        \\[3mm]
        \mathbf{a} \!&=&\! \dfrac{e^{-\Gamma_0 t}}{r^2-1}\left[b_x(r^2-1)\mathbf{e}+\left(r\sqrt{r^2-1}\sinh{2\theta_d}+b_y(r^2-1)\cosh{2\theta_d} + b_z\sqrt{r^2-1}\sinh{2\theta_d}\right)\boldsymbol{\gamma}\right.\nonumber\\
        &&\quad\quad\quad-\left.\left(r(\cosh{2\theta_d}-1)+b_y\sqrt{r^2-1}\sinh{2\theta_d}+b_z(\cosh{2\theta_d}-r^2)\right)\mathbf{e}\times\boldsymbol{\gamma}\right],
\end{eqnarray}
where
\begin{equation}
    \theta_d(t)\: \equiv\: |\mathbf{\mathbf{E}}|\sqrt{r^2-1}\,t\,.
\end{equation}

\subsubsection{The General Case: $\mathbf{E}\not\perp\mathbf{\Gamma}$}

We now consider the general case of unstable two-level quantum systems, for which $\mathbf{E}\not\perp\mathbf{\Gamma}$. In this case,  the vectors $\mathbf{e}$ and $\boldsymbol{\gamma}$ are not orthogonal to each other, neither the mass nor decay eigenstates are degenerate. 

To start with, we define an angle $\theta_{e\gamma}$ between $\mathbf{e}$ and $\boldsymbol{\gamma}$ as $\cos{\theta_{e\gamma}} \equiv \mathbf{e}\cdot\boldsymbol{\gamma}$.
The eigenvalues of the effective Hamiltonian~\eqref{non_hermitian Hamiltonian} are
\begin{equation}
    \lambda_{\pm}\: =\ E_0\,\pm\,\dfrac{\sqrt{2}|\mathbf{E}|r\cos{\theta_{e\gamma}}}{\kappa^{1/2}}\: -\: \frac{i}{2}\left(\Gamma_0\,\pm\,\sqrt{2}\kappa^{1/2}|\mathbf{E}|\right),
\end{equation}
where $\kappa$ is defined as
\begin{equation}
    \kappa \: \equiv\: (1+r^4+2r^2\cos{2\theta_{e\gamma}})^{1/2}+r^2-1\,.
\end{equation}
The two eigenstates corresponding to the eigenvalues in terms of the Bloch vectors are
\begin{equation}
    \mathbf{b}_{\pm}\: =\: \mp\, \dfrac{\sqrt{2}r^2\cos{\theta_{e\gamma}}}{\kappa^{1/2}}\mathbf{n}_1\, +\, \dfrac{2r^2\cos^2{\theta_{e\gamma}}-\kappa}{r\kappa\sin{\theta_{e\gamma}}}\mathbf{n}_2\, \mp\, \dfrac{2r^2\cos^2{\theta_{e\gamma}}-\kappa}{\sqrt{2}r\kappa^{1/2}\sin{\theta_{e\gamma}}}\mathbf{n}_3\,,
\end{equation}
where $\mathbf{n}_1$, $\mathbf{n}_2$, and $\mathbf{n}_3$ are linearly independent and orthonormal basis vectors defined as
\begin{equation}
    \mathbf{n}_1\:\equiv\: \mathbf{e}\,,\qquad \mathbf{n}_2\: \equiv\: \dfrac{\mathbf{e}\times\boldsymbol{\gamma}}{\sin{\theta_{e\gamma}}}\;, \qquad \mathbf{n}_3\: \equiv\: \dfrac{\mathbf{e}\times(\mathbf{e}\times\boldsymbol{\gamma})}{\sin{\theta_{e\gamma}}}\;.
\end{equation}

The time evolution of the unstable qubit can again be computed using \eqref{eq:a0t} and~\eqref{eq:avect}, substituting it assuming an arbitrary initial state, $\mathbf{b}_0$. 
To facilitate our presentation of the analytic results, we define the quantities $u$ and $v$ through:
\begin{equation}
    \begin{aligned}
        |\mathbf{H}_{\text{eff}}|\: &=\:  |\mathbf{E}|\,\bigg(\dfrac{2r\cos{\theta_{e\gamma}}-i\kappa}{(2\kappa)^{1/2}}\bigg)\: \equiv\: |\mathbf{E}|\,(u-iv)\,,\\[2mm]
        |\mathbf{H}_{\text{eff}}|^*\: &=\: |\mathbf{E}|\,\bigg(\dfrac{2r\cos{\theta_{e\gamma}}+i\kappa}{(2\kappa)^{1/2}}\bigg)\: \equiv\: |\mathbf{E}|\,(u+iv)\,,
    \end{aligned}
\end{equation}
as well as the time-dependent angles $\theta_1(t)$ and $\theta_2(t)$,
\begin{equation}
    \theta_1(t)\: \equiv\: 2|\mathbf{E}|u\, t\,, \qquad \theta_2(t)\: \equiv\: 2|\mathbf{E}|v\,t\,.
\end{equation}
With the help of these quantities, the general solution $a^\mu$ for an unstable qubit is given by
\begin{eqnarray}
        a^0 &=& \dfrac{e^{-\Gamma_0 t}}{2z}\Big[\Big(\kappa-2r\mathbf{b}_0\cdot(\mathbf{e\times}\boldsymbol{\gamma})\Big)\cos{\theta_1} + \Big(2z-\kappa+2r\mathbf{b}_0\cdot(\mathbf{e\times}\boldsymbol{\gamma})\Big)\cosh{\theta_2}
        \nonumber\\
        &&\quad\quad\quad
        +\Big(ru\mathbf{b}_0\cdot \boldsymbol{\gamma}+v\mathbf{b}_0\cdot \boldsymbol{e}\Big)\sin{\theta_1}+\Big(ru\mathbf{b}_0\cdot \boldsymbol{\gamma}-v\mathbf{b}_0\cdot \boldsymbol{e}\Big)\sinh{\theta_2}\Big],\\[2mm]
        \mathbf{a} &=& \dfrac{e^{-\Gamma_0 t}}{2z}\Big[\Big((\kappa+2)\mathbf{b}_0+2r\mathbf{e}\times\boldsymbol{\gamma}-2(\mathbf{b}_0\cdot\mathbf{e})\mathbf{e}-2r(\mathbf{b}_0\cdot\boldsymbol{\gamma})\boldsymbol{\gamma}\Big)\cos{\theta_1}\nonumber\\
        &&\quad\quad\quad
        +\Big((\kappa+2r^2)\mathbf{b}_0-2r\mathbf{e}\times \boldsymbol{\gamma}+2(\mathbf{b}_0\cdot\mathbf{e})\mathbf{e}+2r(\mathbf{b}_0\cdot\boldsymbol{\gamma})\boldsymbol{\gamma}\Big)\cosh{\theta_2}\nonumber\\
        &&\quad\quad\quad
        +\Big(u\mathbf{b}_0\times\mathbf{e}-rv\mathbf{b}_0\times\boldsymbol{\gamma}\Big)\sin{\theta_1}+\Big(v\mathbf{b}_0\times\mathbf{e}+ ru\mathbf{b}_0\times\boldsymbol{\gamma}\Big)\sinh{\theta_2}\Big],
\end{eqnarray}
where $z = (1+r^4+2r^2\cos{2\theta_{e\gamma}})^{1/2}$.

\section{Hamiltonians for Critical Unstable Qudits}\label{App:CUQhamilton}

In this appendix, we discuss the form of the non-Hermitian Hamiltonians $\text{H}_{\rm eff}$
in~\eqref{non_hermitian Hamiltonian} for {\em critical unstable multi-level systems}, which we call Critical Unstable Qudits (CUQds). Here, our discussion goes beyond that of Appendix~A in\cite{Karamitros:2023tqr}.

\subsection{The CUQ Hamiltonian}

We consider first the non-Hermitian CUQ Hamiltonian, before embarking on the more general case of effective Hamiltonians describing CUQds.

To start with, we will show that a CUQ Hamiltonian can be represented as a product of two Hermitian matrices. To see this, consider two Hermitian matrices $\mathrm{H}_1$ and $\mathrm{H}_2$. In the Pauli basis, $\mathrm{H}_1$ and $\mathrm{H}_2$ may be conveniently expressed as
\begin{equation}
    \mathrm{H}_{1}\: =\: \mathrm{H}_{1\mu}\sigma^\mu\: =\: \mathrm{H}^0_1\mathbb{1} \,-\, \mathbf{H}_1\cdot\boldsymbol{\sigma}\,,\qquad \mathrm{H}_{2}\: =\: \mathrm{H}_{2\mu}\sigma^\mu\: =\: \mathrm{H}^0_2\mathbb{1}\, -\, \mathbf{H}_2\cdot\boldsymbol{\sigma}\,,
\end{equation}
with $\text{H}^\mu_{1,2} \in \mathbb{R}^4$. 
If these two matrices are non-commuting, i.e.~$[\mathrm{H}_1,\mathrm{H}_2]\neq 0$, then their product is a non-Hermitian matrix, 
\begin{equation}
    \widetilde{\mathrm{H}}_{\text{eff}}\: \equiv\: \mathrm{H}_1\mathrm{H}_2\: \ne\: \widetilde{\mathrm{H}}_{\text{eff}}^\dagger\,.
\end{equation}
The non-Hermitian matrix can be written in the Pauli basis as
\begin{equation}
    \widetilde{\mathrm{H}}_{\text{eff}}\: =\: (\mathrm{H}_1^0\mathrm{H}_2^0+\mathbf{H}_1\cdot\mathbf{H}_2)\,\mathbb{1}\,-\,(\mathrm{H}_1^0\mathbf{H}_2+\mathrm{H}_2^0\mathbf{H}_1)\cdot\boldsymbol{\sigma}\, +\,i(\mathbf{H}_1\times \mathbf{H}_2)\cdot\boldsymbol{\sigma}\, .
\end{equation}
Therefore, $\widetilde{\mathrm{H}}_{\text{eff}}$ can effectively be written in the following form:
\begin{equation}
    \widetilde{\mathrm{H}}_{\text{eff}}\: \equiv\:  \mathrm{E}^0\mathbb{1} -\left(\mathbf{E}-\dfrac{i}{2}\boldsymbol{\Gamma}\right)\cdot\boldsymbol{\sigma},\label{Heffnewform}
\end{equation}
where
\begin{equation}
   \label{HeffprodH1H2}
        \mathbf{E}\,\equiv\, \mathrm{H}_1^0\mathbf{H}_2+\mathrm{H}_2^0\mathbf{H}_1\,,\qquad
        \boldsymbol{\Gamma}\,\equiv\,2\mathbf{H}_1\times \mathbf{H}_2\,,\qquad
        \text{E}^0\,\equiv\,\mathrm{H}_1^0\mathrm{H}_2^0 + \mathbf{H}_1\cdot \mathbf{H}_2\,.
\end{equation}
We observe that the effective CUQ Hamiltonian~\eqref{nonHerm_Hamiltonian_Paulibasis} given by the WW approximation for unstable particles is related to $\widetilde{\mathrm{H}}_{\text{eff}}$ as
\begin{equation}
    \mathrm{H}_{\text{eff}}\: =\: \widetilde{\mathrm{H}}_{\text{eff}}\,-\, \dfrac{i}{2}\Gamma^0\mathbb{1}\,.
\end{equation}
It is easy to verify that $\mathbf{E}\cdot\boldsymbol{\Gamma} =0$ in \eqref{HeffprodH1H2}. Thus, writing an effective Hamiltonian as a product of two Hermitian matrices trivially satisfies the first of the two required conditions for a CUQ stated in~\eqref{CUQcondition}. 

From the above reformulation, we may now identify the value of $r$ as
\begin{equation}
    r = \dfrac{|\boldsymbol{\Gamma}|}{2|\mathbf{E}|} = \left(\dfrac{|\mathbf{H}_1|^2|\mathbf{H}_2|^2 - (\mathbf{H}_1\cdot\mathbf{H}_2)^2}{(\mathrm{H}_1^0)^2|\mathbf{H}_2|^2+(\mathrm{H}_2^0)^2|\mathbf{H}_1|^2+2\mathrm{H}_1^0\mathrm{H}_2^0(\mathbf{H}_1\cdot\mathbf{H}_2)}\right)^{1/2}.\label{prod2cuqr}
\end{equation}
Therefore, we know the form of $\mathrm{H}_{\text{eff}}$ for a choice of Hermitian matrices $\mathrm{H}_1$ and $\mathrm{H}_2$. We now try to see what the form of $\mathrm{H}_1$ and $\mathrm{H}_2$ can be for a given effective non-Hermitian Hamiltonian. Of~course, the number of solutions for a matrix as a product of two other matrices is infinite. We can see this from~\eqref{HeffprodH1H2}, where we have three equations to solve for $\mathbf{H}_1$, $\mathbf{H}_2$, $\mathrm{H}_1^0$, and $\mathrm{H}_2^0$. From~\eqref{HeffprodH1H2}, we also see that $\mathbf{H}_1\cdot\boldsymbol{\Gamma} = \mathbf{H}_2\cdot\boldsymbol{\Gamma} =0$. So, $\mathbf{H}_1$ and $\mathbf{H}_2$ lie in the $\{\mathbf{e},\mathbf{e}\times\boldsymbol{\gamma}\}$-plane and can be written in the form:
\begin{equation}
    \mathbf{H}_{1,2}\: \equiv\: \mathrm{H}_{1,2}^\mathrm{e}\mathbf{e}\, +\, \mathrm{H}_{1,2}^{\mathrm{e}\gamma}\mathbf{e}\times\boldsymbol{\gamma}\,.
\end{equation}
Then, the equations in~\eqref{HeffprodH1H2} can be rewritten as follows:
\begin{subequations}
  \label{CUQprodconstreqn}
    \begin{align}
    \mathrm{H}_2^0\mathrm{H}_{1}^{\mathrm{e}\gamma}\ &=\ -\mathrm{H}_1^0\mathrm{H}_{2}^{\mathrm{e}\gamma}\label{CUQprodconstr1},\\
    \mathrm{H}_2^0\mathrm{H}_{1}^{\mathrm{e} }+\mathrm{H}_1^0 \mathrm{H}_{2}^{\mathrm{e}}\ &=\ |\mathbf{E}|,\label{CUQprodconstr2}\\
    \mathrm{H}_{1}^{\mathrm{e}}\mathrm{H}_{2}^{\mathrm{e}\gamma}-\mathrm{H}_{2}^{\mathrm{e}}\mathrm{H}_{1}^{\mathrm{e}\gamma}\ &=\ \dfrac{|\boldsymbol{\Gamma}|}{2},\label{CUQprodconstr3}\\
    \mathrm{H}_1^0\mathrm{H}_2^0 + \mathrm{H}_1^{\mathrm{e}}\mathrm{H}_2^{\mathrm{e}}+ \mathrm{H}_1^{\mathrm{e}\gamma}\mathrm{H}_2^{\mathrm{e}\gamma}\ &=\ \mathrm{E}^0\,.\label{CUQprodconstr4}
    \end{align}    
\end{subequations}
In this way, we find that there are extra degrees of freedom with four equations and six unknowns that can be constrained by making choices for two constraints. A simple constraint that can be applied to obtain a solution is
\begin{equation}
    \mathrm{H}_2^0\: =\: 0 \quad (\text{or}\quad \mathrm{H}_1^0\: =\: 0),\label{CUQprodchoice1}
\end{equation}
which, when applied to \eqref{CUQprodconstreqn} with an additional choice of $\mathrm{H}_2^\mathrm{e}=1$, gives the following two Hermitian matrices as our solution:
\begin{equation}
    \begin{aligned}
        \mathrm{H}_1\ &=\ |\mathbf{E}|\,\mathbb{1}\: -\:\left(\mathrm{E}^0\mathbf{e}\,-\,\frac{|\boldsymbol{\Gamma}|}{2}\mathbf{e}\times\boldsymbol{\gamma}\right)\cdot\boldsymbol{\sigma}\\
        \mathrm{H}_2\ &=\ -\mathbf{e}\cdot\boldsymbol{\sigma}. 
    \end{aligned}\label{CUQprodchoiceres1}
\end{equation}
Another set of constraints that can be applied to obtain a solution is
\begin{equation}
    \mathrm{H}_1^e = 0\quad (\text{or} \quad \mathrm{H}_2^e = 0). \label{CUQprodchoice2}
\end{equation}
With an additional constraining choice of $\mathrm{H}_1^0 = 1$, we obtain the following two Hermitian matrices as the solution:
\begin{equation}
    \begin{aligned}
        \mathrm{H}_1\ &=\ \mathbb{1}\, +\, r(\mathbf{e}\times\boldsymbol{\gamma})\cdot\boldsymbol{\sigma}\,,\\
        \mathrm{H}_2\ &=\ \frac{\mathrm{E}^0}{1-r^2}\,-\,\left(\mathbf{E}+\frac{\mathrm{E^0}r}{1-r^2}\mathbf{e}\times\boldsymbol{\gamma}\right)\cdot\boldsymbol{\sigma}\,. 
    \end{aligned} \label{CUQprodchoiceres2}
\end{equation}
Note that the matrix $\mathrm{H}_1$ is positive definite when $r< 1$ [cf.~\eqref{CUQcondition}], yielding a non-singular $\mathrm{H}_2$. Hence,  $\widetilde{\mathrm{H}}_{\text{eff}}$ qualifies as a CUQ effective Hamiltonian. This construction of a CUQ Hamiltonian is related to an important theorem which will be presented in the next subsection.

There is another set of solutions obtained by applying the simple symmetric constraints:
\begin{equation}
    \mathrm{H}_1^0\: =\: \mathrm{H}_2^0 \qquad \text{and}\qquad \mathbf{H}_1\cdot\mathbf{H}_2\: =\: 0\label{constraintCUQham}\,.
\end{equation}
The second constraint fixes the degree of freedom in the angle between $\mathbf{H}_1$ and $\mathbf{H}_2$, i.e.~$\mathbf{H}_1\perp \mathbf{H}_2$. With these two constraints, we can identify two Hermitian matrices whose product gives the non-Hermitian matrix~\eqref{Heffnewform}. In detail, we have for the zeroth component of their Pauli four-vectors
\begin{equation}
    \mathrm{H}_1^0\: =\: \mathrm{H}_2^0\: =\: (\mathrm{E}^0)^{1/2},
\end{equation}
and for the corresponding three-vectors 
\begin{equation}
    \begin{aligned}
        \mathbf{H}_1\ &=\ \dfrac{\epsilon\pm\sqrt{\epsilon^2 - 4\mathrm{E}^0r^2}}{2}\,\mathbf{e}\: +\: r (\mathrm{E}^0)^{1/2}\,\mathbf{e}\times\boldsymbol{\gamma}\,,\\
        \mathbf{H}_2\ &=\ \dfrac{\epsilon\mp\sqrt{\epsilon^2 - 4\mathrm{E}^0r^2}}{2}\,\mathbf{e}\: -\: r(\mathrm{E}^0)^{1/2}\,\mathbf{e}\times\boldsymbol{\gamma}\, .
    \end{aligned}\label{hvectorpairsol}
\end{equation}
Here we have introduced the dimensional variable
\begin{equation}
    \epsilon\: \equiv\: \dfrac{|\mathbf{E}|}{(\mathrm{E^0})^{1/2}}\;.
\end{equation}
Thus, after applying all constraints, we still have two solutions for the pair of vectors $\mathbf{H}_{1,2}$ for the same vectors $\mathbf{E}$ and $\boldsymbol{\Gamma}$, given by~\eqref{hvectorpairsol}. Note that the solutions~\eqref{hvectorpairsol} exist only if we have $\epsilon^2 \geq4\mathrm{E}^0r^2$, which leads to real three-vectors. The latter constraint
implies
\begin{equation}
r\: \le\: \dfrac{|\mathbf{E}|}{2\mathrm{E}^0}\;.\label{CUQcontraintonr}
\end{equation}
However, for a proper CUQ that obeys the restrictions~\eqref{CUQcondition}, we must have $r<1$. This in turn translates into the constraint: $|\mathbf{E}|/(2\mathrm{E}^0) < 1$. In addition, 
by squaring the last inequality, we may equivalently derive a lower bound on the four-vector $\mathrm{E}^\mu$ to obtain a valid CUQ:
\begin{equation}
    \mathrm{E}^\mu \mathrm{E}_\mu\: >\: -3(\mathrm{E}^0)^2\,.\label{Ebound CUQ}
\end{equation}
For this symmetric set of constraints~\eqref{constraintCUQham}, a time-like or light-like $\mathrm{E}^\mu$ will always describe a CUQ. Thus, if an effective non-Hermitian Hamiltonian $\mathrm{H}_{\text{eff}}$ of the form \eqref{Heffnewform} can be written as a product of two Hermitian Hamiltonians $\mathrm{H}_1$ and $\mathrm{H}_2$ such that the four-vectors of the Hamiltonians in the Pauli basis satisfy the constraints in \eqref{constraintCUQham}, then $\mathrm{H}_{\text{eff}}$ describes a CUQ provided that $\mathrm{E}^\mu$ satisfies the inequality~\eqref{Ebound CUQ}.

\subsection{The CUQd Hamiltonian}

The above procedure for constructing CUQs becomes extraordinarily more complex when applied to critical unstable $d$-level quantum systems, called CUQds in the beginning of this appendix.
Nevertheless, we have learned from the previous subsection that good candidates for CUQd Hamiltonians can result from the product of two $d\times d$-dimensional Hermitian matrices, $\mathrm{H}_1$ and $\mathrm{H}_2$, where $d\ge 2$. We should clarify that a CUQd Hamiltonian should describe a non-trivial mixing among the $d$ states of the unstable qudit, while all 
$d$ decay widths of the qudit states must be equal:~$\Gamma_1 = \Gamma_2 = \cdots = \Gamma_d \equiv \Gamma_0$.

Before we proceed as outlined above, it is important to state a theorem\cite{Carlson1965OnRE} that allows us to properly develop our approach to constructing CUQd Hamiltonians.
\begin{theorem}
    \begin{shaded}
    Let $\mathrm{A}$ be a complex matrix such that all its eigenvalues are real. Then, there exist two Hermitian matrices $\mathrm{J}$ and $\mathrm{K}$, with $\mathrm{J}$ positive definite and $\mathrm{K}$ nonsingular, for which~$\mathrm{A = J\,K}$.
\end{shaded}\label{theorem1}
\end{theorem}
For example, the choice of Hermitian matrices in~\eqref{CUQprodchoiceres2} that gives a CUQ as a product of two matrices verifies the theorem for $2\times2$ matrices, with $\mathrm{J} = \mathrm{H}_1$ and $\mathrm{K} = \mathrm{H}_2$.

We may now ask the opposite question. {\em Under what conditions does the product of two Hermitian matrices give rise to a CUQd Hamiltonian?} For a two-level system, we know 
from~\eqref{prod2cuqr} that the only additional condition needed to be satisfied for a CUQ is
\begin{equation}
    r\: =\: \dfrac{\sin^2{\theta_{12}}}{x_1^2 + x_2^2 + 2x_1x_2 \cos{\theta_{12}}}\: <\: 1\,,\label{CUQcondprod}
\end{equation}
where $x_{1,2}\equiv \dfrac{\mathrm{H^0_{1,2}}}{|\mathbf{H}_{1,2}|}$ and $\theta_{12} \equiv \cos^{-1}\left({\dfrac{\mathbf{H}_{1}\cdot \mathbf{H}_{2}}{|\mathbf{H}_{1}||\mathbf{H}_{2}|}}\right)$. 

The theorem stated in~\ref{theorem1} implies a useful corollary, which addresses the question posed above.
\begin{corollary}
    \begin{shaded}
    The eigenvalues of the product of two Hermitian matrices are real if at least one of the matrices is positive definite\cite{Carlson1965OnRE}. 
\end{shaded}\label{corollary1}
\end{corollary}
Notice that for our considerations, the real eigenvalues translate to a CUQd system. It is easy to show that for a two-level system, the inequality condition \eqref{CUQcondprod} is satisfied for at least one positive definite matrix. Without loss of generality, let us assume that $\mathrm{H}_1$ is a positive definite matrix. For this to happen, we must have $x_1>1$. This inequality  successively implies
\begin{align}
r\: =\: \dfrac{\sin^2{\theta_{12}}}{x_1^2 + x_2^2 + 2x_1x_2 \cos{\theta_{12}}}\: <\: \dfrac{\sin^2{\theta_{12}}}{1 + x_2^2 + 2x_2 \cos{\theta_{12}}}\: \leq\: \dfrac{\sin^2{\theta_{12}}}{1 - \cos^2{\theta_{12}}}\: =\: 1\,,
\end{align}
as should be for a CUQ Hamiltonian on account of~\eqref{CUQcondition}.

The corollary is known in linear algebra and can be proved for any $d\times d$ matrix whose proof is repeated here. Without loss of generality, consider the product of two $d\times d$ Hermitian matrices $\mathrm{H_1}$ and $\mathrm{H_2}$, where $\mathrm{H_1}$ is positive definite. Let the eigenvalues of the product $\mathrm{H_1H_2}$ be collectively denoted by $\lambda$. Then,
we have
\begin{equation}
    \text{det}(\mathrm{H}_1\mathrm{H}_2 - \lambda\mathbb{1}) \:=\: 0\,.
\end{equation}
Since $\mathrm{H_1}$ is positive definite, we can evaluate the square root A of this matrix such that $\mathrm{A^2 = H_1}$. The matrix $\mathrm{H}_1$, being Hermitian, can be represented in a diagonalized form: $\mathrm{H_1} = U^\dagger \mathrm{D} U$, where $\mathrm{D}$ is a diagonal matrix with positive eigenvalues of $\mathrm{H}_1$ as its diagonal elements and $U$ is a unitary matrix. In this representation, since $\mathrm{H}_1$ is positive definite, its square root has a unique form: $\mathrm{A} = U^\dagger \sqrt{\mathrm{D}}\,U$, where $\sqrt{\mathrm{D}}$ is a positive diagonal matrix. According to this definition, the square-root matrix $\text{A}$ is positive definite and Hermitian as well.
Using judiciously the known property for the product of two determinants, we may successively take the following steps:
\begin{equation}
    \begin{aligned}
      \text{det}(\mathrm{H_1}\mathrm{H}_2 - \lambda\mathbb{1})\ &=\  \text{det}(\mathrm{A}^2\mathrm{H}_2 - \lambda\mathbb{1})\ =\ \text{det}(\mathrm{A})\:\text{det}(\mathrm{A}\mathrm{H}_2 - \lambda\mathrm{A}^{-1})\\
      &=\        \text{det}(\mathrm{A}\mathrm{H}_2 - \lambda\mathrm{A}^{-1})\:\text{det}(\mathrm{A})\ =\ \text{det}(\mathrm{A}\mathrm{H}_2\mathrm{A} - \lambda\mathbb{1})\ =\ 0\, .
    \end{aligned}
\end{equation}
Therefore, the eigenvalues of $\mathrm{H}_1\mathrm{H}_2$ and $\mathrm{AH_2A}$ are equal. Since $\mathrm{A}^\dagger = \mathrm{A}$, we easily deduce that $\mathrm{(AH_2A)}^\dagger = \mathrm{AH_2A}$, i.e.~the product  is a Hermitian matrix. Consequently, the eigenvalues $\lambda$ are all real, thus proving the corollary~\ref{corollary1}.

With the help of theorem~\ref{theorem1}, it is therefore straightforward to construct CUQd Hamiltonians for $d$-level systems. We should only warn the reader that the theorem is a \textit{sufficient} but \textit{not a necessary} condition for constructing a CUQ Hamiltonian. An example of this is the Hermitian matrices
\begin{equation}
\mathrm{H}_1\: = \:\begin{pmatrix}
  6 & 5+7i \\
  5-7i & -1
\end{pmatrix},\qquad
\mathrm{H}_2 \:=\: \begin{pmatrix}
  1 & 4+6i \\
  4-6i & -3
\end{pmatrix}.
\end{equation}
Neither of the two matrices is positive definite, but their product 
\begin{equation}
\widetilde{\mathrm{H}}_{\text{eff}} \:=\: \mathrm{H}_1\mathrm{H}_2 \:=\: \begin{pmatrix}
  68-2i & 9+15i \\
  1-i & 65 +2i
\end{pmatrix} 
\end{equation}
has positive eigenvalues and thus qualifies as a CUQ Hamiltonian. This indicates that there is a set of matrices that can be missed if we simply try to construct a CUQ Hamiltonian by taking the product of two Hermitian matrices, where at least one is positive definite. However, we also note that the theorem \ref{theorem1} tells us that there exists a product with at least one positive definite matrix, and that product is given by the matrices in \eqref{CUQprodchoiceres2}. The theorem, although limited,  does provide a direction for constructing CUQd Hamiltonians that can be extremely helpful for arbitrary $d$-level systems. In such systems, we have an increasing ladder of constraints for $d>2$ that need to be satisfied in order to obtain a CUQd Hamiltonian (cf.~Appendix A of \cite{Karamitros:2023tqr}).

\newpage
\bibliography{bibs-refs}{}
\bibliographystyle{JHEP}

\end{document}

%% file: macros.tex
\usepackage{ifthen}
\usepackage{tikz}
\usepackage{xspace}
\usepackage{physics,bm}

\newcounter{NumArgs}

\newcommand{\eqs}[1]{\setcounter{NumArgs}{0}\foreach\i in{#1}{\stepcounter{NumArgs}}%
	\ifthenelse{\equal{\theNumArgs}{1}}{(\ref{#1})}%
	{\ifthenelse{\equal{\theNumArgs}{2}}%
		{\foreach\i[count=\q]in{#1}{\ifthenelse{\equal{\q}{\theNumArgs}}{and (\ref{\i})}{(\ref{\i})~}}}%
		{\foreach\i[count=\q]in{#1}{\ifthenelse{\equal{\q}{\theNumArgs}}{and (\ref{\i})}{(\ref{\i}),~}}}}}

\newcommand{\refs}[1]{\setcounter{NumArgs}{0}\foreach\i in{#1}{\stepcounter{NumArgs}}%
	\ifthenelse{\equal{\theNumArgs}{1}}{(\ref{#1})}%
	{\ifthenelse{\equal{\theNumArgs}{2}}%
		{\foreach\i[count=\q]in{#1}{\ifthenelse{\equal{\q}{\theNumArgs}}{and (\ref{\i})}{(\ref{\i})~}}}%
		{\foreach\i[count=\q]in{#1}{\ifthenelse{\equal{\q}{\theNumArgs}}{and (\ref{\i})}{(\ref{\i}),~}}}}}

\newcommand{\Figs}[1]{\setcounter{NumArgs}{0}\foreach\i in{#1}{\stepcounter{NumArgs}}%
	\ifthenelse{\equal{\theNumArgs}{1}}{Figure~\ref{#1}}%
	{\ifthenelse{\equal{\theNumArgs}{2}}%
		{Figures~\foreach\i[count=\q]in{#1}{\ifthenelse{\equal{\q}{\theNumArgs}}{and \ref{\i}}{\ref{\i}~}}}%
		{Figures~\foreach\i[count=\q]in{#1}{\ifthenelse{\equal{\q}{\theNumArgs}}{and \ref{\i}}{\ref{\i},~}}}}}